\documentclass[a4paper,11pt]{article}
\pdfoutput=1 

\usepackage{amssymb}
\usepackage{dsfont}
\usepackage[T1]{fontenc} 
\usepackage{physics}
\usepackage{comment}
\usepackage{amsmath}
\usepackage{graphicx}
\usepackage{array, makecell}
\usepackage{float}
\setcellgapes{3pt}
\usepackage{xcolor}
\usepackage{amssymb}
\usepackage[normalem]{ulem}
\usepackage[numbers,sort&compress]{natbib}
\usepackage{geometry}
\usepackage{pdflscape}
\usepackage{mathtools}
\usepackage{multirow}
\usepackage{caption}
\usepackage{subcaption}
\usepackage{setspace}
\usepackage{ulem}

\definecolor{azure}{rgb}{0.0, 0.5, 1.0}
\definecolor{darkblue}{rgb}{0.15,0.35,0.7}
\definecolor{reddish}{rgb}{0.65, 0.2, 0.2}
\definecolor{brandeisblue}{rgb}{0.0, 0.44, 1.0}
\definecolor{ceruleanblue}{rgb}{0.16, 0.32, 0.75}
\definecolor{indigo(dye)}{rgb}{0.0, 0.25, 0.42}
\definecolor{dgrey}{rgb}{0.3,0.3,0.3}
\definecolor{grey}{rgb}{0.9,0.9,0.9}

\usepackage[linktocpage=true]{hyperref}
\hypersetup{
colorlinks=true,
citecolor=ceruleanblue,
linkcolor=ceruleanblue,
urlcolor=ceruleanblue,
pdfauthor={},
pdftitle={},
pdfsubject={}
}

\usepackage{cleveref}
\usepackage{bm}
\crefname{lem}{lemma}{lemmas}
\crefname{thm}{theorem}{theorems}
\crefname{cor}{corollary}{corollaries}
\crefname{rem}{remark}{remarks}
\crefname{prop}{proposition}{propositions}

\usepackage{colortbl}
\definecolor{dgreen}{rgb}{0, 0.55, 0}
\definecolor{llightyellow}{rgb}{1.0, 0.95, 0.7}
\definecolor{llightblue}{rgb}{0.7, 0.9, 1.0}
\definecolor{llightpink}{rgb}{1.0, 0.85, 0.95}
\definecolor{llightgreen}{rgb}{0.7, 1.0, 0.4}
\colorlet{lightyellow}{llightyellow!50!white}
\colorlet{lightblue}{llightblue!50!white}
\colorlet{lightgreen}{llightgreen!50!white}
\colorlet{lightpink}{llightpink!50!white}

\usepackage[framemethod=TikZ]{mdframed} 
\usepackage{tikz-cd} 
\usetikzlibrary{arrows,snakes,shapes.arrows,decorations.markings}
     \tikzset{>=triangle 90}
     \tikzstyle{bbc}=[draw,circle,fill=black,scale=.75]
     \tikzstyle{rc}=[circle,fill=red,scale=.6]
     \tikzstyle{wc}=[draw,circle,scale=.75]

\usetikzlibrary{cd}

\tikzset{snake it/.style={decorate, decoration=snake}}

\tikzset{
	on each segment/.style={
		decorate,
		decoration={
			show path construction,
			moveto code={},
			lineto code={
				\path [#1]
				(\tikzinputsegmentfirst) -- (\tikzinputsegmentlast);
			},
			curveto code={
				\path [#1] (\tikzinputsegmentfirst)
				.. controls
				(\tikzinputsegmentsupporta) and (\tikzinputsegmentsupportb)
				..
				(\tikzinputsegmentlast);
			},
			closepath code={
				\path [#1]
				(\tikzinputsegmentfirst) -- (\tikzinputsegmentlast);
			},
		},
	},
	mid arrow/.style={postaction={decorate,decoration={
				markings,
				mark=at position .5 with {\arrow[#1]{stealth}}
	}}},
}

\usetikzlibrary{automata,positioning}
\usetikzlibrary{decorations.markings}

\tikzset{line/.style={line width=0.25mm},
curve/.style={line,smooth,tension=1},
->-/.style={decoration={
  markings,
  mark=at position #1 with {\arrow[>=stealth]{>}}},postaction={decorate}},
-<-/.style={decoration={
  markings,
  mark=at position #1 with {\arrow[>=stealth]{<}}},postaction={decorate}},
}

\tikzset{bg/.style={opacity=.5}}

\tikzset{
    partial ellipse/.style args={#1:#2:#3}{
        insert path={+ (#1:#3) arc (#1:#2:#3)}
    }
}

\makeatletter
\renewcommand\section{\@startsection {section}{1}{\z@}%
                               {-3.5ex \@plus -1ex \@minus -.2ex}
                               {2.3ex \@plus.2ex}%
                               {\normalfont\large\bfseries}}
\renewcommand\subsection{\@startsection{subsection}{2}{\z@}%
                                 {-3.25ex\@plus -1ex \@minus -.2ex}%
                                 {1.5ex \@plus .2ex}%
                                 {\normalfont\bfseries}}
\makeatother

\let\non\nonumber

\newfont{\goth}{ygoth.tfm scaled 1200}                   

\numberwithin{equation}{section}

\newcommand{\be}{\begin{equation}}
\newcommand{\ee}{\end{equation}}
\newcommand{\bee}{\begin{equation} \begin{aligned}}
\newcommand{\eee}{\end{aligned} \end{equation}}

\newcommand{\CA}{\mathcal{A}}
\newcommand{\CC}{\mathcal{C}}
\newcommand{\CD}{\mathcal{D}}

\newcommand{\CN}{\mathcal{N}}

\newcommand\doubleC{\mathbb{C}}

\newcommand\doubleZ{\mathbb{Z}}

\newcommand{\VEC}{\mathrm{Vec}}

\newcommand\TYpo{\mathrm{TY}_{p,o}}
\newcommand\TYpd{\mathrm{TY}_{p,d}}

\newcommand{\dsi}{\mathds{1}}
\newcommand{\IZ}{\mathbb{Z}}

\newcommand{\ii}{\mathsf{i}}

\newcommand\Rep{\mathrm{Rep}}
\newcommand\TY{\mathrm{TY}}

\newcommand{\figref}[1]{Figure \ref{#1}}

\newcommand{\appref}[1]{Appendix\,\ref{#1}}

\newcommand{\Aut}{\operatorname{Aut}}

\newcommand\SPT{\operatorname{SPT}}

\newcommand{\tX}{\widetilde{X}}
\newcommand{\tZ}{\widetilde{Z}}
\newcommand{\tC}{\widetilde{C}}
\newcommand{\wt}[1]{\widetilde{#1}}
\newcommand{\wh}[1]{\widehat{#1}}

\newcommand{\kw}{\operatorname{KW}}
\newcommand{\tCC}{\widetilde{\mathcal{C}}}

\begin{document} 

\begin{titlepage}
\begin{center}

\hfill         \phantom{xxx}  

\vskip 2 cm {\Large \bf Intrinsic NISPT Phases, igNISPT Phases, and Mixed Anomalies of Non-Invertible Symmetries} 

\vskip 1.25 cm {\bf Da-Chuan Lu${}^{1,2}$ ~, Zhengdi Sun${}^{3}$}\non 

\vskip 0.2 cm
{\it ${}^{1}$ Department of Physics, Harvard University, Cambridge, MA 02138, USA}
\vskip 0.2 cm
{\it ${}^{2}$ Department of Physics and Center for Theory of Quantum Matter, University of Colorado, Boulder, Colorado 80309, USA}
\vskip 0.2 cm
{\it ${}^{3}$ Mani L. Bhaumik Institute for Theoretical Physics, Department of Physics and Astronomy, University of California Los Angeles, CA 90095, USA}

\end{center}
\vskip 1.5 cm

\begin{abstract}
\noindent A bosonic non-invertible Symmetry Protected Topological (NISPT) phase in (1+1)-dim is referred to as \textit{intrinsic} if it cannot be mapped, under discrete gauging, to a gapped phase with any invertible symmetry, that is, if it is protected by a non-group-theoretical fusion category symmetry. We construct the intrinsic NISPT phases by performing discrete gauging in a partial SSB phase with a fusion category symmetry that has a certain mixed anomaly. Sometimes, the anomaly of that symmetry category can be alternatively understood as a self-anomaly of a proper categorical sub-symmetry; when this is the case, the same gauging provides an anomaly resolution of this anomalous categorical sub-symmetry. This allows us to construct intrinsic gapless SPT (igSPT) phases, where the anomalous faithfully acting symmetry is non-invertible; and we refer to such igSPT phases as igNISPT phases. We provide two concrete lattice models realizing an intrinsic NISPT phase and an igNISPT phase, respectively. We also generalize the construction of intrinsic NISPT phases to (3+1)-dim.

\baselineskip=18pt

\end{abstract}
\end{titlepage}

\tableofcontents

\flushbottom

\newpage

\section{Introduction}
In recent years, the notion of symmetry has been generalized in a number of ways, including to so-called \textit{non-invertible symmetries} to incorporate symmetry operators without an inverse. Bosonic finite non-invertible symmetries were originally studied in two dimensions and they are described by the mathematical structure known as the \textit{fusion categories}, for which there is a rather extensive literature, see e.g. \cite{verlinde1988fusion,Petkova:2000ip,Fuchs:2002cm,Bhardwaj:2017xup,Chang:2018iay,Lin:2022dhv,Komargodski:2020mxz,Tachikawa:2017gyf, Frohlich:2004ef, Frohlich:2006ch, Frohlich:2009gb, Carqueville:2012dk, Brunner:2013xna,symtft2019XGW, Huang:2021zvu, Thorngren:2019iar, Thorngren:2021yso, Lootens:2021tet, Huang:2021nvb, Inamura:2022lun, Kaidi:2025hyr}. For this reason, these symmetries are also referred to as \textit{fusion category symmetries} or \textit{(fusion) categorical symmetries}. Recent progress has also extended these results to higher dimensions  \cite{Kaidi:2021xfk,Choi:2021kmx, Koide:2021zxj,Choi:2022zal,Hayashi:2022fkw,Arias-Tamargo:2022nlf,Roumpedakis:2022aik,Bhardwaj:2022yxj,Kaidi:2022uux,Choi:2022jqy,Cordova:2022ieu,Antinucci:2022eat,Bashmakov:2022jtl,Damia:2022seq,Damia:2022bcd,Choi:2022rfe,Lu:2022ver,Bhardwaj:2022lsg,Lin:2022xod,Bartsch:2022mpm,Apruzzi:2022rei,GarciaEtxebarria:2022vzq, Benini:2022hzx, Wang:2021vki, Chen:2021xuc, DelZotto:2022ras, Heckman:2022muc,Kaidi:2022cpf,Bashmakov:2022uek,Bonetti:2024etn,Kim:2025zdy}.

A powerful tool in the study of non-invertible symmetries is discrete gauging, which often allows one to map a complicated non-invertible symmetry to a simpler symmetry without changing the dynamics of the theory \cite{Aasen:2016dop,Aasen:2020jwb,Bhardwaj:2017xup,Arias-Tamargo:2022nlf,Bartsch:2022ytj,Fechisin:2023odt,Lootens:2021tet,Lu:2024lzf,Lu:2024ytl,Li:2024fhy,Perez-Lona:2023djo,Diatlyk:2023fwf,Choi:2024rjm,Seifnashri:2024dsd,sal2024mod,Vanhove:2024lmj,Vancraeynest-DeCuiper:2025wkh,Jacobson:2024muj}. As an example, for the categorical symmetry in 2-dim, in the most extreme and well-studied cases, there are a non-trivial amount of fusion category symmetries $\mathcal{C}$ can be mapped to an invertible symmetry, and they are referred to as being \textit{group-theoretical} \cite{ostrik2002module}. Consequently, those symmetries are rather easy to study by exploiting the corresponding discrete gauging. 

In the study of non-invertible symmetries, it is natural to ask if the familiar notions in ordinary symmetries still hold. As an example, the 't Hooft anomaly of an ordinary symmetry has two implications--it obstructs the gauging of the symmetries, as well as prevents any theory with the symmetry from flowing to a trivially gapped phase under the RG while preserving the symmetry. For non-invertible symmetries, however, the two statements are no longer equivalent \cite{Choi:2023xjw}, and one defines a non-invertible symmetry $\mathcal{C}$ to be \textit{non-anomalous} if there exists a $\mathcal{C}$-symmetric gapped phase with a unique ground state\cite{Thorngren:2019iar}. Following the notion of ordinary symmetries, we refer to such phases as non-invertible symmetry protected topological (NISPT) phases.

Recently, NISPT phases in two dimensions have been explored in \cite{Seifnashri:2024dsd,Thorngren:2019iar,Inamura:2024jke,Lu:2025gpt,Lu:2024lzf,Cao:2025qhg,Bhardwaj:2023idu,Meng:2024nxx,Lu:2025rwd,aswin2025nispt}. In all known examples, the associated categorical symmetry is group-theoretical and can be mapped, via discrete gauging, to an invertible symmetry. Consequently, the corresponding NISPT phases can be mapped to gapped phases with invertible symmetries, and their classification and characterization can be deduced from the well-understood framework of gapped phases of invertible symmetries \cite{Seifnashri:2024dsd,Lu:2025gpt,Cao:2025qhg,ostrik2002module,natale2017grptheo}\footnote{Alternatively, one can also achieve the same result via SymTFT, by exploiting the fact that the SymTFTs for those symmetries are relatively simple Dijkgraaf-Witten theories, see e.g. \cite{Bhardwaj:2025piv,Bhardwaj:2024qiv,Bhardwaj:2023idu}.}.

Closely related is the phenomenon of spontaneous symmetry breaking (SSB) transitions governed by non-invertible symmetries.\footnote{Such transitions are also referred to as “order–disorder transitions’’ (or “ferromagnetic–paramagnetic’’ transitions), where the ordered (ferromagnetic) phase corresponds to complete symmetry breaking, and the disordered (paramagnetic) phase corresponds to a fully symmetric gapped phase with a unique ground state.} A non-invertible SSB transition between the fully broken phase and the fully symmetric phase can be mapped, under discrete gauging, to a transition involving an ordinary (invertible) symmetry but possibly endowed with a ’t Hooft anomaly. On the one hand, such non-invertible SSB transitions occur only for anomaly-free non-invertible symmetries, i.e., when a fully symmetric gapped phase with a unique ground state exists. On the other hand, their dual descriptions in the invertible-symmetry frame can realize “beyond-Landau’’ critical points—transitions between two symmetry-broken phases with mutually incompatible unbroken subgroups.

However, there remains a question of whether there exist NISPT phases that cannot be mapped to any gapped phases of invertible symmetries, which will be referred to as \textit{intrinsic} NISPT phases. And consequently, the SSB transition of such \textit{intrinsic} anomaly-free non-invertible symmetry cannot be mapped to any phase transition of an ordinary invertible symmetry with a possible 't Hooft anomaly. The question of finding an intrinsic NISPT is equivalent to finding the anomaly-free non-group-theoretical fusion category symmetry and whose fiber functors or the rank-1 module categories will give rise to the intrinsic NISPT phases.

To guide our search for intrinsic NISPT phases, it is useful to understand why the majority of NISPT phases constructed so far are non-intrinsic. The most well-studied non-invertible symmetries are the Tambara-Yamagami fusion category symmetries $\TY(A,\chi,\epsilon)$ consisting of an Abelian finite group $A$ symmetry together with a non-invertible Kramers–Wannier duality defect describing the invariance under gauging $A$-symmetry. However, one can show using SymTFTs that a necessary condition for $\TY$ to admit an SPT phase is that it must be non-intrinsic. Furthermore, the same result holds for any $G$-extension of $\VEC_A$ symmetry, which is the generalization of $\TY$-fusion categories and is the easiest way to construct non-invertible symmetries systematically. 

While at first glance, this seems dis-encouraging as it suggests that to look for such intrinsic NISPT phase we need to look for more exotic categorical symmetry. However, the following two observations allow us to actually make progress using the simple construction mentioned above.
\begin{enumerate}
    \item Being non-group-theoretical is a property unchanged by discrete gauging.
    \item The discrete gauging may trivialize the mixed anomaly.
\end{enumerate}
The first one simply follows from the definition. The second observation has a well-known invertible symmetry example. Consider $\mathbb{Z}_2^a \times \mathbb{Z}_2^b$ in 2d with a type II mixed anomaly, then either $\mathbb{Z}_2^a$ or $\mathbb{Z}_2^b$ is gaugeable. The gauging leads to the dual symmetry $\mathbb{Z}_4$ with the trivial anomaly. In this work, we will demonstrate that similar phenomena occur for non-invertible symmetries as well.  

Combining the two observations, it motivates us to consider an anomalous $\mathbb{Z}_2\times \mathbb{Z}_2$-extension of $\VEC_{\mathbb{Z}_p\times\mathbb{Z}_p}$, which we denote as $\mathcal{C}_p$ (for simplicity, we restrict $p$ to be an odd prime). This category contains $\mathbb{Z}_p\times \mathbb{Z}_p$ invertible symmetry, a $\TY$-duality defect $\mathcal{N}_o$ associated with the off-diagonal bicharacter, a swap symmetry $t$ exchanging two $\mathbb{Z}_p$ factors, as well as another $\TY$-duality defect $\mathcal{N}_d = \mathcal{N}_o t$ associated with the diagonal bicharacter. $\mathcal{C}_p$ is non-group-theoretical and, therefore, anomalous. However, its anomaly can be interpreted as a mixed anomaly, roughly speaking, between $\mathcal{N}_o$ and $t$, which will be removed by gauging $\mathbb{Z}_2^t$ symmetry. The resulting dual categorical symmetry $\widetilde{\mathcal{C}}_p$ is anomaly-free and non-group-theoretical, and its SPT phases can be classified via the discrete gauging map. This allows us to find intrinsic NISPT phases protected by $\widetilde{\mathcal{C}}_p$, and we provide a concrete lattice realization of a specific intrinsic NISPT phase we find. Notice that the desired symmetry $\widetilde{\mathcal{C}}_p$ indeed realizes a partial duality under gauging some complicated non-invertible symmetry in $\Rep S_3\times \Rep\mathbb{Z}_3$.

\

However, this is the end of the story of mixed anomalies and discrete gauging, as they also play an important role in the anomaly resolution and intrinsic gapless SPT phases. Given an anomalous categorical symmetry $\mathcal{D}$ of some theory, the idea of \textit{anomaly resolution} is to extend $\mathcal{D}$ to a larger anomaly-free symmetry of the same theory with a trivially acting kernel. Applying this idea to the RG flow leads to the notion of intrinsic gapless SPT phases: starting with a UV theory $\mathcal{T}_{UV}$ with anomaly-free symmetry $\mathcal{C}_{UV}$, under the RG, it may flow to a gapless theory $\mathcal{T}_{IR}$ with a unique ground state (therefore it is referred to as a gapless SPT phase). In $\mathcal{T}_{IR}$, some symmetries in $\mathcal{C}_{UV}$ may act trivially; and upon quotienting out this part, one acquires the faithfully acting symmetry $\mathcal{C}_{IR}$, which may develop an emergent 't Hooft anomaly (which then fits into the role of anomalous category $\mathcal{D}$ in the anomaly resolution). The categorical quotient is captured by the short exact sequence of fusion categories \cite{bruguieres2011exact}
\begin{equation}\label{eq:cq}
    \Rep \mathrm{H} \rightarrow \mathcal{C}_{UV} \rightarrow \mathcal{C}_{IR} ~,
\end{equation}
where the trivially acting part must be free of anomaly and be a representation category $\Rep \mathrm{H}$ of some Hopf algebra $H$. The anomaly then obstructs us from trivially gapping the IR theory $\mathcal{T}_{IR}$ while preserving $\mathcal{C}_{IR}$. In this case, such an igSPT phase is referred to as intrinsic. 

The first example of igSPT phase is constructed via the symmetry resolution of the anomalous $\mathbb{Z}_2$-symmetry extended by another $\mathbb{Z}_2$-symmetry in 2d via the group extension \cite{Thorngren:2020wet}:
\begin{equation}\label{eq:intro_Z2e}
    \mathbb{Z}_2 \rightarrow \mathbb{Z}_4 \rightarrow \mathbb{Z}_2 ~.
\end{equation}
There have been many generalizations: for instance, in \cite{Perez-Lona:2025ncg,Bhardwaj:2024qrf,Antinucci:2024ltv,Robbins:2025puq}, the anomaly resolution of anomalous invertible symmetries by some non-invertible symmetries has been studied; and the anomaly resolution of anomalous non-invertible symmetries has been constructed in \cite{Antinucci:2024ltv} in 4d. Yet, it remains to find an example of anomaly resolution of some non-invertible symmetry in 2d. 

To make progress, we first notice that the anomaly resolution of $\mathbb{Z}_2$ via the non-trivial group extension \eqref{eq:intro_Z2e} has a dual interpretation after gauging the $\mathbb{Z}_2 \subset \mathbb{Z}_4$, where the anomaly resolution is captured by the mixed anomaly between two $\mathbb{Z}_2$ symmetries that can absorb the self-anomaly of a $\mathbb{Z}_2$-factor by a shift in basis. This phenomenon again generalizes to non-invertible symmetries and is realized in some of $\mathcal{C}_p$ categories constructed before. This allows us to find an example of anomaly resolution of non-invertible symmetries in 2d, which also provides a natural physical realization of the $G$-equivariantization $(\mathcal{C}_{IR})^G$ of the category $\mathcal{C}_{IR}$ and the associated short exact sequence
\begin{equation}\label{eq:seGe}
    \Rep G\rightarrow (\mathcal{C}_{IR})^G \rightarrow \mathcal{C}_{IR} ~.
\end{equation}
Starting with a seed fixed-point theory $\mathcal{X}$ with an anomalous symmetry $\mathcal{C}_{IR}$, assuming $\mathcal{X}$ is also invariant under some anomaly-free invertible automorphism symmetry $G$ of $\mathcal{C}_{IR}$\footnote{It is important to emphasize that the full symmetry $\mathcal{C}_{IR}\rtimes G$ is still anomalous, as its subcategory $\mathcal{C}_{IR}$ is anomalous.}, then one can stack $\mathcal{X}$ with a gapped theory (say the gap is $\Delta$) realizing a $G$-SBB phase. The total theory now has the symmetry $\mathcal{C}_{IR}\rtimes G$, where $G$ here is the diagonal group of the $G$-symmetry in $\mathcal{X}$ and the $G$-symmetry in the stacked $G$-SSB phase. Gauging $G$ leads to a theory $\widetilde{\mathcal{X}}$ and the dual symmetry is precisely the $G$-equivariantization $(\mathcal{C}_{IR})^G$ \footnote{Equivariantization is also used in a slightly different context on gauging the 0-form global symmetry of symmetry enriched topological order in 2+1d \cite{Barkeshli:2014cna,Cui:2018hxz}.}. Furthermore, in the theory $\widetilde{\mathcal{X}}$, below $\Delta$, the stacked sector becomes trivial due to the $G$-gauging, and we simply recover the seed theory $\mathcal{X}$ where $\Rep G$ acts trivially and the faithfully acting symmetry in $(\mathcal{C}_{IR})^G$ becomes $\mathcal{C}_{IR}$; in other words, the simple RG flow $\widetilde{\mathcal{X}} \rightarrow \mathcal{X}$ realizes the exact sequence \eqref{eq:seGe} in the sense discussed before\footnote{Notice that one can also just gauge the $G$-symmetry in $\mathcal{X}$ without the stacking, which also leads to the dual symmetry $(\mathcal{C}_{IR})^G$. But generically this leads to a different theory in the IR and the dual symmetry $\Rep G$ does not act trivially.}. In the case where $(\mathcal{C}_{IR})^G$ is anomaly-free, the UV theory $\widetilde{\mathcal{X}}$ provides an anomaly resolution of the anomalous symmetry $\mathcal{C}_{IR}$, as one can find some deformation in $\widetilde{\mathcal{X}}$ to trivially gap the theory. Furthermore, if the seed theory $\mathcal{X}$ is a gapless theory with a unique ground state, then this construction leads to an igSPT phase. Since the igSPT phase is protected by the anomalous IR non-invertible symmetry $\mathcal{C}_{IR}$, we refer to it as an igNISPT phase (where ``NI'' stands for ``non-invertible'')\footnote{The Type II igSPT phase in the context of duality symmetries in \cite{Antinucci:2024ltv} is an example of the igNISPT phase defined here.}. We will provide a concrete lattice example resolving the anomalous $\mathcal{C}_{IR} = \TY(\mathbb{Z}_3\times \mathbb{Z}_3,\chi_d,+1)$-symmetry to explicitly demonstrate the above construction.

It is worth noting that the lattice realization of the igNISPT typically flows to a non-relativistic gapless phase that is scale-invariant but lacks conformal symmetry. Our construction includes the 3-state clock model as a special case \cite{Fateev:1985mm,Fateev:1985ig,Mong:2014ova}, and for certain parameter choices reduce to the chiral 3-state clock model \cite{cardy1993chiralpotts,paul2012chiralclock,taylor2015chiralclock,subir2018chiralclock,mila2022chiralclock}, realizing Lifshitz-type or other non-relativistic critical points with the dynamical exponent $z \neq 1$. For the $\widetilde{\mathcal{C}}_p$ igNISPT, the infrared theory becomes a $p$-state clock or Potts model with non-invertible self-duality symmetry, which may realize weakly first-order transitions or complex critical points \cite{Ma:2018euv,Tang:2024blm,Jacobsen:2024jel}. These models offer examples of studying the non-invertible symmetry in scaling invariant theory or complex conformal field theory.

\

Finally, despite the construction of $\widetilde{\mathcal{C}}_p$ having appeared in the mathematical literature—where this non-group-theoretical fusion category of order $4p^2$ (with the smallest example of order $36$ for $p=3$) is well known \cite{nikshych2008non,Gelaki:2009blp,natale2010hopf,cuadra2017orders} and has recently been generalized to order $q^2p^2$ for certain $p,q$ \cite{galindo2024integral}—our approach, emphasizing the mixed anomaly and discrete gauging in $\mathcal{C}_p$, further enables the classification of $\widetilde{\mathcal{C}}_p$-SPT phases, the construction of anomaly resolutions for non-invertible symmetries, and generalization to higher dimensions.

\

The paper is organized as follows. In Section \ref{sec:review}, we briefly review some background materials on $\TY$-fusion categories, classifications of their SPT phases, and some results on $\TY$-categories acquired from SymTFTs. In Section \ref{sec:intrinNISPT}, we construct the intrinsic NISPT phases using the mixed anomaly and the discrete gauging, and provide a concrete lattice model realizing an intrinsic NISPT phase. In Section \ref{sec:igNISPT}, we provide a method of searching for anomaly resolution of generic categorical symmetries via mixed anomaly, and discover the anomaly resolution of some anomalous $\TY$-fusion categories. We then construct an explicit lattice model realizing an igNISPT phase related to the anomaly resolution we find. In Section \ref{sec:4d}, we propose a generalization of our construction of intrinsic NISPT phases to 4-dim.

\section{Review}\label{sec:review}

\subsection{$\TY$-fusion category symmetries and their generalizations}
In (1+1)-d QFTs, topological defect lines (TDLs) are line operators commuting with the stress-energy tensor, and various correlation functions are invariant under local deformations of them. In this work, we focus on TDLs which satisfy the mathematical axioms of (unitary) fusion categories \cite{Fuchs:2002cm,Bhardwaj:2017xup,Chang:2018iay}. Such topological lines generate bosonic finite categorical symmetries in 1+1d.

Given two TDLs $\mathcal{L}_a$ and $\mathcal{L}_b$, we can fuse them by putting them close to each other and this generates a TDL, which then in general decomposes into a finite sum of other topological lines,
\begin{equation}
    \mathcal{L}_a \otimes \mathcal{L}_b = \bigoplus_{c} N^{c}_{ab} \mathcal{L}_c, \quad N^c_{ab} \in \doubleZ_{\geq 0}.
\end{equation}
The \textit{simple} topological lines are those that cannot be written as a sum of at least two other lines. We denote the trivial topological line as $\dsi$.

When $N_{ab}^c \ne 0$, two topological lines $\mathcal{L}_a$ and $\mathcal{L}_b$ can join each other locally and become the line $\mathcal{L}_c$ at a trivalent junction. The set of topological junction operators form a vector space whose complex dimension is given by $N_{ab}^c$. We always fix a basis of the junction vector space:
\begin{equation}
    \begin{tikzpicture}[scale=0.8,baseline={([yshift=-.5ex]current bounding box.center)},vertex/.style={anchor=base,
    circle,fill=black!25,minimum size=18pt,inner sep=2pt},scale=0.50]
    \draw[thick, black] (-2,-2) -- (0,0);
    \draw[thick, black] (+2,-2) -- (0,0);
    \draw[thick, black] (0,0) -- (0,2);
    \draw[thick, black, -stealth] (-2,-2) -- (-1,-1);
    \draw[thick, black, -stealth] (+2,-2) -- (1,-1);
    \draw[thick, black, -stealth] (0,0) -- (0,1);
    \filldraw[thick, black] (0,0) circle (3pt);
    \node[black, right] at (0,0) {\footnotesize $\mu$};
    \node[black, below] at (-2,-2) {$\mathcal{L}_a$};
    \node[black, below] at (2,-2) {$\mathcal{L}_b$};
    \node[black, above] at (0,2) {$\mathcal{L}_c$};
    
\end{tikzpicture}, \quad \mu = 1,2,\cdots, N^{c}_{ab} \,.
\end{equation}
There are two possible ways where three TDLs can fuse together, and they are related by the so-called associativity map. With the explicit basis chosen, the associativity map is characterized by a set of $\mathbb{C}$-numbers known as the $F$-symbols,
\begin{equation}
\begin{tikzpicture}[baseline={([yshift=-1ex]current bounding box.center)},vertex/.style={anchor=base,
    circle,fill=black!25,minimum size=18pt,inner sep=2pt},scale=0.7]
	\draw[thick, black, -<-=0.5] (0,0) -- (-0.75,-0.75);
	\draw[thick, black, -<-=0.5] (0,0) -- (+0.75,-0.75);
	\draw[thick, black, ->-=0.5] (0,0) -- (+0.75,+0.75);
	\draw[thick, black, ->-=0.5] (+0.75,+0.75) -- (1.5,1.5);
	\draw[thick, black, -<-=0.5] (0.75,0.75) -- (2.25,-0.75);
	
	\node[below, black] at (-0.75,-0.75) {\scriptsize $\mathcal{L}_a$};
	\node[below, black] at (+0.75,-0.75) {\scriptsize $\mathcal{L}_b$};
	\node[below, black] at (+2.25,-0.75) {\scriptsize $\mathcal{L}_c$};
	\node[right, black] at (0.25,0.2) {\scriptsize $\mathcal{L}_e$};
 	\node[left, black] at (0,0) {\scriptsize $\mu$};
    \node[left, black] at (0.75,0.75) {\scriptsize $\nu$};
	\node[above, black] at (1.5,1.5) {\scriptsize $\mathcal{L}_d$};
	\filldraw[black] (0,0) circle (2pt);
	\filldraw[black] (0.75,0.75) circle (2pt);
	
	\node[black] at (5,0.25) {$\displaystyle = \sum\limits_{f,\rho,\sigma} \left[F^{abc}_d\right]_{(e,\mu,\nu),(f,\rho,\sigma)}$};
	
	\draw[thick, black,->-=0.5] (9-0.75,-0.75) -- (9+0.75,0.75);
	\draw[thick, black,->-=0.5] (9+0.75,+0.75) -- (9+1.5,1.5);
        \draw[thick, black,->-=0.5] (9+0.75,-0.75) -- (9+1.5,0.);
        \draw[thick, black,->-=0.5] (9+2.25,-0.75) -- (9+1.5,0.);
        \draw[thick, black,->-=0.5] (9+1.5,0.) -- (9+0.75,+0.75);
	
	\node[below, black] at (9-0.75,-0.75) {\scriptsize $\mathcal{L}_a$};
	\node[below, black] at (9+0.75,-0.75) {\scriptsize $\mathcal{L}_b$};
	\node[below, black] at (9+2.25,-0.75) {\scriptsize $\mathcal{L}_c$};
	\node[right, black] at (9+1.1,0.5) {\scriptsize $\mathcal{L}_f$};
	\node[above, black] at (9+1.5,1.5) {\scriptsize $\mathcal{L}_d$};

    \node[left, black] at (9+1.5,0) {\scriptsize $\rho$};
	\node[left, black] at (9+0.75,+0.75) {\scriptsize $\sigma$};
	\filldraw[black] (9+0.75,+0.75) circle (1.5pt);
	\filldraw[black] (9+1.5,0) circle (1.5pt);	
\end{tikzpicture} ~,
\end{equation}
and the $F$-symbols are constrained by the so-called pentagon equations.

\

The Tambara-Yamagami fusion category $\TY(A,\chi,\epsilon)$ is a well-known categorical symmetry (see \cite{Thorngren:2019iar} for more detailed reviews), specified by a finite Abelian group $A$, a symmetric non-degenerate bicharacter $\chi:A\times A\rightarrow U(1)$, and a sign $\epsilon = \pm 1$. It contains an anomaly-free finite Abelian $A$ and a non-invertible duality line $\mathcal{N}_{\chi,\epsilon}$, arising from the invariance under $A$-gauging. At the level of twisted partition function, this means any theory $\mathcal{X}$ with $\TY(A,\chi,\epsilon)$ symmetry must satisfy\footnote{Here, since $U(1) \simeq \mathbb{R}/\mathbb{Z}$, we can view $\chi:A\times A\rightarrow \mathbb{R}/\mathbb{Z}$, and on 2-simplex $(ijk)$ in a triangulation, $\chi(a^{(1)},A^{(1)})_{ijk} := \chi((a^{(1)})_{ij},(A^{(1)})_{jk})\in \mathbb{R}/\mathbb{Z}$.}
\begin{equation}\label{eq:TY_gauging}
    Z_{\mathcal{X}}[A^{(1)}] = \frac{1}{\sqrt{|H^1(\mathcal{M}_2,A)|}}\sum_{a^{(1)}} Z_{\mathcal{X}}[a^{(1)}] \exp\left(2\pi i\int_{\mathcal{M}_2} \chi(a^{(1)},A^{(1)})\right) ~,
\end{equation}
where $A^{(1)},a^{(1)}$ are the discrete gauge fields in $H^1(\mathcal{M}_2,A)$. For simplicity of the presentation, we will drop the normalization factor from now on, and it is straightforward to recover it. The fact that $\chi$ is symmetric and non-degenerate implies that the gauging operation \eqref{eq:TY_gauging} on any given theory $\mathcal{X}$ with a non-anomalous $A$-symmetry is an order-$2$ operation. The sign $\epsilon = \pm 1$ is understood as the Forbenius-Schur (FS) indicator of the duality line $\mathcal{N}_{\chi,\epsilon}$.

The fusion rules involving the duality line $\mathcal{N}_\chi$ are given by
\begin{equation}
    \mathcal{N}_{\chi,\epsilon} \otimes a = a \otimes \mathcal{N}_{\chi,\epsilon} = \mathcal{N}_{\chi,\epsilon} ~, \quad \mathcal{N}_{\chi,\epsilon} \otimes \mathcal{N}_{\chi,\epsilon} = \sum_{a\in A}a ~.
\end{equation}
And the non-trivial $F$-symbols are given by
\begin{equation}
    F^{a,\mathcal{N}_{\chi,\epsilon}, b}_{\mathcal{N}_{\chi,\epsilon}} = F^{\mathcal{N}_{\chi,\epsilon}, a, \mathcal{N}_{\chi,\epsilon}}_b = \chi(a,b) ~, \quad \left[F_{\mathcal{N}_{\chi,\epsilon},\mathcal{N}_{\chi,\epsilon},\mathcal{N}_{\chi,\epsilon}}^{\mathcal{N}_{\chi,\epsilon}}\right] = \frac{\epsilon}{\sqrt{|A|}} \frac{1}{\chi(a,b)} ~, \quad a,b \in A ~,
\end{equation}
where $|A|$ denotes the order of the Abelian group $A$.

\

Notice that the $\TY$ categorical symmetry is a special case of $G$-extended fusion categories. A categorical symmetry $\mathcal{C}$ is $G$-graded if it admits a direct sum decomposition
\begin{equation}
    \mathcal{C} = \bigoplus_{g\in G} \mathcal{C}_g
\end{equation}
and grading is preserved by the fusion $\mathcal{C}_g \otimes \mathcal{C}_{g'} \subset \mathcal{C}_{gg'}$. Given a categorical symmetry $\mathcal{D}$, its $G$-extension is a $G$-graded fusion category $\mathcal{C}$ such that the trivial grading component $\mathcal{C}_1 = \mathcal{D}$. The $\TY$-fusion category can be thought of as a $\mathbb{Z}_2$-graded extension of $\VEC_{A}$\footnote{We use $\VEC_G^\omega$ to denote the categorical symmetry of a finite group $G$ with 't Hooft anomaly $[\omega] \in H^3(G,U(1))$.}; and the non-trivial grading component is $\{\mathcal{N}_{\chi,\epsilon}\}$. Notice that a $G$-extension does not necessarily lead to non-invertible symmetry, for instance, let $G$ be a subgroup of $\Aut(A)$, then $\VEC_{A\rtimes G}$ is also a $G$-extension of $\VEC_{A}$ but it contains only invertible symmetries.

\subsection{Classification of SPT phases of certain $\TY$-fusion categories}\label{sec:SPTgrpt}
Given a gapped phase $\mathcal{B}$ of some categorical symmetry $\mathcal{C}$ in (1+1)-dim, discrete gauging\footnote{Generically, a discrete gauging is specified by a symmetric $\Delta$-separate Frobenius algebra in $\mathcal{C}$ (for more details, see e.g. \cite{Choi:2023xjw}). In this work, we will only make use of the discrete gauging of some finite group $H$ in $\mathcal{C}$. In this case, the gauging is specified by an anomaly subgroup $H$ in $\mathcal{C}$ and a discrete torsion in $H^2(H,U(1))$.} leads to a dual gapped phase $\widehat{\mathcal{B}}$ of the dual categorical symmetry $\widehat{\mathcal{C}}$. Furthermore, there always exists an inverse gauging in $\widehat{\mathcal{C}}$, under which $\widehat{\mathcal{B}}$ is mapped back to $\mathcal{B}$. The two discrete gaugings thus establish the bijection between the gapped phases of $\mathcal{C}$ and the gapped phases of $\widehat{\mathcal{C}}$. This provides a powerful tool for classifying gapped phases of generalized symmetry, which allows us to attack the problem in a frame where symmetry becomes simpler. 

One can apply the above discrete gauging technique in the classification of SPT phases of $\TY$-categorical symmetries. In the $\TY$-frame, the computation of SPT phases can be organized into two steps, using the observation that a $\TY$-SPT phase must be a $A$-SPT phase invariant under the corresponding gauging operation in the first place. The first step is to find all such $A$-SPT phases as candidates; and the second step is to enrich each candidate $A$-SPT phase with additional data describing the insertion of generic networks of defects (which contain local junctions of $\mathcal{N}$'s). Generically, this is not always possible and also not always unique when possible. The first step is straightforward; and the second step, while it has been established for generic $\TY$-categories in \cite{tambara2000representations}, is not so obvious and can be simplified if converted to the invertible symmetry frame. As we will review shortly in Section \ref{sec:SymTFT_review}, converting to an invertible symmetry frame in the second step is always possible if a candidate $A$-SPT exists in the first step. 

\

Let's demonstrate this explicitly for two $\TY$-fusion categories used in this work.\footnote{Notice that there are many ways to acquire the same results. Other approaches (as well as some detailed computations of the method here) are discussed in Appendix \ref{app:ff}, which reproduces the same results as a consistency check.} The first one is $\TY(\mathbb{Z}_p \times \mathbb{Z}_p,\chi_d,+1)$ where $p$ is an odd prime. Here, $\chi_o$ denotes the off-diagonal bicharacter given by $\chi_o(g,h) = e^{\frac{2\pi i}{p} (\mathrm{g}_1 \mathrm{h}_2 + \mathrm{g}_2 \mathrm{h}_1)}$ where we parameterize $g \in \mathbb{Z}_p \times \mathbb{Z}_p$ by $(\mathrm{g}_1,\mathrm{g}_2)$ in an additive notation. For simplicity, we will also denote $\TY(\mathbb{Z}_p\times \mathbb{Z}_p,\chi_{o},+1)$ as $\TYpo$ and its duality defect as $\mathcal{N}_o$. Any 2d QFT $\mathcal{X}$ admitting $\TYpo$-symmetry is invariant under the following gauging:
\begin{equation}\label{eq:o_gg1}
    Z_{\mathcal{X}}[A^{(1)},B^{(1)}] = \sum_{a^{(1)},b^{(1)}} Z_{\mathcal{X}}[a^{(1)},a^{(2)}] \exp\left(\frac{2\pi i}{p}\int_{\mathcal{M}_2}a^{(1)} \cup B^{(1)} + b^{(1)} \cup A^{(1)} \right)  ~,
\end{equation}
where $a^{(1)}, b^{(1)}, A^{(1)}, B^{(1)} $ are discrete $\mathbb{Z}_p$-gauge fields in $H^1(\mathcal{M}_2,\mathbb{Z}_p)$. 

The first step is to search for $\mathbb{Z}_p \times \mathbb{Z}_p$-SPT phase invariant under \eqref{eq:o_gg1}. A generic $\mathbb{Z}_p\times \mathbb{Z}_p$ SPT phase is labeled by $k \in H^2(\mathbb{Z}_p\times\mathbb{Z}_p,U(1)) \simeq \mathbb{Z}_p$ and is captured by the following partition function:
\begin{equation}\label{eq:SPT_k}
    Z_{\SPT_k}[A^{(1)},B^{(1)}] = \exp\left(\frac{2\pi i k}{p} \int_{\mathcal{M}_2} A^{(1)} \cup B^{(1)} \right) ~, \quad k \in \mathbb{Z}_p ~.
\end{equation}
It is straightforward to check that $\SPT_{\pm 1}$ are the only two $\mathbb{Z}_p \times \mathbb{Z}_p$-SPT phases satisfying \eqref{eq:o_gg1}. 

In the second step, we must classify the possible enrichment of $\SPT_{\pm 1}$ to include the duality defect $\mathcal{N}_o$, which can be easily done via the following \textit{off-diagonal} gauging:
\begin{equation}\label{eq:o_dg1}
    Z_{\widetilde{\mathcal{X}}}[A^{(1)},B^{(1)}] = \sum_{a^{(1)}\in H^1(\mathcal{M}_2,\mathbb{Z}_p)} Z_{\mathcal{X}}[a^{(1)},B^{(1)}] \exp\left(\frac{2\pi i}{p}\int_{\mathcal{M}_2} a^{(1)}\cup A^{(1)} \right) ~,
\end{equation}
with the inverse gauging:
\begin{equation}\label{eq:o_dgi1}
    Z_{\mathcal{X}}[A^{(1)},B^{(1)}] = \sum_{a^{(1)}\in H^1(\mathcal{M}_2,\mathbb{Z}_p)} Z_{\widetilde{\mathcal{X}}}[a^{(1)},B^{(1)}] \exp\left(\frac{2\pi i}{p}\int_{\mathcal{M}_2} a^{(1)}\cup A^{(1)} \right) ~.
\end{equation}
Under \eqref{eq:o_dg1}, the non-invertible defect $\mathcal{N}_o$ in theory $\mathcal{X}$ is mapped to an invertible automorphism $\mathbb{Z}_2$-symmetry in theory $\widetilde{\mathcal{X}}$, as the relation \eqref{eq:o_gg1} becomes
\begin{equation}
    Z_{\widetilde{\mathcal{X}}}[A^{(1)},B^{(1)}] = Z_{\widetilde{\mathcal{X}}}[B^{(1)},A^{(1)}] ~.
\end{equation}
This means the non-invertible $\TYpo$ symmetry in $\mathcal{X}$ is mapped to the invertible symmetry $(\mathbb{Z}_p^r \times \mathbb{Z}_p^s)\rtimes \mathbb{Z}_2^t$ with trivial 't Hooft anomaly\footnote{To see this, first notice that $\mathbb{Z}_p^r\times \mathbb{Z}_p^s$ is clearly anomaly free. Furthermore, there can not be any mixed anomaly between $\mathbb{Z}_2^t$ and $\mathbb{Z}_p^r\times \mathbb{Z}_p^s$ as $p$ coprime with $2$. Finally, there is no self-anomaly of $\mathbb{Z}_2^t$ as the FS indicator of $\mathcal{N}_o$ is $+1$.}, where $t$ acts as a swap symmetry between $r$ and $s$. Here, $\mathbb{Z}_p^r$ and $\mathbb{Z}_p^s$ correspond to the background gauge field $A^{(1)}$ and $B^{(1)}$ in the LHS of \eqref{eq:o_dg1} respectively.

\begin{figure}[!tbp]
	\centering
	\begin{tikzpicture}[scale = 1.5]
	\filldraw[black] (-1,0) circle (1.5pt);
	\node[above] at (-1,0) {$\SPT_{+1}$};
    \draw[red, thick, smooth, ->-=1] (-1.2,-0.2) arc (120:420:0.4);
    \node[red, below] at (-1,-1) {$\mathcal{N}_o$: \eqref{eq:o_gg1}};
    \draw[dashed, thick] (-2,+0.75) -- (0,+0.75) -- (0,-1.5) -- (-2,-1.5) -- (-2,+0.75);

    \draw[very thick, ->-=1] (0.2,-0.4) -- (1.8,-0.4);
    \node[above] at (1,-0.4) {\scriptsize gauging \eqref{eq:o_dg1}};
    \draw[very thick, ->-=1] (1.8,-0.6) -- (0.2,-0.6);
    \node[below] at (1,-0.6) {\scriptsize ungauging \eqref{eq:o_dgi1}};
    
    \filldraw[black] (3,0) circle (1.5pt);
    \draw[red, thick, smooth, ->-=1] (2.8,-0.2) arc (120:420:0.4);
    \node[red, below] at (3,-1) {$t: r \leftrightarrow s$};
    \draw[dashed, thick] (2,+0.75) -- (4,+0.75) -- (4,-1.5) -- (2,-1.5) -- (2,+0.75);
    \node[above, align = center] at (3,0) {\setstretch{1}\footnotesize $\mathbb{Z}_p^r\times \mathbb{Z}_p^s$-SSB with \\ \footnotesize $\langle rs^{-1}\rangle$  unbroken};

    \draw[dashed, very thick, ->-=1] (-1,-1.75) -- (-1,-2.75); 
    \node[left] at (-1,-2.25) {? $\exists$ enrichment};

    \filldraw[black] (-1,-3) circle (1.5pt);
    \node[below] at (-2,-3.2) {$\TYpo$-SPT phase: $\mathcal{F}_{o,+}$};

    \draw[very thick, ->-=1] (3,-1.75) -- (3,-2.75); 
    \node[right] at (3,-2.25) {$\exists !$ enrichment};

    \filldraw[black] (3,-3) circle (1.5pt);
    \node[below, align = center] at (3.2,-3.2) {\setstretch{1}{\scriptsize $(\mathbb{Z}_p^r \times \mathbb{Z}_p^s)\rtimes \mathbb{Z}_2^t$-}{\footnotesize SSB with} \\ \footnotesize $\langle rs^{-1},t\rangle$ unbroken};

    \draw[very thick, ->-=1] (0.2,-2.9) -- (1.8,-2.9);
    \node[above] at (1,-2.9) {\scriptsize gauging \eqref{eq:o_dg1}};
    \draw[very thick, ->-=1] (1.8,-3.1) -- (0.2,-3.1);
    \node[below] at (1,-3.1) {\scriptsize ungauging \eqref{eq:o_dgi1}};
	
	\end{tikzpicture}
    
    \caption{The fact that there is a unique enrichment from the $\mathbb{Z}_p\times \mathbb{Z}_p$-SPT $\SPT_{+1}$ to the $\TYpo$-SPT $\mathcal{F}_{o,+}$ can be established by considering the discrete gauging \eqref{eq:o_dg1} under which the non-invertible $\mathcal{N}_o$ becomes the invertible swap symmetry $t$ exchanging $\mathbb{Z}_p^r$ and $\mathbb{Z}_p^s$. The $\SPT_{+1}$ under \eqref{eq:o_dg1} is mapped to a $\mathbb{Z}_p^r$ and $\mathbb{Z}_p^s$-partial SSB phase with $\langle rs^{-1}\rangle$ unbroken. There exists a unique enrichment on the invertible symmetry frame turning it into the $(\mathbb{Z}_p^r \times \mathbb{Z}_p^s)\rtimes \mathbb{Z}_2^t$-partial SSB phase with $\langle rs^{-1},t\rangle$ unbroken. This means there exists a unique enrichment in the $\TYpo$ frame.}
    \label{fig:eSPTvg}
\end{figure}
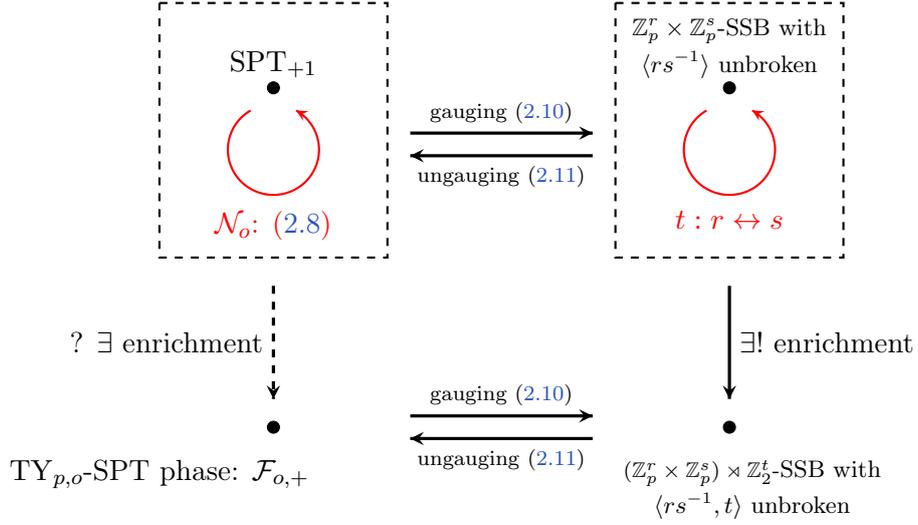

Under \eqref{eq:o_dg1}, the $\SPT_{+1}$ is mapped to  $\widetilde{\SPT_{+1}}$\footnote{Here, the tilde is on the entire $\SPT_{+1}$, by which we mean the dual phase we get from applying \eqref{eq:o_dg1} on $\SPT_{+1}$, which is not an SPT phase of the dual symmetry. Later, we will also use $\widetilde{\SPT}_{\cdots}$ where the tilde is only on SPT. By that, we denote the SPT phase of the dual symmetry labeled by $\cdots$.}, which is a partial SSB phase of the dual $\mathbb{Z}_p^r\times \mathbb{Z}_p^s$-symmetry where the subgroup $\langle rs^{-1}\rangle$ remains unbroken, as
\begin{equation}
    Z_{\widetilde{\SPT_{+1}}} [A^{(1)},B^{(1)}] = \delta(A^{(1)} + B^{(1)}) ~.
\end{equation}
Clearly, $\widetilde{\SPT_{+1}}$ is invariant under the swap symmetry $t$. Furthermore, the classification of enriching $\SPT_{+1}$ to a $\TYpo$-SPT phase becomes the classification of enriching $\widetilde{\SPT_{+1}}$ to a partial SSB phase of the full invertible symmetry $(\mathbb{Z}_p^r \times \mathbb{Z}_p^s)\rtimes \mathbb{Z}_2^t$ without changing the ground state degeneracy. It is straightforward to verify the only possible enrichment leads to the partial SSB phase with $\langle rs^{-1},t\rangle \simeq D_{2p}$ unbroken. Hence, we conclude in the $\TYpo$-frame, there is a unique way of enriching $\SPT_{+1}$ and denoting the resulting $\TYpo$-SPT phase as $\mathcal{F}_{o,+}$. This process is illustrated in Figure \ref{fig:eSPTvg}.

Similarly, $\SPT_{-1}$ under discrete gauging is mapped to the partial SSB phase where $\langle rs\rangle$ is unbroken, as under \eqref{eq:o_dg1}
\begin{equation}
    Z_{\widetilde{\SPT_{-1}}} [A^{(1)},B^{(1)}] = \delta(A^{(1)} - B^{(1)}) ~.
\end{equation}
And there is a unique enrichment which turns it to be the $(\mathbb{Z}_p^r \times \mathbb{Z}_p^s)\rtimes \mathbb{Z}_2^{t}$-partial SSB phase with $\langle rst \rangle \simeq \mathbb{Z}_{2p}$ unbroken. This implies there is a unique enrichment turning $\SPT_{-1}$ to a $\TYpo$-SPT phase denoted as $\mathcal{F}_{o,-}$.

\

As another example, let us consider $\TY(\mathbb{Z}_p\times \mathbb{Z}_p,\chi_{d},+1)$ (where $p$ is an odd prime) with the diagonal bicharacter $\chi_d(g,h) = e^{\frac{2\pi i}{p} (\mathrm{g}_1 \mathrm{h}_1 + \mathrm{g}_2 \mathrm{h}_2)}$. Similarly, we denote this category as $\TYpd$ and the duality defect as $\mathcal{N}_d$ for simplicity. Any 2d QFT $\mathcal{X}$ admitting $\TYpd$-symmetry is invariant under the following \textit{diagonal} gauging:
\begin{equation}\label{eq:d_gg1}
    Z_{\mathcal{X}}[A^{(1)},B^{(1)}] = \sum_{a^{(1)},b^{(1)}} Z_{\mathcal{X}}[a^{(1)},a^{(2)}] \exp\left(\frac{2\pi i}{p}\int_{\mathcal{M}_2}a^{(1)} \cup A^{(1)} + b^{(1)} \cup B^{(1)} \right)  ~.
\end{equation}
And the first step is to search for $\mathbb{Z}_p\times \mathbb{Z}_p$-SPT phase satisfying \eqref{eq:d_gg1}. Using \eqref{eq:SPT_k}, we find $\SPT_0$ is mapped to $\mathbb{Z}_p\times \mathbb{Z}_p$-SSB phase while $\SPT_k$ ($k \neq 0$) is mapped to $\SPT_{-k^{-1}}$ where $k^{-1}$ denotes the mod $p$ inverse of $k$. This implies the self-dual SPT exists only when
\begin{equation}
    k = -k^{-1} \mod p \quad \Longleftrightarrow \quad k^2 = -1 \mod p
\end{equation}
admitting solutions. This happens when $p = 1 \mod 4$ and the solutions are given by $\pm x$ where \footnote{$\left[\left(\frac{p-1}{2}\right)!\right]^2 = -1 \mod p$ follows directly from the Wilson's theorem: $(p-1)! = -1 \mod p$ when $p$ is a prime number.}
\begin{equation}
    x = \left(\frac{p-1}{2}\right)! ~.
\end{equation}
This means there are no $\TYpd$ SPT phases when $p = 3 \mod 4$. In other words, $\TYpd$ is anomalous when $p = 3 \mod 4$.

When $p = 1 \mod 4$, we still need to go through the second step. Under the discrete gauging 
\begin{equation}\label{eq:d_dg1}
    Z_{\widetilde{\mathcal{X}}'}[A^{(1)},B^{(1)}] = \sum_{a^{(1)}}Z_{\mathcal{X}}[a^{(1)},x a^{(1)} + B^{(1)}]\exp\left(\frac{2\pi i}{p} \int_{\mathcal{M}_2} a^{(1)} \cup A^{(1)} + \frac{p+1}{2}x A^{(1)}\cup B^{(1)}\right) ~, 
\end{equation}
one can turn the non-invertible $\mathcal{N}_d$ in $\mathcal{X}$ into the following automorphism symmetry of the dual $\mathbb{Z}_p\times \mathbb{Z}_p$ in $\widetilde{\mathcal{X}}'$ which acts as
\begin{equation}
    Z_{\widetilde{\mathcal{X}}}[A^{(1)},B^{(1)}] = Z_{\widetilde{\mathcal{X}}}[x B^{(1)},-xA^{(1)}] ~.
\end{equation}
The full symmetry in the dual theory $\widetilde{\mathcal{X}}'$ is 
\begin{equation}
    G' = (\mathbb{Z}_p^{r'} \times \mathbb{Z}_p^{s'}) \rtimes' \mathbb{Z}_2^{t'} = \langle r',s',t'| r'^p = s'^p = t'^2 = 1 ~, r's' = s'r'~, t's't' = r'^{-x}\rangle ~,
\end{equation}
with trivial 't Hooft anomaly. $\mathbb{Z}_p^{r'} ,\mathbb{Z}_p^{s'}$ corresponds to the gauge field $A^{(1)},B^{(1)}$ on the LHS of \eqref{eq:d_dg1} respectively. Under \eqref{eq:d_dg1}, $\SPT_{+x}$ is mapped to the $\mathbb{Z}_p^{r'} \times \mathbb{Z}_p^{s'}$-partial SSB with $\langle r's'^{-x}\rangle$ unbroken. There is a unique way of enriching it which leads to $G'$-partial SSB phase with $\langle r's'^{-x} t'\rangle$ unbroken. This means there is a unique enrichment turning $\SPT_{+x}$ into a $\TYpd$-SPT, which we denote as $\mathcal{F}_{d,+}$. Similarly, $\SPT_{-x}$ is mapped to the $\mathbb{Z}_p^{r'} \times \mathbb{Z}_p^{s'}$-partial SSB with $\langle r's'^{x}\rangle$ unbroken, which has a unique enrichment into the $G'$-SSB phase with $\langle r's'^{x},t'\rangle$ unbroken.  This means there is a unique enrichment turning $\SPT_{-x}$ into a $\TYpd$-SPT, which we denote as $\mathcal{F}_{d,-}$.

\subsection{Duality defects from SymTFT}\label{sec:SymTFT_review}
Symmetry topological field theory (SymTFT) is a powerful tool for studying generalized symmetries, and it allows us to separate the generic symmetry data from the dynamical data specific to a given QFT \cite{Kitaev:2011dxc,symtft2019XGW,symtft2020XGW2,symtft2021XGW3,symtft2023XGW4,symtft2022XGW5,symtft2021Gaiotto,symtft2021Sakura,moradi2023topoholo,symtft2022Apruzzi,Freed:2022qnc,Kaidi:2022cpf,symtft2022Kulp,symtft2023kaidi2,Brennan:2024fgj,Bonetti:2024cjk,DelZotto:2024tae,Argurio:2024oym,Franco:2024mxa,Putrov:2024uor,Huang:2024ror,Freed:2022qnc,Lin:2022dhv,Bhardwaj:2024igy,Antinucci:2024ltv,Bhardwaj:2024qiv,Bhardwaj:2024qrf,Bhardwaj:2023bbf,Bhardwaj:2023ayw,Bhardwaj:2023wzd,Sun:2023xxv,Zhang:2023wlu,Copetti:2024onh,Antinucci:2023ezl,Choi:2024wfm,Choi:2024tri,Cordova:2023bja,Antinucci:2024zjp,Bhardwaj:2024ydc,Chen:2023qnv,Cui:2024cav}. It plays a crucial role in the study of duality defects and their generalizations.

Let's demonstrate the SymTFT by considering the anomaly-free $\mathbb{Z}_p \times \mathbb{Z}_p$ (where $p$ is an odd prime number) 0-form symmetry in 2d QFT $\mathcal{X}$. The SymTFT for this symmetry is a 3d TFT given by the $\mathbb{Z}_p \times \mathbb{Z}_p$-gauge theory specified by the action 
\begin{equation}
    \frac{2\pi i}{p} \int_{\mathcal{M}_3} a_1^{(1)} \cup \delta \widehat{a}_1^{(1)} + a_2^{(1)}\cup \delta \widehat{a}_2^{(1)} ~,
\end{equation}
where $a_1^{(1)},a_2^{(1)},\widehat{a}_1^{(1)},\widehat{a}_2^{(1)}$ are discrete gauge fields in $C^1(\mathcal{M}_3,\mathbb{Z}_p)$. The theory contains $p^4$ invertible topological line operators, generated by the following $\mathbb{Z}_p$ Wilson lines:
\begin{equation}
    e_1 = e^{\frac{2\pi i}{p} \oint a_1^{(1)}} ~, \quad e_2 = e^{\frac{2\pi i}{p}\oint a_2^{(1)}} ~, \quad m_1 = e^{\frac{2\pi i}{p} \oint \widehat{a}_1^{(1)}} ~, \quad m_2 = e^{\frac{2\pi i}{p}\oint \widehat{a}_2^{(1)}} ~,
\end{equation}
with non-trivial braiding relations:
\begin{equation}
    B(e_1,m_1) = B(e_2,m_2) = e^{\frac{2\pi i}{p}} ~.
\end{equation}
Those lines generate 1-form symmetry $\mathbb{Z}_p^4$, and the braiding phases mean the 1-form symmetry has 't Hooft anomaly. However, one can always choose a maximal anomaly-free subgroup $L$; and gauging this subgroup will remove all the lines and lead to the trivial theory. Such a subgroup $L$ is called a Lagrangian subgroup or a Lagrangian algebra. They correspond to topological boundaries of the SymTFT: one can consider gauging the symmetry on half space with the Dirichlet boundary condition, and this creates a topological interface between the SymTFT and the trivial theory, which is equivalent to a topological boundary of the SymTFT, and we will denote it as $\mathcal{B}_L$. Notice that the lines in $L$ can terminate on this topological boundary. The Lagrangian algebras for this particular SymTFT are given by
\begin{equation}\label{eq:Zpp_LA}
    \langle e_1, e_2 \rangle ~, \quad \langle m_1, e_2 \rangle ~, \quad \langle m_2 m_1^k, e_1 e_2^{-k}\rangle ~, \quad \langle m_1 e_2^{-k}, m_2 e_1^{k}\rangle ~, \quad k \in \mathbb{Z}_p ~.
\end{equation}

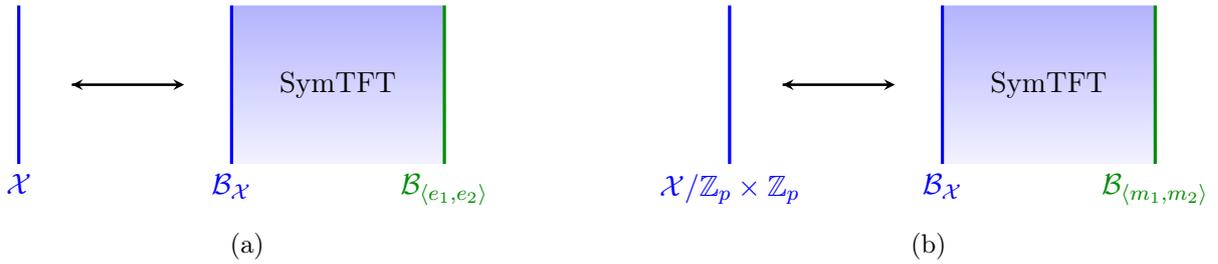
\begin{figure}[!tbp]
	\centering
    \begin{subfigure}{0.45\textwidth}
    \centering
	\begin{tikzpicture}[scale=0.7]
	\shade[line width=2pt, top color=blue!30, bottom color=blue!5] 
	(-3,0) to [out=90, in=-90]  (-3,3)
	to [out=0,in=180] (1,3)
	to [out = -90, in =90] (1,0)
	to [out=180, in =0]  (-3,0);
	
	\draw[very thick, blue] (-7,0) -- (-7,3);
	\node[blue, below] at (-7,0) {$\mathcal{X}$};
	\draw[thick, ->-=1] (-4,1.5) -- (-6, 1.5);
    \draw[thick, -<-=0.05] (-4,1.5) -- (-6, 1.5);
	
	\draw[very thick, blue] (-3,0) -- (-3,3);
	\draw[very thick, dgreen] (1,0) -- (1,3);
	\node at (-1,1.5) {SymTFT};
	\node[blue, below] at (-3,0) {$\mathcal{B}_{\mathcal{X}}$};
	\node[dgreen, below] at (1,0) {$\mathcal{B}_{\langle e_1, e_2\rangle}$}; 
	
	\end{tikzpicture}
    \caption{}
    \label{fig:SymTFT1}
    \end{subfigure}
    \hfill
    \begin{subfigure}{0.45\textwidth}
    \centering
 	\begin{tikzpicture}[scale=0.7]
	\shade[line width=2pt, top color=blue!30, bottom color=blue!5] 
	(-3,0) to [out=90, in=-90]  (-3,3)
	to [out=0,in=180] (1,3)
	to [out = -90, in =90] (1,0)
	to [out=180, in =0]  (-3,0);
	
	\draw[very thick, blue] (-7,0) -- (-7,3);
	\node[blue, below] at (-7,0) {$\mathcal{X}/\mathbb{Z}_p\times \mathbb{Z}_p$};
	\draw[thick, ->-=1] (-4,1.5) -- (-6, 1.5);
    \draw[thick, -<-=0.05] (-4,1.5) -- (-6, 1.5);
	
	\draw[very thick, blue] (-3,0) -- (-3,3);
	\draw[very thick, dgreen] (1,0) -- (1,3);
	\node at (-1,1.5) {SymTFT};
	\node[blue, below] at (-3,0) {$\mathcal{B}_{\mathcal{X}}$};
	\node[dgreen, below] at (1,0) {$\mathcal{B}_{\langle m_1, m_2\rangle}$}; 
	
	\end{tikzpicture}
    \caption{}
    \label{fig:SymTFT2}
    \end{subfigure}
    \caption{SymTFT for $\mathbb{Z}_p \times \mathbb{Z}_p$ 0-form symmetry in 2d QFT $\mathcal{X}$. In Figure \ref{fig:SymTFT1}, choosing the topological boundary $\mathcal{B}_{\langle e_1,e_2\rangle}$ and reducing along the interval recovers the original 2d theory $\mathcal{X}$. In Figure \ref{fig:SymTFT2}, replacing $\mathcal{B}_{\langle e_1,e_2\rangle}$ with $\mathcal{B}_{\langle m_1, m_2\rangle}$ and then reducing along the interval leads to gauging $\mathbb{Z}_p\times \mathbb{Z}_p$ with trivial discrete torsion in $\mathcal{X}$.}
    \label{fig:SymTFT}
\end{figure}

The basic idea is to place the SymTFT on an interval, as shown in Figure \ref{fig:SymTFT}. The left boundary is endowed with a generically non-topological boundary $\mathcal{B}_{\mathcal{X}}$ capturing the dynamics of 2d QFT $\mathcal{X}$, which we will refer to as the physical boundary; while the right boundary (referred to as the symmetry boundary) is the topological boundary specified by the pure electric Lagrangian subgroup $\langle e_1, e_2\rangle$. Since the bulk is topological, the length of the interval is unimportant; and one can take the zero-length limit upon which one recovers the 2d QFT $\mathcal{X}$. One can, of course, choose different topological boundary conditions on the symmetry boundary and then take the interval reduction. This corresponds to performing discrete gauging in the 2d theory $\mathcal{X}$. For instance, choosing $\mathcal{B}_{\langle m_1, e_2 \rangle}$ and $\mathcal{B}_{\langle m_2 m_1^k, e_1 e_2^{-k}\rangle}$  corresponds to gauging a $\mathbb{Z}_p$ subgroup in $\mathcal{X}$; while choosing boundary conditions $\mathcal{B}_{\langle m_1 e_2^{-k}, m_2 e_1^k \rangle}$ corresponds to gauging the entire $\mathbb{Z}_p\times \mathbb{Z}_p$-symmetry with discrete torsions specified by $k$ in $\mathcal{X}$.

The SymTFT can be used to classify gapped phases of a given symmetry. Notice that the physical boundary can also be topological. In this case, the reduced 2d theory is topological; therefore, it describes a symmetric gapped phase. Different gapped phases of a given symmetry can be acquired by holding fixed the symmetry boundary, and placing different topological boundaries on the physical boundary. For instance, let's fix the symmetry boundary to be $\mathcal{B}_{\langle e_1, e_2 \rangle}$, and choosing different topological boundary conditions on the physical boundary leads to different gapped phases of $\mathbb{Z}_p \times \mathbb{Z}_p$ symmetry: 
\begin{enumerate}
    \item $\mathcal{B}_{\langle m_1 e_2^{-k}, m_2 e_1^{k}\rangle}$ leads to $\mathbb{Z}_p\times \mathbb{Z}_p$-SPT phases, and the Lagrangian algebras $\langle m_1 e_2^{-k}, m_2 e_1^k\rangle$ are referred to as the magnetic Lagrangian algebras;
    \item $\mathcal{B}_{\langle e_1, e_2\rangle}$ leads to the $\mathbb{Z}_p \times \mathbb{Z}_p$ SSB phase;
    \item  $\mathcal{B}_{\langle m_1,e_2\rangle},\mathcal{B}_{\langle m_1 m_2^k,e_1 e_2^{-k}\rangle}$ leads to a partial SSB phase where some $\mathbb{Z}_p$ subgroup remains unbroken. 
\end{enumerate}

\

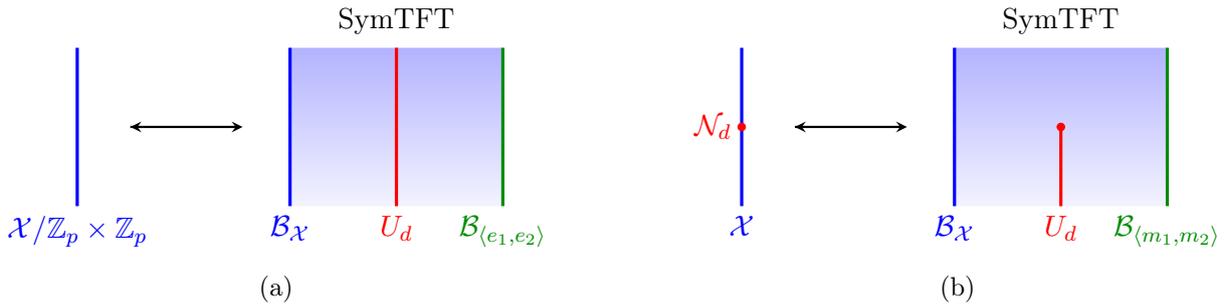
\begin{figure}[!tbp]
	\centering
    \begin{subfigure}{0.45\textwidth}
    \centering
	\begin{tikzpicture}[scale=0.7]
	\shade[line width=2pt, top color=blue!30, bottom color=blue!5] 
	(-3,0) to [out=90, in=-90]  (-3,3)
	to [out=0,in=180] (1,3)
	to [out = -90, in =90] (1,0)
	to [out=180, in =0]  (-3,0);
	
	\draw[very thick, blue] (-7,0) -- (-7,3);
	\node[blue, below] at (-7,0) {$\mathcal{X}/\mathbb{Z}_p\times \mathbb{Z}_p$};
	\draw[thick, ->-=1] (-4,1.5) -- (-6, 1.5);
    \draw[thick, -<-=0.05] (-4,1.5) -- (-6, 1.5);
	
	\draw[very thick, blue] (-3,0) -- (-3,3);
	\draw[very thick, dgreen] (1,0) -- (1,3);
	\node at (-1,3.5) {SymTFT};

    \draw[very thick, red] (-1,0) -- (-1,3);
    \node[below, red] at (-1,0) {$U_d$};
    
	\node[blue, below] at (-3,0) {$\mathcal{B}_{\mathcal{X}}$};
	\node[dgreen, below] at (1,0) {$\mathcal{B}_{\langle e_1, e_2\rangle}$}; 
	
	\end{tikzpicture}
    \caption{}
    \label{fig:SymTFT_td1}
    \end{subfigure}
    \hfill
    \begin{subfigure}{0.45\textwidth}
    \centering
 	\begin{tikzpicture}[scale=0.7]
	\shade[line width=2pt, top color=blue!30, bottom color=blue!5] 
	(-3,0) to [out=90, in=-90]  (-3,3)
	to [out=0,in=180] (1,3)
	to [out = -90, in =90] (1,0)
	to [out=180, in =0]  (-3,0);
	
	\draw[very thick, blue] (-7,0) -- (-7,3);
	\node[blue, below] at (-7,0) {$\mathcal{X}$};
	\draw[thick, ->-=1] (-4,1.5) -- (-6, 1.5);
    \draw[thick, -<-=0.05] (-4,1.5) -- (-6, 1.5);

    \draw[very thick, red] (-1,0) -- (-1,1.5);
    \filldraw[red] (-1,1.5) circle (2pt);
    \node[red, below] at (-1,0) {$U_d$};
    
	\draw[very thick, blue] (-3,0) -- (-3,3);
	\draw[very thick, dgreen] (1,0) -- (1,3);
	\node at (-1,3.5) {SymTFT};
	\node[blue, below] at (-3,0) {$\mathcal{B}_{\mathcal{X}}$};
	\node[dgreen, below] at (1,0) {$\mathcal{B}_{\langle m_1, m_2\rangle}$}; 
    \filldraw[red] (-7,1.5) circle (2pt);
    \node[red, left] at (-7,1.5) {$\mathcal{N}_d$};
    
	\end{tikzpicture}
    \caption{}
    \label{fig:SymTFT_td2}
    \end{subfigure}
    \caption{Relation between the bulk symmetry $U_d$ and the duality defect $\mathcal{N}_d$ in $\mathcal{X}$. In Figure \ref{fig:SymTFT_td1}, the interval reduction with $U_d$ insertion leads to the theory $\mathcal{X}/\mathbb{Z}_p\times \mathbb{Z}_p$ generically. In Figure \ref{fig:SymTFT_td2}, the bulk symmetry $U_d$ may end on a topological line operator in the bulk (known as the twist defect). Interval reduction with the $U_d$ twist defect insertion leads to an topological interface between theory $\mathcal{X}$ and $\mathcal{X}/\mathbb{Z}_p\times \mathbb{Z}_p$. If $\mathcal{B}_\mathcal{X}$ is invariant under $U_d$, in other words $\mathcal{X} \simeq \mathcal{X}/\mathbb{Z}_p\times \mathbb{Z}_p$ via the identification of the symmetry specified by $U_d$, then this topological interface becomes the duality defect $\mathcal{N}_d$.}
    \label{fig:SymTFT_td}
\end{figure}

SymTFT is extremely powerful in the study of duality-like non-invertible symmetry. This is because the non-invertible $\TY$ duality defect $\mathcal{N}$ in the 2d theory $\mathcal{X}$ corresponds to the invertible symmetry permuting the topological lines in the SymTFT. For instance, the $\mathcal{N}_o$ and $\mathcal{N}_d$ duality defects in $\TYpo$ and $\TYpd$ mentioned previously correspond to the following bulk symmetries:
\begin{equation}
    U_o = \begin{pmatrix} 0 & 0 & 0 & 1 \\ 0 & 0 & 1 & 0 \\ 0 & 1 & 0 & 0 \\ 1 & 0 & 0 & 0\end{pmatrix} ~, \quad U_d = \begin{pmatrix} 0 & 0 & 1 & 0 \\ 0 & 0 & 0 & 1 \\ 1 & 0 & 0 & 0 \\ 0 & 1 & 0 & 0\end{pmatrix} ~,
\end{equation}
where the matrix acts on the charge vector $(q_1,q_2,\widehat{q}_1,\widehat{q}_2)^T$ of the Wilson line $e^{\frac{2\pi i}{p}\oint q_1 a_1+q_2 a_2 + \widehat{q}_1 \widehat{a}_1 + \widehat{q}_2 \widehat{a}_2}$. The relation between the bulk symmetry and the boundary duality defect is illustrated in Figure \ref{fig:SymTFT_td}. To describe the relation and distinguish between $U_d$ and $U_o$, one can consider a simplified presentation for the interval reduction, where two boundaries are represented as states of the TQFT Hilbert space\footnote{Strictly speaking, the bare boundary condition $\mathcal{B}_{\langle e_1,e_2\rangle}$ corresponds to $A_1^{(1)} = A_2^{(1)} = 0$, and generic states $|A_1^{(1)}, A_2^{(1)}\rangle$ are acquired by pushing magnetic lines $m_i$ to the bare boundary.}:
\begin{equation}
    \mathcal{B}_\mathcal{X} \rightarrow \langle \mathcal{X}| = \sum_{a^{(1)}_i \in H^1(\mathcal{M}_2,\mathbb{Z}_p)} \langle a^{(1)}_1,a^{(1)}_2| Z_{\mathcal{X}}[a^{(1)}_1,a^{(1)}_2] ~, \quad \quad  \mathcal{B}_{\langle e_1, e_2 \rangle} \rightarrow | A_1^{(1)}, A_2^{(1)}\rangle ~,
\end{equation}
where $Z_{\mathcal{X}}[a^{(1)},a^{(2)}]$ is the partition function of $\mathcal{X}$ coupled to background fields $a_i^{(1)}$, and $| a_1^{(1)}, a_2^{(2)}\rangle$ is an orthonormal basis of states diagonalizing the electric lines $e_i$:
\begin{equation}
    e_i(\gamma)|a_1^{(1)},a_2^{(1)}\rangle = e^{\frac{2\pi i}{p}\oint_\gamma a_i^{(1)}} |a_1^{(1)}, a_2^{(1)} \rangle ~, \quad m_i(\gamma) |a_1^{(1)},a_2^{(1)}\rangle = |a_1^{(1)} - \delta_{1i}[\gamma], a_2^{(1)} - \delta_{2i} [\gamma]\rangle ~,
\end{equation}
where $[\gamma] \in H^1(\mathcal{M}_2,\mathbb{Z}_p)$ denotes the Poincar\'e dual of $\gamma$. The interval reduction is then captured as the inner product
\begin{equation}
    \langle \mathcal{X}|A^{(1)}_1,A^{(1)}_2\rangle  = Z_{\mathcal{X}}[A^{(1)}_1,A^{(1)}_2] ~,
\end{equation}
which reproduces the partition function of $\mathcal{X}$.

Both $U_o$ and $U_d$ map $\mathcal{B}_{\langle e_1,e_2\rangle}$ to $\mathcal{B}_{\langle m_1, m_2\rangle}$, meaning they all implement $\mathbb{Z}_p \times \mathbb{Z}_p$ gauging when pushing to the symmetry boundary, as shown in Figure \ref{fig:SymTFT_td1}. Their difference can be seen from the above inner product picture--the interval reduction with $U$'s inserted can be seen as computing $\langle \mathcal{X}|U|A^{(1)}_1,A^{(1)}_2\rangle$, and we find
\begin{equation}\label{eq:TY_gaugingbk}
\begin{aligned}
    \langle \mathcal{X}|U_o|A^{(1)}_1, A^{(1)}_2\rangle &= \frac{1}{|H^1(\mathcal{M}_2,\mathbb{Z}_p)|} \sum_{a^{(1)}_i} Z_{\mathcal{X}}[a_1^{(1)},a_2^{(1)}] \exp\left(\frac{2\pi i}{p}\int_{\mathcal{M}_2} a_1^{(1)}\cup A_1^{(1)} + a_2^{(1)} \cup A_2^{(1)}\right) ~, \\
    \langle \mathcal{X}|U_d|A^{(1)}_1, A^{(1)}_2\rangle &= \frac{1}{|H^1(\mathcal{M}_2,\mathbb{Z}_p)|} \sum_{a^{(1)}_i} Z_{\mathcal{X}}[a_1^{(1)},a_2^{(1)}] \exp\left(\frac{2\pi i}{p}\int_{\mathcal{M}_2} a_1^{(1)}\cup A_2^{(1)} + a_2^{(1)} \cup A_1^{(1)}\right) ~, \\
\end{aligned}
\end{equation}
where we used
\begin{equation}
\begin{aligned}
    & U_d|A_1^{(1)},A_2^{(1)}\rangle = \frac{1}{|H^1(\mathcal{M}_2,\mathbb{Z}_p)|} \sum_{b^{(1)}_i} \exp\left(\frac{2\pi i}{p}\int_{\mathcal{M}_2} a_1^{(1)}\cup A_1^{(1)} + a_2^{(1)} \cup A_2^{(1)} \right) |b_1^{(1)},b_2^{(1)}\rangle ~, \\
    & U_o|A_1^{(1)},A_2^{(1)}\rangle = \frac{1}{|H^1(\mathcal{M}_2,\mathbb{Z}_p)|} \sum_{b^{(1)}_i} \exp\left(\frac{2\pi i}{p}\int_{\mathcal{M}_2} b_1^{(1)}\cup A_2^{(1)} + b_2^{(1)} \cup A_1^{(1)} \right) |b_1^{(1)},b_2^{(1)}\rangle ~. \\
\end{aligned}
\end{equation}
We see that both $U_d$ and $U_o$ lead to $\mathbb{Z}_p \times \mathbb{Z}_p$-gauging, however, the coupling to dual symmetry background fields is different. 

In the bulk, the symmetry operators can terminate on topological line operators, and the configuration is known as the twist defect. As illustrated in Figure \ref{fig:SymTFT_td2}, interval reduction with a twist defect will generically lead to an interface between theory $\mathcal{X}$ and $\mathcal{X}/\mathbb{Z}_p\times \mathbb{Z}_p$. However, in the special case where $\mathcal{B}_\mathcal{X}$ is invariant under the $U_d$ (or $U_o$), that is, $\langle \mathcal{X}|U_{d,o}|A_1^{(1)},A_2^{(1)}\rangle = \langle \mathcal{X}|A_1^{(1)},A_2^{(1)}\rangle$ for any $A_1,A_2$ respectively, the interface will become a topological defect inside the theory $\mathcal{X}$. This is precisely the corresponding duality defect $\mathcal{N}_d$ (or $\mathcal{N}_o$)\footnote{Notice that the FS indicator of $\mathcal{N}$ cannot be determined at this stage.}. 

Therefore, any $\mathbb{Z}_p\times \mathbb{Z}_p$-SPT phases may admit the $\mathcal{N}_o$ (or $\mathcal{N}_d$) duality defect only if the corresponding magnetic Lagrangian algebra is invariant (or stable) under the $U_o$ (or $U_d$) symmetry. In other words, a necessary condition for $\mathcal{N}_{o}$ (or $\mathcal{N}_d$) to admit an SPT phase is that there exists a stable magnetic Lagrangian algebra under $U_o$ (or $U_d$) in the SymTFT of $\mathbb{Z}_p\times \mathbb{Z}_p$.

Notice that one can consider the same construction in Figure \ref{fig:SymTFT_td2} for other bulk invertible symmetries, and it leads to an invertible symmetry on the boundary if the bulk symmetry leaves $\mathcal{B}_{\langle e_1,e_2\rangle}$ invariant. One can ask if the duality defect $\mathcal{N}$ can be made invertible under discrete gauging. Since any discrete gauging can be realized by choosing a gapped boundary $\mathcal{B}_{L}$ for some Lagrangian algebra $L$, this means the $\mathcal{N}$ can be made invertible if there exists a Lagrangian algebra stable under the corresponding bulk symmetry $U$\footnote{Notice that this argument only shows that the existence of $U$-stable Lagrangian algebra is a sufficient condition for the corresponding boundary line to be group-theoretical. It is further proved that this condition is necessary for fusion category symmetry in 2d \cite{Gelaki:2009blp}, and it is conjectured to hold for higher dimension as well, see \cite{Sun:2023xxv}.}. Combined with the previous result, we see that a $\TY$-fusion category admits an SPT only if the duality defect can be made invertible, that is, it is group-theoretical. Notice that the same statement holds for the fusion category constructed from any bulk symmetry $G$ in the SymTFT for an Abelian group $A$ \cite{Gelaki:2009blp}.

\section{Constructing of intrinsic NISPT from mixed anomaly}\label{sec:intrinNISPT}
As reviewed in Section \ref{sec:SymTFT_review}, any $G$-extension of $\VEC_{A}$ may admit an SPT phase only if it is group-theoretical. To construct an intrinsic NISPT phase, we must bypass this no-go theorem. The most straightforward way is to consider $G$-extension of fusion categories other than $\VEC_A$, but the bulk symmetry of a generic SymTFT is hard to enumerate and we do not know what category we should start with in the first place.

To make progress, we make the following two observations. First, if a categorical symmetry is non-group-theoretical, then any dual categorical symmetry acquired from some discrete gauging must also be non-invertible and non-group-theoretical. Otherwise, by composing two gaugings, one could map the initial categorical symmetry to a finite group symmetry, which leads to a contradiction. Second, if the anomaly is a mixed anomaly, even though there will be ground state degeneracy, the spontaneously broken symmetry can still be gauged due to the lack of self-anomaly. This will remove the ground state degeneracy and lead to an SPT phase of the dual symmetry.

Combining the two observations, we find that we could start with some non-group-theoretical $G$-extension $\mathcal{C}$ of $\VEC_{A}$. While this does imply that $\mathcal{C}$ is anomalous, we could choose $G$ such that the anomaly of $\mathcal{C}$ is a mixed anomaly. Starting with the partial SSB phase and gauging the anomaly-free, spontaneously broken symmetry will exactly lead to an intrinsic NISPT phase. In the following, we will first describe a $\mathbb{Z}_2 \times \mathbb{Z}_2$-extension of $\VEC_{\mathbb{Z}_p\times \mathbb{Z}_p}$ with a mixed anomaly in Section \ref{sec:mixed_anom_ext}. Then, we discuss the anomaly-free non-group-theoretical dual categorical symmetry acquired by gauging and the classification of its SPT phases in Section \ref{sec:iNISPTclassify}. Finally, we provide a lattice model realizing a specific intrinsic NISPT phase in Section \ref{sec:iNISPT_lattice}.

\subsection{A $\doubleZ_2 \times \doubleZ_2$-graded extension of $\VEC_{\doubleZ_p \times \doubleZ_p}$}\label{sec:mixed_anom_ext}
We now describe the $\doubleZ_2 \times \doubleZ_2$-graded extension of $\VEC_{\doubleZ_p\times \doubleZ_p}$ denoted as $\CC_p$. For simplicity, we restrict ourselves to the case where $p$ is an odd prime number. As alluded to before, we want each $\doubleZ_2$-graded extension to be anomaly-free, and there is only a mixed anomaly between them. So we take the first $\doubleZ_2$-extension factor to be the off-diagonal duality defect $\CN_o$ with $+1$-FS indicator, and the second $\mathbb{Z}_2$-extension factor to be the swap symmetry exchanging two copies of $\mathbb{Z}_p\times \mathbb{Z}_p$ with no self-anomaly. 

The resulting fusion category $\mathcal{C}_p$ has $2p^2$ invertible lines which form the group $(\doubleZ_p \times \doubleZ_p)\rtimes \doubleZ_2^t \simeq D_{2p} \times \doubleZ_p$. Furthermore, it contains two $\TY$-duality defects $\CN_{d}$ and $\CN_{o}$ associated with diagonal gauging and off-diagonal gauging respectively, and both have FS indicator $+1$. Parameterizing its simple objects as
\begin{equation}
    g ~, \quad \mathcal{N}_{o} ~, \quad gt ~, \quad \mathcal{N}_{o} t \equiv \CN_d ~, \quad g \equiv (\mathrm{g}_1,\mathrm{g}_2) \in \mathbb{Z}_p \times \mathbb{Z}_p ~, 
\end{equation}
the fusion rule is given by
\begin{equation}
    gt^\alpha  \otimes ht^\beta =  g ({}^\alpha h) t^{\alpha + \beta}~, \quad gt^\alpha  \otimes \mathcal{N}_{o} t^\beta = \mathcal{N}_{o} t^{\alpha+\beta} ~, \quad \mathcal{N}_{o}t^\alpha \otimes g t^\beta = \mathcal{N}_{o}t^{\alpha+\beta} ~, \quad \mathcal{N}_{o}t^\alpha \otimes \mathcal{N}_{o} t^{\beta} = \left(\sum_g g\right)t^{\alpha+\beta} ~,
\end{equation}
where we use the $g = (\mathrm{g}_1,\mathrm{g}_2) \in \doubleZ_p \times \doubleZ_p$, and ${}^\beta g = g$ if $\beta = 0$ and ${}^\beta g = (\mathrm{g}_2,\mathrm{g}_1)$ if $\beta = 1$. The $F$-symbols are given by:
\begin{equation}
    F^{at^\alpha, \mathcal{N}_o t^\beta,  c t^\gamma}_{\mathcal{N}_ot^{\alpha+\beta+\gamma}} = \chi_o(a,{}^{\alpha+\beta}c) ~, \quad F^{\mathcal{N}_o t^\alpha, bt^\beta , \mathcal{N}_o t^\gamma}_{d t^{\alpha+\beta+\gamma}} = \chi_o({}^\alpha b,d) ~, \quad \left[F^{t^\alpha\mathcal{N}_o,t^\beta\mathcal{N}_o,t^\gamma\mathcal{N}_o}_{t^{\alpha+\beta+\gamma}\mathcal{N}_o}\right]_{at^{\alpha+\beta},bt^{\beta+\gamma}} = \frac{1}{p} \frac{1}{\chi_o(a,{}^\alpha b)} ~,
\end{equation}
where $\chi_o(a,b) = e^{\frac{2\pi i}{p}(\mathrm{a}_1 \mathrm{b}_2 + \mathrm{a}_2 \mathrm{b}_1)}$ is the off-diagonal bicharacter. It is useful to notice that when $\alpha = 1$, $\chi_o(a,{}^\alpha b) = e^{\frac{2\pi i}{p} (\mathrm{a}_1 \mathrm{b}_1 + \mathrm{a}_2 \mathrm{b}_2)} = \chi_d(a,b)$ becomes the diagonal bicharacter of $\mathbb{Z}_p\times \mathbb{Z}_p$. This implies that all the invertible lines together with $\CN_d = t\CN_o$ generate $\TYpd$ with diagonal bicharacter.

For future use, we want to point out that this fusion category admits two constructions as crossed product categories. Let us first recall the concept of crossed product categories. Consider a fusion category $\CC$ together with a $G$-action on it, the crossed product $\CC\rtimes G$ is defined as the Abelian category $\CC\boxtimes \VEC_G$ equipped with the following tensor product
\begin{equation}
    (X\boxtimes g) \otimes (Y\boxtimes h) := (X\otimes T_g(Y)) \boxtimes gh ~, \quad X,Y \in \CC ~, \quad g,h \in G ~,
\end{equation}
where $T_g$ denotes the action of $g\in G$ and the associativity constraints come from those in $\CC$. Notice that the above fusion rules are in reminiscence of the definition of semi-direct product of finite groups. 

The group of tensor autoequivalences for TY fusion category $\TY(A,\chi,\epsilon)$ can be computed straightforwardly, and is given by the group $\Aut(A,\chi)$, that is the subgroup of group automorphisms of $A$ preserving the bicharacter $\chi$:
\begin{equation}
    \chi({}^g a, {}^g b) = \chi(a,b) ~, \quad g \in \Aut(A,\chi) ~,
\end{equation}
where ${}^g a$ denotes the $g$ action on $a \in A$. The action $T_g$ on $\TY$ as
\begin{equation}
    T_g(a) = {}^ga ~, \quad T_g(\CN) = \CN ~,
\end{equation}
and it does not introduce any additional phase on all local fusion junctions.

It is straightforward to check that the fusion category $\CC_p$ can be written as a crossed product in the following two ways:
\begin{equation}
    \CC_p = \TYpd \rtimes \mathbb{Z}_2^t = \TYpo \rtimes \mathbb{Z}_2^t ~.
\end{equation}
This also means that one can view $\CC_p$ as $\mathbb{Z}_2^t$-graded extension of $\TYpo$ or $\TYpd$. 

\subsubsection*{Proof of non-group-theoretical}
We are now ready to prove an important property of the fusion category $\CC_p$, namely, it is non-group-theoretical. For this, we can use the result stated in Section \ref{sec:SymTFT_review} to show that there is no Lagrangian algebra stable under the bulk $\mathbb{Z}_2 \times \mathbb{Z}_2$ symmetry in the SymTFT. First, notice that generators of the $\mathbb{Z}_2\times \mathbb{Z}_2$ symmetry act as
\begin{equation}
    U_t = \begin{pmatrix} 0 & 1 & 0 & 0 \\ 1 & 0 & 0 & 0 \\ 0 & 0 & 0 & 1 \\ 0 & 0 & 1 & 0 \end{pmatrix} ~, \quad U_{o} = \begin{pmatrix} 0 & 0 & 0 & 1 \\ 0 & 0 & 1 & 0 \\ 0 & 1 & 0 & 0 \\ 1& 0 & 0 & 0 \end{pmatrix} ~.
\end{equation}
With the full list of Lagrangian algebras given in \eqref{eq:Zpp_LA}, it is straightforward to check there is no Lagrangian algebra invariant under both $U_t$ and $U_o$. Then, by the theorem given in \cite{Gelaki:2009blp}, the fusion category $\mathcal{C}_p$ (as well as any dual symmetry acquired by discrete gauging) is non-group-theoretical.

\subsubsection*{Anomaly structure of $\CC_p$}
An implication of the category $\mathcal{C}_p$ being non-group-theoretical (in the context that it is an $\mathbb{Z}_2 \times \mathbb{Z}_2$-extension of $\VEC_{\doubleZ_p\times \doubleZ_p}$) is that it is automatically anomalous. Furthermore, its anomaly can be understood as a mixed anomaly between two of its subcategories.

Generically, let $\mathcal{C}$ be a fusion category symmetry and let $\mathcal{D}_i$'s be subcategories of $\CC$ such that $\mathcal{C}$ is generated by $\mathcal{D}_i$'s. We say there is a \textit{mixed anomaly} among $\mathcal{D}_i$'s if each $\mathcal{D}_i$ admits a trivially gapped phase, but not the entire $\mathcal{C}$. This generalizes the notion of mixed anomaly of invertible symmetries. However, while for finite group $G$ we use the terminology mixed anomaly when $G$ splits into (semi-)direct product of $H_i$ most of the time, for categorical symmetry $\mathcal{C}$ generically it does not split as finite groups do.

The category $\mathcal{C}_p$ is an example of the mixed anomaly between the $\TY$-fusion category $\TYpo$ and the $(\mathbb{Z}_p\times\mathbb{Z}_p)\rtimes \mathbb{Z}_2^t$ invertible symmetries. To see this, we can show that for one subcategory, any of its SPT phases is not invariant under the global transformation of the other subcategory. Let's consider the SPT phases $\mathcal{F}_{o,\pm}$ of $\TY(\mathbb{Z}_p\times \mathbb{Z}_p, \chi_{o},+1)$. As shown explicitly in Appendix \ref{app:ff}, its SPT phases have the twisted partition functions coupled to $\mathbb{Z}_p\times \mathbb{Z}_p$ background fields 
\begin{equation}
    Z_{\mathcal{F}_{o,\pm}}[A^{(1)},B^{(1)}] = \exp\left(\pm \frac{2\pi i}{p}\int_{\mathcal{M}_2} A^{(1)}\cup B^{(1)}\right) ~, \quad A^{(1)},B^{(1)} \in H^1(\mathcal{M}_2,\mathbb{Z}_p) ~.
\end{equation}
Under the swap symmetry, $A^{(1)}\leftrightarrow B^{(1)}$,
\begin{equation}
    t: e^{\pm \frac{2\pi i}{p} \int A^{(1)} \cup B^{(1)}} \mapsto e^{\pm \frac{2\pi i}{p} \int B^{(1)}\cup A^{(1)}} = e^{\mp \frac{2\pi i}{p} \int A^{(1)}\cup B^{(1)}} ~,
\end{equation}
hence $\mathcal{F}_{o,\pm}$ are exchanged under the $t$ action, meaning the full symmetry cannot admit an SPT phase. 

Alternatively, $(\mathbb{Z}_p\times\mathbb{Z}_p)\rtimes \mathbb{Z}_2^t$ admits a unique SPT phase\footnote{This can be shown by directly computing the $H^2((\mathbb{Z}_p\times\mathbb{Z}_p)\rtimes \mathbb{Z}_2^t,U(1))$ using the LHS spectral sequence.} which has the twisted partition function when coupling to the $\mathbb{Z}_p\times \mathbb{Z}_p$ background fields:
\begin{equation}
    Z[A^{(1)},B^{(1)}] = 1 ~,
\end{equation}
but it will not be invariant under the off-diagonal gauging, as
\begin{equation}
    \CN_o: 1 \mapsto \frac{1}{\sqrt{|H^1(\mathcal{M}_2,\mathbb{Z}_p\times \mathbb{Z}_p)|}}\sum_{a^{(1)},b^{(1)} \in H^1(\mathcal{M}_2,\mathbb{Z}_p)} 1 \cdot e^{\frac{2\pi i}{p}\int a^{(1)}\cup B^{(1)} + b^{(1)}\cup A^{(1)}} = \delta(A) \delta(B) \neq 1 ~.
\end{equation}

To conclude, we see both subcategories $\TYpo$ and $(\mathbb{Z}_p\times \mathbb{Z}_p)\rtimes \mathbb{Z}_2^t$ are anomaly-free as they admit SPT phases. However, the full symmetry $\mathcal{C}_p$ is anomalous as it does not admit an SPT phase.

\subsection{Anomaly-free intrinsic non-invertible symmetries and classification of its SPTs}\label{sec:iNISPTclassify}
We are now ready to construct the anomaly-free non-group-theoretical categorical symmetry and classify its SPT phases. This is done by gauging the $\mathbb{Z}_2^t$ subgroup of $\CC_p$, and we will show the resulting symmetry $\widetilde{\CC}_p$ is anomaly-free and classify its SPT phases by relating to the SSB phases of $\CC_p$ symmetries generalizing the techniques used in \cite{Seifnashri:2024dsd,Cao:2025qhg}.

\subsubsection*{Dual Symmetry $\widetilde{\CC}_p$}\label{sec:nongrpcat}
Let us first describe the dual symmetry $\widetilde{\CC}_p$ one gets from gauging the $\mathbb{Z}_2^t$-subgroup of $\CC_p$. Here, we will only present the simple symmetry lines and their fusion rules, while the $F$-symbols can be computed following e.g. \cite{Choi:2023vgk}.

The dual symmetry admits a $\doubleZ_2$-grading inherent from the $\mathbb{Z}_2^o$-grading in $\CC_p$. The trivial grading component is acquired from the dual symmetry of gauging $\mathbb{Z}_2^t$ of the fusion subcategory $\VEC_{(\mathbb{Z}_p^a\times \mathbb{Z}^b_p) \rtimes \doubleZ_2^t} \subset \CC_p$. Notice that one can re-parameterize $\mathbb{Z}_p^a \times \mathbb{Z}_p^b$ as 
\begin{equation}
    \widetilde{a} = ab^{-1} ~, \quad \widetilde{b} = ab ~,
\end{equation}
which is valid because $p$ is odd. We see that $t$ commutes with $\widetilde{b}$, but acts on $\widetilde{a}$ as charge conjugation $t\widetilde{a} t^{-1} = \widetilde{a}^{-1}$ therefore $\mathbb{Z}_p^{\widetilde{a}}\rtimes \mathbb{Z}_2\simeq D_{2p}$. Thus, $\VEC_{(\mathbb{Z}_p^a\times \mathbb{Z}^b_p) \rtimes \doubleZ_2^t} = \VEC_{D_{2p}}\boxtimes \VEC_{\mathbb{Z}_p^{\widetilde{b}}}$, and gauging $\mathbb{Z}_2^t$ does not affect $\mathbb{Z}_p^{\widetilde{b}}$, but will change $\VEC_{D_{2p}}$ to $\Rep D_{2p}$\footnote{This can be seen as follows: gauging $D_{2p}$ in $\VEC_{D_{2p}}$ leads to $\Rep D_{2p}$, but this gauging can be decomposed as first gauging $\mathbb{Z}_p$ subgroup and then gauge $\mathbb{Z}_2^t$. But gauging $\mathbb{Z}_p$ in $\VEC_{D_{2p}}$ still leads to $\VEC_{D_{2p}}$, hence gauging $\mathbb{Z}_2^t$ in $\VEC_{D_{2p}}$ leads to $\Rep D_{2p}$.}.

$\Rep D_{2p}$ admits a $\doubleZ_2^{\widetilde{t}}$ subgroup symmetry arising with the dual symmetry of $\doubleZ_2^t$, as well as $\frac{p-1}{2}$ non-invertible lines $X_i$ with quantum dimension $2$, which intuitively arise from
\begin{equation}
    X_i = a^i b^{-i} + a^{-i} b^i ~, \quad 1 \leq i \leq \frac{p-1}{2} ~, \quad X_i = X_{-i} = X_{i+p} ~.
\end{equation}
The fusion rules are given by
\begin{equation}
    \widetilde{t} \times X_i = X_i \times \widetilde{t} = X_i ~, \quad X_i \times X_j = \begin{cases} X_{i+j} + X_{i-j} ~, \quad \quad i + j \neq 0 \mod p ~, \quad i-j\neq 0 \mod p ~, \\ 1 + \widetilde{t} + X_{i-j} ~, \quad \quad i + j = 0 \mod p ~, \quad i -j \neq 0 \mod p ~, \\
    1+\widetilde{t} + X_{i+j} ~, \quad \quad i + j \neq 0 \mod p ~, \quad i -j = 0 \mod p~. \quad \end{cases} 
\end{equation}
For the $\Rep \doubleZ_p$ invertible symmetry, we denote its generator as $s$ and $s$ comes from $a^{-1}b^{-1} \in \mathcal{C}_p$ which commutes with $t$. 

The non-trivial grading component contains two partial duality defects $\widetilde{\CN}$ and $\widetilde{t}\widetilde{\CN} = \widetilde{\CN} \widetilde{t}$ with the fusion rules given by
\begin{equation}
    \widetilde{\CN}^2 = 1 + s + s^2 + \cdots + s^{p-1} + \sum_{j = 0}^{p-1} \sum_{i=1}^{\frac{p-1}{2}} s^j X_i ~.
\end{equation}
By the general argument in \cite{Diatlyk:2023fwf}, this fusion rule implies that $\widetilde{\CN}$ describes a self-duality under gauging the algebra object given by 
\begin{equation}
    \CA = 1 + s + s^2 + \cdots + s^{p-1} + \sum_{j = 0}^{p-1} \sum_{i=1}^{\frac{p-1}{2}} s^j X_i ~,
\end{equation}
in $\Rep D_{2p}\boxtimes \Rep\doubleZ_p$, that is, it is a (partial) self-duality under gauging non-invertible symmetries. As discussed in the beginning of this section, even though this fusion category is not group-theoretical, it could be anomaly-free as this is a $\mathbb{Z}_2$-extension of a category other than $\VEC_A$. We will now prove that it is indeed anomaly-free by constructing and classifying its SPT phases explicitly.

\subsubsection*{Classification of SPT phases from SSB phases of the mixed-anomalous symmetry}
We now prove the existence of the SPT phase of the dual non-invertible symmetries as well as classify them by generalizing the techniques in the group-theoretical case. We start by noticing that any SPT phase of the dual symmetry $\widetilde{\mathcal C}_p$ is acquired by an SSB phase of $\mathcal{C}_p$ with exactly two degenerate ground states and the $\doubleZ_2^t$ swap symmetry spontaneously broken. Therefore, to classify the SPT phases of the dual symmetry is equivalent to classifying such SSB phases.

To understand such SSB phases, we consider restricting the symmetry to be $\mathbb{Z}_p\times \mathbb{Z}_p$ invertible symmetry inside $\mathcal{C}_p$. The two degenerate ground states must each realize an $\mathbb{Z}_p \times \mathbb{Z}_p$-SPT phase, and the two SPT phases are exchanged by the $t$-symmetry. Furthermore, the two $\mathbb{Z}_p \times \mathbb{Z}_p$-SPT phases combined must be closed under the $\mathcal{N}_o$ and $\mathcal{N}_d$ actions. And we find the following two cases possible:
\begin{equation}
    \begin{tikzpicture}[baseline=0, square/.style={regular polygon,regular polygon sides=4}]
	\filldraw[black] (-1,0) circle (2pt);
    \filldraw[black] (+1,0) circle (2pt);
	\node[above] at (-1.5,0) {\footnotesize $\SPT_{+1}$};
    \node[above] at (+1.5,0) {\footnotesize $\SPT_{-1}$};

    \draw[thick, ->-=1] (-0.8,0) -- (0.8,0);
    \draw[thick, -<-=0.1] (-0.8,0) -- (0.8,0);
    \node[above] at (0,0) {\scriptsize $\mathcal{N}_d,t$};

    \draw[thick, smooth, ->-=1] (-1.2,-0.2) arc (120:420:0.4);
    \node[below] at (-1,-1) {\scriptsize $\mathcal{N}_o$};

    \draw[thick, smooth, ->-=1] (0.8,-0.2) arc (120:420:0.4);
    \node[below] at (1,-1) {\scriptsize $\mathcal{N}_o$};

    \node[below] at (0,-1.5) {(I)};
	\end{tikzpicture} ~, \quad \begin{tikzpicture}[baseline=0, square/.style={regular polygon,regular polygon sides=4}]
	\filldraw[black] (-1,0) circle (2pt);
    \filldraw[black] (+1,0) circle (2pt);
	\node[above] at (-1.5,0) {\footnotesize $\SPT_{+x}$};
    \node[above] at (+1.5,0) {\footnotesize $\SPT_{-x}$};

    \draw[thick, ->-=1] (-0.8,0) -- (0.8,0);
    \draw[thick, -<-=0.1] (-0.8,0) -- (0.8,0);
    \node[above] at (0,0) {\scriptsize $\mathcal{N}_o,t$};

    \draw[thick, smooth, ->-=1] (-1.2,-0.2) arc (120:420:0.4);
    \node[below] at (-1,-1) {\scriptsize $\mathcal{N}_d$};

    \draw[thick, smooth, ->-=1] (0.8,-0.2) arc (120:420:0.4);
    \node[below] at (1,-1) {\scriptsize $\mathcal{N}_d$};

    \node[below] at (0,-1.5) {(II)};
	\end{tikzpicture}
\end{equation}
where in the case (II) $x$ satisfies $x^2 = -1$ mod $p$, which only exists when $p = 1 \mod 4$ and can be represented as $x = \left(\frac{p-1}{2}\right)!$.

Let's study the case (I) first. In this case, we must enrich the $\SPT_{\pm 1}$ on each vacuum to become a $\TY_{p,o}$-SPT phase. As discussed in Section \ref{sec:SPTgrpt}, there is a unique enrichment for each case which turns $\SPT_{\pm}$ to $\TYpo$-SPT phases denoted as $\mathcal{F}_{o,\pm}$ respectively. Thus, the case (I) becomes
\begin{equation}\label{eq:2d_erc_I}
    \text{(I)}:\quad \quad  
    \begin{tikzpicture}[baseline=0, square/.style={regular polygon,regular polygon sides=4}]
	\filldraw[black] (-1,0) circle (2pt);
    \filldraw[black] (+1,0) circle (2pt);
	\node[above] at (-1.5,0) {\footnotesize $\SPT_{+1}$};
    \node[above] at (+1.5,0) {\footnotesize $\SPT_{-1}$};

    \draw[thick, ->-=1] (-0.8,0) -- (0.8,0);
    \draw[thick, -<-=0.1] (-0.8,0) -- (0.8,0);
    \node[above] at (0,0) {\scriptsize $\mathcal{N}_d,t$};

    \draw[thick, smooth, ->-=1] (-1.2,-0.2) arc (120:420:0.4);
    \node[below] at (-1,-1) {\scriptsize $\mathcal{N}_o$};

    \draw[thick, smooth, ->-=1] (0.8,-0.2) arc (120:420:0.4);
    \node[below] at (1,-1) {\scriptsize $\mathcal{N}_o$};

	\end{tikzpicture} \xrightarrow[on ~ each ~ vacuum]{enrich ~ with ~ \mathcal{N}_o} \begin{tikzpicture}[baseline=0, square/.style={regular polygon,regular polygon sides=4}]
	\filldraw[black] (-1,0) circle (2pt);
    \filldraw[black] (+1,0) circle (2pt);
	\node[above] at (-1.5,0) {\footnotesize $\mathcal{F}_{o,+}$};
    \node[above] at (+1.5,0) {\footnotesize $\mathcal{F}_{o,-}$};

    \draw[thick, ->-=1] (-0.8,0) -- (0.8,0);
    \draw[thick, -<-=0.1] (-0.8,0) -- (0.8,0);
    \node[above] at (0,0) {\scriptsize $\mathcal{N}_d,t$};


	\end{tikzpicture} ~.
\end{equation}
Then, taking the direct sum of two $\TYpo$-SPT phases
\begin{equation}
    \mathcal{F}_{o,+} \bigoplus \mathcal{F}_{o,-} ~,
\end{equation}
we acquire a desired partial SSB phase of the full $\mathcal{C}_p$-symmetry.

Next, let us consider the case (II) which exists only when $p = 1$ mod $4$. We then need to enrich the $\SPT_{\pm x}$ on each vacuum to become a $\TYpd$-SPT phase. As discussed in Appendix \ref{app:TYd}, again there is a unique enrichment for each case which turns $\SPT_{\pm}$ to $\TYpd$-SPT phases denoted as $\mathcal{F}_{d,\pm}$ respectively. Taking the direct sum of two $\mathcal{F}_{d,\pm}$, again we get a desired SSB phase.

Careful readers may notice that in the above analysis, we implicitly use the fact that specifying the SPTs of the unbroken symmetry on each vacuum uniquely determines a SSB phase of the full symmetry. Indeed, this follows from the classification theorem of module categories of the graded fusion categories in \cite{meir2012module}\footnote{The theorem states that:
\textit{An indecomposable module category over a graded fusion category $\displaystyle \CC = \bigoplus_{g\in G} \CC_g$ is given by a tuple $(N,H,\Phi,v,\beta)$, where $N$ is an indecomposable module category over $\CD = \CC_{\dsi}$, $H$ is a subgroup of $G$ which acts trivially on $N$, $\Phi:H \rightarrow Aut(Aut_{\CD}(N))$ is a homomorphism, $v$ belongs to a torsor over $H^1(H,Z(Aut_{\CD}(\CN)))$, and $\beta$ belongs to a torsor over $H^2(H,k^*)$.} In both scenarios, the indecomposable module category $N$ is a SPT phase of the $\TY(\doubleZ_p\times \doubleZ_p,\chi_{d},+1)$ and $\TY(\doubleZ_p\times \doubleZ_p,\chi_o,+1)$ respectively. And the swap symmetry provides an $\doubleZ_2$-extension over this $\TY$. Since the swap symmetry always acts non-trivially on the $\TY$-SPTs, the subgroup $H$ is trivial. Thus, both the obstruction class and potential higher data $\Phi, v,\beta$ are trivial.}. 

To summarize, by mapping to the SSB phases of $\CC_p$ under $\mathbb{Z}_2^t$-gauging, we classify the SPT phases of $\widetilde{\CC}_p$, where there is a single SPT phase when $p = 3\mod 4$ and there are two SPT phases when $p = 1 \mod 4$.

It is interesting to mention that $\wt \CC_3$ is the representation category of an order 36 self-dual Hopf algebra \cite{cuadra2017orders}, therefore, $\wt \CC_3$ is self-dual under gauging itself, similar to the case of $\Rep(H_8)$ \cite{Choi:2023vgk,Diatlyk:2023fwf,Perez-Lona:2024sds}. Moreover, our construction of SPT phase of $\wt \CC_3$ corresponds to its fiber functor and can be used to construct the gauging map of $\wt \CC_3$.

\subsection{Lattice model}\label{sec:iNISPT_lattice}
The smallest interesting case is $p=3$ in the previous discussion, and we will construct a lattice realization for it. To construct the NISPT of the intrinsic non-invertible symmetry $\tCC_3$, we start with the $\IZ_2^t$ SSB phase of $\CC_3 \equiv \TY(\IZ_3^a\times \IZ_3^b,\chi_o,+1)\rtimes \IZ_2^t$ and then gauge the $\IZ_2^t$ symmetry.

In the lattice model we construct, it is convenient to present $\mathcal{C}_3$ and $\widetilde{\CC}_3$ in a slightly different basis for the $\mathbb{Z}_3^a \times \mathbb{Z}_3^b$ invertible symmetry. For the convenience of the readers, we will briefly summarize the $p = 3$ with the new basis we used. To begin, we introduce two new generators $r,s$ related to the old ones $a,b$ via
\begin{equation}
    r = a^{-1}b ~, \quad s = a^{-1} b^{-1} ~.
\end{equation}
Then the invertible part $(\mathbb{Z}_3^a \times\mathbb{Z}_3^b)\rtimes \mathbb{Z}_2^t \subset \mathcal{C}_3$ becomes $(\mathbb{Z}_3^r\rtimes \mathbb{Z}_2^t)\times \mathbb{Z}_3^s \simeq S_3 \times \mathbb{Z}_3$ in the new parameterization, where $t$ acts as charge conjugation on $r$ but leaves $s$ invariant. The off-diagonal duality defect $\mathcal{N}_o$ implements the following gauging in the new basis
\begin{equation}\label{eq:o_ggrs}
    Z_{\mathcal{X}}[A_r^{(1)},B_{s}^{(1)}] = \sum_{a^{(1)}_r, b^{(1)}_s} Z_{\mathcal{X}}[a^{(1)}_r,b^{(1)}_s] \exp\left(\frac{2\pi i}{3}\int_{\mathcal{M}_2} - a^{(1)}_{r} \cup A^{(1)}_r + b_s^{(1)} \cup B_s^{(1)} \right) ~,
\end{equation}
while the gauging corresponding to $\mathcal{N}_d$ remains unchanged.

Let us also write down the simple lines and their fusion rules in $\widetilde{\mathcal{C}}_3$. First, $\mathbb{Z}_3^s$ symmetry will survive the gauging and $\mathbb{Z}_3^r$ gives rise to the unique $X$-type line $X = r + r^{-1}$. For convenience, let us also denote $Y = s X = X s$ and $\overline{Y} = s^{-1}X = X s^{-1}$. Together with $\widetilde{t}$, they generate the $\Rep S_3 \times \Rep \mathbb{Z}_3 \subset \widetilde{\mathcal{C}}_3$. Finally, there is the duality line $\widetilde{\CN}$ as well as $\widetilde{t}\widetilde{\CN} = \widetilde{\CN}\widetilde{t}$. And the other fusion rules are given by
\begin{equation}\label{eq:tC3fusion}
\begin{aligned}
    &\widetilde{t} \times X = X \times \widetilde{t} = X ~, \quad X \times X = 1+ \widetilde{t} + X ~, \quad  \\
    &\widetilde{\mathcal{N}} \times \widetilde{\mathcal{N}} = (\widetilde{t} \widetilde{\mathcal{N}}) \times (\widetilde{t}\widetilde{\mathcal{N}}) = (1+s+s^2)(1+X) ~,
\end{aligned}
\end{equation}
while the other fusion rules can be derived from the above. Note that $\wt \CN$ is the product of the self-duality line in $\TY(\IZ_3,+1)$ and a quantum dimension $\sqrt{3}$ line in $SU(2)_4$ fusion category \cite{eck2024reps3,xgw2024reps3,Barkeshli:2014cna}.

\

We consider the following $S_3 \times \IZ_3$ symmetric Hamiltonian with $2$ ground state degeneracy, which is the variation of the model analyzed in \cite{xgw2025nispt}. To manifest the $\TY_{3,o}$ symmetry, we introduce the $\sigma$ spins on both integer sites and half-integer sites,
\begin{equation}\label{eq:TYxz2ham}
    H_{\text{$\IZ_2^t$ SSB}}=-\sum_i \tZ_i X_i^{\sigma^z_{i+\frac{1}{2}}} \tZ_{i+1}^\dagger+Z^\dagger_{i-1} \tX_i^{\sigma^z_i} Z_{i}+h.c. -J\sum_{j\in\IZ/2} \sigma^z_{j} \sigma_{j+\frac{1}{2}}^z
\end{equation}
where $X_i^{\sigma_j^z}\equiv X_i \frac{1+\sigma^z_j}{2} +X_i^\dagger \frac{1-\sigma^z_j}{2}$ \footnote{More generally, we adopt $X_i^{A}\equiv X_i \frac{1+A}{2}+X_i^\dagger\frac{1-A}{2}$, where $A$ is any operator that $A^2 = 1$.}. The local Hilbert space is $\doubleC^3\otimes \doubleC^3\otimes \doubleC^2\otimes \doubleC^2$ acted by, say, $X_i\otimes \tX_i \otimes \sigma^x_{i}\otimes \sigma^x_{i+\frac{1}{2}}$ and other matrices \footnote{This is reminiscent of the construction in \cite{xgw2024reps3}, where two qubits are needed for the manifestation of the self-duality symmetry.}. The global symmetry of the above Hamiltonian \eqref{eq:TYxz2ham} is generated by,
\begin{align}\label{eq:latts3z3}
    U_r = \prod_{i \in \mathbb{Z}} \tX_i ~, \quad U_s =\prod_{i\in\mathbb{Z}} X_i ~, \quad U_t = \prod_{i\in\mathbb{Z}} \tC_i \sigma^x_i \sigma^x_{i+\frac{1}{2}} ~,
\end{align} 
where $\widetilde{C}_i$ is the unitary charge conjugation acting only on the $\widetilde{X}_i$ and $\widetilde{Z}_i$ variables, $\wt C^\dagger (\wt X,\wt Z) \wt C =(\wt X^\dagger,\wt Z^\dagger) $. In the $J\gg 1$ limit, the $\sigma^z$ spins are spontaneously broken, and the 2 ground states support the two distinct $\IZ_3^r\times \IZ_3^s$ SPTs given by the stabilizer Hamiltonians,
\begin{equation}\label{eq:z3z3spts}
\begin{aligned}
    & \sigma_j^z = \uparrow~\forall j\in \IZ/2~, \quad H_{\text{$\IZ_2^t$ SSB, $\uparrow$}} =  -\sum_i \tZ_i X_i \tZ_{i+1}^\dagger+Z^\dagger_{i-1} \tX_i Z_{i} + h.c. ~, \\
    & \sigma_j^z = \downarrow~\forall j\in \IZ/2 ~,\quad H_{\text{$\IZ_2^t$ SSB, $\downarrow$}} =  -\sum_i \tZ_i^\dagger X_i \tZ_{i+1}+Z_{i-1} \tX_i Z_{i}^\dagger+h.c. ~.
\end{aligned}
\end{equation}

The Hamiltonian \eqref{eq:TYxz2ham} is further invariant under gauging the $\IZ^r_3\times \IZ^s_3$ symmetry corresponding to \eqref{eq:o_ggrs}. To be specific, we consider the diagonal gauging of both $\IZ_3$ symmetries followed by charge conjugation $U_{\tC} = \prod_i \tC_i$ applied to only the $\tX,\tZ$ variables, 
\begin{equation}
\begin{aligned}
    & Z_{i} Z_{i+1}^\dagger \rightarrow X_{i+1} ~, \quad X_i\rightarrow Z_i Z_{i+1}^\dagger ~, \\
    & \tZ_{i} \tZ_{i+1}^\dagger \rightarrow \tX^\dagger_{i+1} ~,\quad \tX_i\rightarrow \tZ_i^\dagger \tZ_{i+1} ~.
\end{aligned}
\end{equation}
This is equivalent to the gauging \eqref{eq:o_ggrs} which is implemented by the diagonal gauging of both $\IZ_3$ symmetries followed by the charge conjugation action $t$ on $\mathbb{Z}_3^r$. This self-duality is generated by,
\begin{equation}\label{eq:lattNo}
    U_{\CN_o} = T^\sigma_{\frac{1}{2}} U_{\tC} \kw \wt{\kw},
\end{equation}
where $T^\sigma_{\frac{1}{2}}$ is the translation operator acting on the $\sigma$ variables by $\frac{1}{2}$ lattice site, $\kw$ and $\wt{\kw}$ are the $\IZ_3$ Kramers-Wannier duality operator on the corresponding variables, and the explicit form is in \cite{xgw2024reps3,zhang2025z3kw}. The unitary charge conjugation symmetry $U_{\tC}$ commutes with the $\wt \kw$, since $\wt \kw$ projects the system to $\IZ_3$ symmetric sector, one can alternatively check through the explicit algebra. The fusion rule of $U_{\CN_o}$ with itself is,
\begin{equation}\label{eq:kw3fusion}
    U_{\CN_o} \times U_{\CN_o} = (1+U_r+U_{r^2})(1+U_s+U_{s^2})T
\end{equation}
where $T$ shifts all variables by 1 lattice site. Although the non-invertible symmetry operator mixes with translation, we will argue that translation acts trivially in the infrared. Recall that \eqref{eq:TYxz2ham} has two ground states \eqref{eq:z3z3spts}, which are inequivalent NISPTs of $\TY_{3,o}$ and permuted by $U_t$. To be specific, the two $\CN_o$-symmetric ground states, each uniquely stabilized by the commuting-projector Hamiltonian. The energy gap between the ground and excited states remains finite in the thermodynamic limit, so translation does not generate an anomalous symmetry under RG flow. Consequently, in the infrared, the translation symmetry becomes trivial; therefore, the symmetry operators \eqref{eq:latts3z3} and \eqref{eq:lattNo} give the lattice realization of the $\CC_3$ symmetry\footnote{Notice that the FS indicator for $\mathcal{N}_o$ must be $+1$. This is because the $\mathcal{C}_3$-like category with FS indicator $-1$ for $\mathcal{N}_o$ does not admit an SPT phase in the first place because the $\mathcal{N}_o$ with $-1$ FS indicator is anomalous and obstructs \eqref{eq:2d_erc_I}.}. 

Let us now discuss an alternative presentation of the Hamiltonian, acquired by conjugating \eqref{eq:TYxz2ham} by $\prod_i \mathrm{CNOT}^\sigma_{i,i+1/2}$ acting on the $\sigma$ variables. This simply implements the following change of variables on $\sigma$'s, 
\begin{equation}
    \sigma_i^z \mapsto \sigma_i^z ~, \quad \sigma^z_{i+\frac{1}{2}} \mapsto \sigma_i^z \sigma_{i+\frac{1}{2}}^z ~, \quad \sigma_i^x \mapsto \sigma^x_{i} \sigma^x_{i+\frac{1}{2}} ~, \quad \sigma_{i+\frac{1}{2}}^x \mapsto \sigma_{i+\frac{1}{2}}^x ~.
\end{equation}
And the Hamiltonian becomes
\begin{equation}\label{eq:TYxz2ham2}
    H'_{\text{$\IZ_2^t$ SSB}}=-\sum_i \tZ_i X_i^{\sigma_i^z\sigma^z_{i+\frac{1}{2}}} \tZ_{i+1}^\dagger+Z^\dagger_{i-1} \tX_i^{\sigma^z_i} Z_{i}+h.c. -J\sum_{i\in\IZ}  \sigma_{i+\frac{1}{2}}^z + \sigma^z_{i} \sigma^z_{i+1} \sigma_{i+\frac{1}{2}}^z ~,
\end{equation}
and the symmetries become
\begin{equation}
    U_r = \prod_{i \in \mathbb{Z}} \tX_i ~, \quad U_s =\prod_{i\in\mathbb{Z}} X_i ~, \quad U'_t = \prod_{i\in\mathbb{Z}} \tC_i \sigma^x_i ~,\quad U'_{\CN_o} = U^\sigma_{\mathrm{CNOT}} \times T^\sigma_{\frac{1}{2}} U_{\tC} \kw \wt{\kw}, 
\end{equation}
where $U^\sigma_{\mathrm{CNOT}}=\prod_{i\in \IZ}\mathrm{CNOT}^\sigma_{i,i+1/2}\prod_{i\in \IZ}\mathrm{CNOT}^\sigma_{i-1/2,i}$. Since the symmetry operators are unitarily related to the original ones, their fusion rules remain unchanged. Also, notice that $\sigma_{j+1/2}^z$ commutes with any other terms in the Hamiltonian \eqref{eq:TYxz2ham2}; therefore, it corresponds to a decoupled sector. The Hamiltonian in \cite{xgw2025nispt} corresponds to restricting \eqref{eq:TYxz2ham} to the $\sigma_{i+\frac{1}{2}}^z = 1$, where the non-invertible symmetry $U_{\CN_o}$ is no longer manifested.

\

As discussed previously, once gauging the $\IZ^t_2$ symmetry in $\CC_3$, the dual symmetry $\tCC_3$ is a non-group-theoretical fusion category but anomaly-free, and there is no obstruction to a symmetric gapped phase with a unique ground state. In the lattice model \eqref{eq:TYxz2ham}, the $\IZ_2^t$ symmetry is generated $U_t=\prod_{i\in\mathbb{Z}} \tC_i\sigma^x_i\sigma^x_{i+\frac{1}{2}}$, which can be gauged to get $\tCC_3$. Alternatively, we use the unitarily equivalent model \eqref{eq:TYxz2ham2}, and gauge $U'_t = \prod_{i\in\mathbb{Z}} \tC_i \sigma^x_i$ to get the $\tCC_3$. Following the procedure in \cite{xgw2024reps3} and also as described in \appref{app:gauging}, the gauged Hamiltonian is
\begin{equation}\label{eq:tc3ham}
\begin{aligned}
    & \wh H_\text{NISPT}' = -\sum_i Z_{i-1}^\dagger\tX_i Z_i + \frac{1}{2}(\tZ_{i}^{\mu_{i+1}^x} X_{i} \tZ_{i+1}^\dagger+\tZ_{i}^{-\mu_{i+1}^x} X_{i} \tZ_{i+1}) \\
    & \quad \quad \quad \quad \quad \quad \quad \quad + \frac{1}{2}(\tZ_{i} X_{i} \tZ_{i+1}^{-\mu_{i+1}^x}-\tZ_{i}^\dagger X_{i} \tZ_{i+1}^{\mu_{i+1}^x})\sigma^z_{i+\frac{1}{2}} +h.c. -J\sum_i \sigma^z_{i+\frac{1}{2}}+\mu^x_i \sigma^z_{i+\frac{1}{2}} ~.
\end{aligned}
\end{equation}

Let us first check when $J$ is large, the above Hamiltonian has a unique ground state. First, in the low energy we will have $\sigma_{i + \frac{1}{2}}^z = 1$ and $\mu_i^x = 1$, which simplifies the above Hamiltonian to
\begin{equation}\label{eq:nongrpham}
    \wh H_\text{NISPT}' \rightarrow -\sum_i Z_{i-1}^\dagger\tX_i Z_i + \tZ_{i} X_i \tZ_{i+1}^\dagger + h.c.  ~,
\end{equation}
which has a unique ground state given by the $\IZ_3\times \IZ_3$ cluster state. In particular, as $J\rightarrow +\infty$, the ground state wavefunction is given by the tensor product of the product state $\otimes\ket{+}$ for the $\mu^x$ variable, the product state $\otimes \ket{\uparrow}$ for the $\sigma_{i+\frac{1}{2}}^z$ variable, and the $\IZ^r_3\times \IZ^s_3$ cluster state.

The Hamiltonian \eqref{eq:tc3ham} clearly will manifest the non-group-theoretical fusion category symmetry $\widetilde{\mathcal{C}}_3$. Once gauging the $\IZ^t_2$ symmetry generated by $U_t$, the dual symmetry is 
\begin{equation}
    U_{\wt{t}} = \prod_{i\in\mathbb{Z}} \mu^x_i ~.
\end{equation}
The $\IZ^s_3$ symmetry is generated by $U_s$ which commutes with $\IZ_2^t$ and remains unchanged. For $\mathbb{Z}_3^r$, $U_r$ does not commute with $U_t$ and the invariant combination $U_r + U_{r^2}$ becomes the line $U_X$ with quantum dimension 2 and its explicit form can be acquired using the $\mathbb{Z}_2^t$-gauging map in Appendix \ref{app:gauging}:
\begin{equation}
    U_X = \frac{1}{2}\left(1+\prod_{j=1}^L \mu_j^x \right) \left(\prod_{j=1}^L \tX_j^{\prod_{k=2}^j \mu_k^x}+\prod_{j=1}^L \tX_j^{-\prod_{k=2}^j \mu_k^x}\right),
\end{equation}
and the fusion of $U_X$ is,
\begin{equation}
    U_X \times U_X = 1+U_{\wt{t}}+U_X ~.
\end{equation}
Since $U_s$ is untouched through the gauging procedure, $U_s,U_{s^2}$ lines remain, and,
\begin{equation}
    U_X \times U_s = U_s \times U_X = U_Y,\quad U_X \times U_{s^2} = U_{s^2} \times U_X = U_{\overline{Y}} ~,
\end{equation}
and all the above lines generate the non-invertible symmetries $\Rep S_3 \times \Rep \mathbb{Z}_3 \subset \widetilde{\mathcal{C}}_3$.

Last but not least, the self-duality line after the $\IZ_2^t$ gauging can be acquired by replacing each term via the gauging map, but it is rather cumbersome to write down the explicit form. We denote the dual self-duality line as $U_{\wt{\CN}}$ and it has the fusion rule
\begin{equation}
  U_{\wt{\CN}} \times U_{\wt{\CN}} = (1+U_X)(1+U_s+U_{s^2})T ~.
\end{equation}
Therefore, we constructed all the lattice symmetry lines that correspond to the simple objects in the intrinsic non-invertible symmetry $\tCC_3$ and can be matched with the categorical data in \eqref{eq:tC3fusion}. The $\tCC_3$ symmetric Hamiltonian \eqref{eq:nongrpham} realizes the NISPT phase. Note that the invertible subcategory in $\tCC_3$ is $\IZ_6$ generated by $U_{\wt t}U_s$, and $\tCC_3$ contains the $\Rep(S_3)\times \IZ_3$ as the largest group-theoretical fusion category, where $\Rep(S_3)$ is generated by $U_Z,U_{\wt t}$ and $\IZ_3$ is generated by $U_s,U_{s^2}$. As a consistent check, gauging $\Rep(S_3)\times \IZ_3$ or any algebraic objects of $\Rep(S_3)\times \IZ_3$ won't make the whole fusion category invertible; therefore, $\tCC_3$ is non-group-theoretical and the NISPT phase is intrinsic.

\section{Anomaly resolution of non-invertible symmetries and igNISPT phase}\label{sec:igNISPT}
In this section, we observe that the mixed anomaly plays an important role in the anomaly resolution. This allows us to search for anomaly resolutions for non-invertible symmetries, which we demonstrate with the anomalous $\TYpd$-category for $p = 3$ mod 4 in Section \ref{sec:resg}. This then allows us to construct intrinsic gapless SPT phases protected by the non-invertible symmetry (igNISPT) in $2$-dim, and we provide a lattice model example in Section \ref{sec:igNISPTlatt}.

\subsection{igSPT phases, anomaly resolution, and mixed anomalies}\label{sec:resg}
Starting with a UV theory $\mathcal{T}_{UV}$ with fusion category symmetry $\CC_{UV}$, after the RG flow, any anomaly-free fusion subcategory symmetry of $\CC$ may act trivially in the IR theory $\mathcal{T}_{IR}$. We use $\Rep \mathrm{H}$ to denote such trivially acting fusion subcategory, since any anomaly-free fusion category is the representation category of some Hopf algebra $\mathrm{H}$. The faithfully acting fusion category symmetry $\CC_{IR}$ is captured, roughly speaking, by the quotient category ``$\mathcal{C}_{UV}/\Rep\mathrm{H}$'', and more precisely is captured by the short exact sequence of fusion categories \cite{bruguieres2011exact} (see also \cite{Perez-Lona:2025ncg,Antinucci:2025fjp}):
\begin{equation}
    \Rep \mathrm{H} \rightarrow \CC_{UV} \rightarrow \CC_{IR} ~.
\end{equation}
Clearly, that $\CC_{IR}$ is anomaly-free will imply that $\CC_{UV}$ is anomaly-free, but not the other way around. In other words, it is possible that the faithfully acting $\mathcal{C}_{IR}$ may have an emergent anomaly under the RG flow. When this happens, one can have an intrinsic gapless SPT(igSPT) phase of $\CC_{UV}$ \cite{Scaffidi:2017ppg, Thorngren:2020wet, Wen:2022tkg, Li:2023knf, Huang:2023pyk, Wen:2023otf, Bhardwaj:2024qrf, Perez-Lona:2025ncg}: under the RG, the theory $\mathcal{T}_{UV}$ will flow to a gapless theory $\mathcal{T}_{IR}$ with a unique ground state with faithfully acting $\mathcal{C}_{IR}$-symmetry. Due to the anomaly of $\mathcal{C}_{IR}$, one cannot deform the IR theory $\mathcal{T}_{IR}$ to a trivially gapped phase without breaking the symmetry $\mathcal{C}_{IR}$, therefore, justifying being \textit{intrinsic}\footnote{Notice that the ``intrinsic'' here has a different meaning compared with intrinsic NISPT phase discussed previously. The latter ``intrinsic'' means the NISPT phase cannot be mapped to a gapped phase of invertible symmetries under discrete gauging.}. Notice that it is possible to deform the $\mathcal{T}_{UV}$ theory to a trivially gapped phase while preserving the symmetry; however, such deformation is not accessible in $\mathcal{T}_{IR}$. The procedure of extending an anomalous $\mathcal{C}_{IR}$ symmetry into a larger anomaly-free symmetry of the same theory with a trivially-acting kernel $\Rep \mathrm{H}$ is known as the \textit{anomaly resolution} of $\mathcal{C}_{IR}$.

The above discussion leads to a natural question of given an anomalous $\mathcal{C}_{IR}$, how to find its anomaly resolution generically. This is not so straightforward as $\mathcal{C}_{IR}$ will generically appear as a quotient category of $\mathcal{C}_{UV}$ instead of as a subcategory; this makes the direct search complicated. In the following, we will present a generic method\footnote{Of course, we are not claiming any anomaly resolution will be acquired in this way.} of constructing anomalous resolution related to the mixed anomaly studied previously, where $\mathcal{C}_{IR}$ is embedded as a subcategory instead. This simplifies the search and allows us to find the anomaly resolution of non-invertible symmetries and to construct igSPT phases.

\

Let us begin with demonstrating how the idea works for the classic example where the igSPT phase corresponding to the anomaly resolution of the anomalous $\mathbb{Z}_2^a$ symmetry $\VEC_{\mathbb{Z}_2^a}^\omega$ (where $\omega(a^{i_1},a^{i_2},a^{i_3}) = (-1)^{i_1 i_2 i_3}$. The short exact sequence describing the anomaly resolution is
\begin{equation}\label{eq:arZ2}
    \Rep \doubleZ_2 \rightarrow \VEC_{\doubleZ_4} \rightarrow \VEC_{\doubleZ_2^a}^\omega ~.
\end{equation}
This anomaly resolution may be acquired as follows via discrete gauging. Let us first enlarge the anomalous symmetry $\VEC_{\mathbb{Z}_2^a}$ with an additional $\mathbb{Z}_2^b$ symmetry, where $\mathbb{Z}_2^b$ has no self-anomaly but has a mixed anomaly with $\VEC_{\mathbb{Z}_2^a}^\omega$. Then, the full category $\VEC_{\mathbb{Z}_2^a \times \mathbb{Z}_2^b}^{\omega'}$ has the anomaly
\begin{equation}
    \omega'(a^{i_1} b^{j_1}, a^{i_2} b^{j_2}, a^{i_3} b^{j_3}) = (-1)^{i_1 i_2 i_3 + j_1 i_2 i_3} ~,
\end{equation}
where the second term represents the mixed anomaly. It is interesting to notice that the anomaly of $\VEC_{\mathbb{Z}_2^a \times \mathbb{Z}_2^b}^{\omega'}$, under a different decomposition as $\mathbb{Z}_2^{ab} \times \mathbb{Z}_2^b$, can be understood as just a mixed anomaly:
\begin{equation}
    \omega'( (ab)^{i_1} b^{j_1}, (ab)^{i_2} b^{j_2}, (ab)^{i_3} b^{j_3}) = (-1)^{i_1 j_2 j_3} ~.
\end{equation}
Similar to the case discussed in Section \ref{sec:intrinNISPT}, gauging $\mathbb{Z}_2^b$ will trivialize the mixed anomaly and lead to the anomaly-free dual symmetry $\VEC_{\mathbb{Z}_4}$. When the anomaly-free symmetry $\mathbb{Z}_2^{\widetilde{b}}$ generated by the Wilson line of $b$-gauge field acts trivially, it will lead to the extension \eqref{eq:arZ2}.

The above construction can be generalized to categorical symmetries. Let $\mathcal{C}_{IR}$ be an anomalous categorical symmetry, as pointed out before, given a $G$-action on $\mathcal{C}_{IR}$, one can construct the crossed product category $\mathcal{C}_{IR}\rtimes G$. In the above $\mathcal{C}_{IR} = \VEC_{\mathbb{Z}_2^a}^\omega$ example, $G = \mathbb{Z}_2$ acts trivially on the lines but acts non-trivially on the $a \otimes a\rightarrow 1$ fusion junction by a $(-1)$ phase. Clearly, $\mathcal{C}_{IR}$ is a subcategory of $\mathcal{C}_{IR} \rtimes G$, and as a result $\mathcal{C}_{IR} \rtimes G$ is automatically anomalous. But sometimes it is possible to re-interpret the total anomaly of $\mathcal{C}_{IR} \rtimes G$ as just a mixed anomaly, roughly speaking, between $G$ and some subcategory other than $\mathcal{C}_{IR}$. This mixed anomaly can then be trivialized by $G$-gauging. When this is the case, one acquires an anomaly resolution of $\mathcal{C}_{IR}$, described by the sequence
\begin{equation}
    \Rep G \rightarrow \mathcal{C}_{UV} \rightarrow \mathcal{C}_{IR} ~, 
\end{equation}
where the $\mathcal{C}_{UV} = (\mathcal{C}_{IR} \rtimes G)/G$ is also known as the $G$-equivariantization of $\mathcal{C}_{IR}$ denoted by $(\mathcal{C}_{IR})^G$. This construction provides a simple way to look for anomaly resolution of non-invertible symmetries.

Indeed, we have an example from the previous discussion. Recall that $\TYpd$ as a subcategory of $\mathcal{C}_p = \TYpd \rtimes \mathbb{Z}_2^t$ is anomalous when $p = 3$ mod $4$. However, for any $p$, the anomaly of $\mathcal{C}_p$ can be understood as a mixed anomaly between $\TYpo$ and $\mathbb{Z}_2^t$, therefore, it will be trivialized in the dual category $\widetilde{C}_p$ after gauging $\mathbb{Z}_2^t$. We then get an anomaly resolution of the anomalous, non-invertible symmetry $\TYpd$ (when $p = 3$ mod $4$) described via the sequence
\begin{equation}
    \Rep \mathbb{Z}_2^t \rightarrow \widetilde{\mathcal{C}}_p \rightarrow \TYpd ~, \quad p = 3 \mod 4 ~.
\end{equation}
It is straightforward to generalize this construction to trivializing the anomaly of a large class of $\TY$-fusion categories, if the anomaly arises from bicharacter (that is, there is no self-dual SPT under the corresponding discrete gauging) and there exists a non-anomalous bicharacter. 

This allows us to construct an intrinsic gapless SPT phase where the faithfully acting IR symmetry $\mathcal{C}_{IR}$ is non-invertible, and we denote such a phase as an intrinsic gapless non-invertible symmetry protected topological (igNISPT) phase.

\subsection{Anomaly resolution of non-invertible symmetries and igNISPT}\label{sec:igNISPTlatt}
We now provide a lattice realization of the igNISPT phase associated with the above anomaly resolution of $\TY_{3,d}$. Recall that the igSPT phases describe the emergent anomaly below a certain energy scale $\Delta$, while above which the symmetry is anomaly-free. To be specific, in our setting, the $\Rep \mathbb{Z}_2^t$-symmetry will act only on certain gapped degrees of freedom which can be ignored below the energy scale $\Delta$. As a result, the faithfully acting symmetry on the low-energy degrees of freedom (below $\Delta$) is the anomalous $\TY_{3,d}$-symmetry. And the anomaly-free symmetry $\widetilde{\mathcal{C}}_3 \equiv (\TY_{3,d})^{\mathbb{Z}_2^t}$ acts on both gapped and gapless degrees of freedom.

It is instructive to phrase the construction of intrinsic NISPT phase in the general process of anomaly resolution of $\TY_{3,d}$-symmetric theory $\mathcal{X}$. For simplicity, we will require $\mathcal{X}$ also admits the swap symmetry to have a universal construction, although this condition could be removed as we will comment later. Similar to the construction of the $\widetilde{\mathcal{C}}_3$-symmetric NISPT phase, we first construct the theory with $\CC_3 = \TY_{3,d}\rtimes \IZ_2^t$ symmetry\footnote{Note here we consider the diagonal bicharacter $\chi_d$.} by stacking $\mathcal{X}$ with a gapped theory (say the energy gap is $\Delta$) realizing the SSB phase of some non-anomalous $\mathbb{Z}_2^{\eta}$ symmetry. The $\mathbb{Z}_2^t$ in $\mathcal{C}_3$ is actually given by the diagonal symmetry between $\eta$ and the swap symmetry in $\mathcal{X}$, then gauging the $\IZ_2^t$ symmetry removes the 2-fold degeneracy due to the SSB of $\mathbb{Z}_2^\eta$, and results in a gapless theory $\widetilde{\mathcal{X}}$ realizing the dual symmetry $\tCC_3$. However, below the energy scale $\Delta$, this operation is essentially trivial and the faithfully acting symmetry will still be the anomalous symmetry $\TY_{3,d}$. However, in the full theory, the symmetry is actually the anomaly-free symmetry $\widetilde{\mathcal{C}}_3$ and generically one can trivially gap the theory. In particular, if we choose the seed theory $\mathcal{X}$ to be a gapless theory with a unique ground state, then we engineer an igSPT phase described by $\widetilde{\mathcal{X}}$ with the anomalous faithfully acting symmetry being $\TY_{3,d}$.

Let us now demonstrate how this works in concrete lattice models with two $\mathbb{Z}_3$-spins. Our seed Hamiltonian can be any Hamiltonian realizing a $\TY_{3,d}$-symmetric\footnote{Here, we again switch to the $\mathbb{Z}_3^r\times \mathbb{Z}_3^s$ basis used in Section \ref{sec:iNISPT_lattice}, where the relavent $\mathbb{Z}_2$ automorphism becomes the charge conjugation on $\mathbb{Z}_3^r$.} gapless phase with a unique ground state, which also admits the automorphism symmetry $\widetilde{C}: \widetilde{X} \rightarrow \widetilde{X}^\dagger$. To find the desired seed Hamiltonian, consider $\IZ_3\times \IZ_3$ symmetric gapped phases; under $\CN_d$ they transform as
\begin{equation}
    \begin{array}{ccccc}
        \text{SSB} & ZZ^\dagger+\tZ\tZ^\dagger &\leftrightarrow &\text{Sym}&  X+\tX \\
        \text{PSB} & ZZ^\dagger+\tX &\leftrightarrow &\text{PSB'}&  X+\tZ \tZ^\dagger \\
        \text{PSBD} & ZZ^\dagger\tZ^\dagger \tZ+X\tX &\leftrightarrow &\text{PSBA}&  Z Z^\dagger \tZ \tZ^\dagger +X^\dagger \tX \\
        \text{SPT1} & \tZ X \tZ^\dagger+Z^\dagger \tX Z &\leftrightarrow &\text{SPT2}&  \tZ^\dagger X \tZ + Z \tX Z^\dagger \\
    \end{array}
\end{equation}
It is possible to construct $\CN_d$ invariant Hamiltonians using the pair of $(\text{SSB},\text{Sym})$ and/or $(\text{PSB},\text{PSB'})$. In the following, we will consider the following Hamiltonian acquired by using $(\text{SSB},\text{Sym})$ as an example 
\begin{equation}
    H_{\CN_d\text{ symmetric}}=-\sum_ie^{\ii\phi} (Z_iZ_{i+1}^\dagger+ X_i) +e^{\ii \wt \phi} (\tZ_i\tZ_{i+1}^\dagger + \tX_i) +h.c. ~,
\end{equation}
although clearly the construction works for more general cases. Then, requiring the above Hamiltonian to be $\widetilde{C}$ invariant allows us to fix $\widetilde{\phi} = 0$. 

As discussed previously, once the seed Hamiltonian is chosen, we further add a $\sigma$ variable as before and construct a Hamiltonian by adding the SSB phase of the $\sigma$-variables
\begin{align}
    H_{\text{$\IZ_2^t$ SSB}}=&-\sum_ie^{\ii\phi} (Z_iZ_{i+1}^\dagger+ X_i) + (\tZ_i\tZ_{i+1}^\dagger+\tX_i)+h.c.  -\Delta \sum_{j \in \IZ/2}\sigma_j^z \sigma_{j+\frac{1}{2}}^z ~,
\end{align}
and the $\mathcal{C}_3 = \TY_{3,d}\rtimes \IZ_2^{t}$ takes exactly the same form as in \eqref{eq:latts3z3} and \eqref{eq:lattNo}:
\begin{equation}\label{eq:igSPTUt}
    U_{ t} = U_{\wt C} \prod_i \sigma^x_i \sigma^x_{i+1/2},\quad U_{\CN_d}= T^\sigma_{\frac{1}{2}} \kw \wt{\kw} \prod_i \sigma^x_i \sigma^x_{i+1/2}.
\end{equation}
where $U_{\CN_d} = U_{\CN_o} U_{t}$.

When $\Delta \gg 1$, the theory has two vacua, labeled by $\sigma^z =\pm1$ and the states of the qutrit variables are the same, given by,
\begin{align}
    \sigma^z =\uparrow/\downarrow,\quad  H_{\uparrow/\downarrow} = -\sum_ie^{\ii\phi} (Z_iZ_{i+1}^\dagger+ X_i) + (\tZ_i\tZ_{i+1}^\dagger+\tX_i) +h.c. ~.
\end{align}
Similarly to the previous section, we can simplify the Hamiltonian and the symmetry action by conjugating $\prod_i \mathrm{CNOT}^\sigma_{i,i+1/2}$ which acts on the $\sigma$ variables, 
\begin{align}\label{eq:TYdZ2Ham1}
H_\text{$\IZ_2^t$ SSB}'=&-\sum_ie^{\ii\phi} (Z_iZ_{i+1}^\dagger+ X_i) + (\tZ_i\tZ_{i+1}^\dagger+\tX_i)+h.c. -\Delta\sum_{i\in\IZ}  \sigma_{i+\frac{1}{2}}^z + \sigma^z_{i} \sigma^z_{i+1} \sigma_{i+\frac{1}{2}}^z 
\end{align} 
and the symmetries are given by,
\begin{equation}
    U_r = \prod_{i \in \mathbb{Z}} \tX_i ~, \quad U_s =\prod_{i\in\mathbb{Z}} X_i ~, \quad U'_{ t} = \prod_{i\in\mathbb{Z}} \tC_i \sigma^x_i ~,\quad U'_{\CN_o} = U^\sigma_{\mathrm{CNOT}} T^\sigma_{\frac{1}{2}} \kw \wt{\kw}\prod_i \sigma^x_i ~, 
\end{equation}
where $U^\sigma_{\mathrm{CNOT}}=\prod_{i\in \IZ}\mathrm{CNOT}^\sigma_{i,i+1/2}\prod_{i\in \IZ}\mathrm{CNOT}^\sigma_{i-1/2,i}$. The symmetry is manifested in \eqref{eq:TYdZ2Ham1}. Once gauging the $\IZ_2^{ t}$ symmetry generated by $U'_{ t} = U_{\wt C} \prod_i \sigma^x_i $, we get
\begin{align}\label{eq:igNISPTH}
H_\text{igNISPT}=&-\sum_ie^{\ii\phi} (Z_iZ_{i+1}^\dagger+ X_i)+(\tZ_i^{\mu^x_{i+1}}\tZ_{i+1}^\dagger+\tX_i)+h.c. - \Delta \sum_i \sigma_{i+\frac{1}{2}}^z+ \sigma_{i+\frac{1}{2}}^z\mu_i^x
\end{align} 
which is invariant under the anomaly-free intrinsic non-invertible symmetry $\widetilde{\mathcal{C}}_3$ discussed in \eqref{sec:iNISPT_lattice}, clearly it is possible to add the deformation to a symmetric gapped phase, i.e. drive to the NISPT Hamiltonian \eqref{eq:nongrpham}. 

On the other hand, when $\Delta\gg1$, $\mu^x_i = \sigma^z_{i + \frac{1}{2}} = 1$, the Hamiltonian in the low energy sector simply reduces to the seed Hamiltonian we start with:
\begin{equation}\label{eq:igNISPT_IR}
    H_\text{igNISPT,IR} =-\sum_ie^{\ii\phi} (Z_iZ_{i+1}^\dagger+ X_i) + (\tZ_i\tZ_{i+1}^\dagger+\tX_i)+h.c. ~,
\end{equation}
indicating the faithfully acting symmetry is the anomalous $\TY_{3,d}$ symmetry. This means the UV Hamiltonian \eqref{eq:igNISPTH} indeed resolves the anomalous $\TY_{3,d}$ symmetry in \eqref{eq:igNISPT_IR}.

To ensure that we do construct an igSPT phase, we need to check the seed Hamiltonian \eqref{eq:igNISPT_IR} has a unique ground state, which, if indeed true, will combine with the anomalous $\TY_{3,d}$ to imply the seed theory is gapless. The seed Hamiltonian is a direct product of the critical chiral 3-state clock model of variable $X$ and the critical 3-state clock model of the variable $\tX$ \cite{cardy1993chiralpotts,paul2012chiralclock,taylor2015chiralclock,subir2018chiralclock,mila2022chiralclock}. The critical 3-state clock model is described by 3-state Potts CFT \cite{Fateev:1985mm,Fateev:1985ig,Mong:2014ova}. The behavior of the critical chiral 3-state clock model depends on the phase factor $\phi$, when $\phi$ is small, it is described by the 3-state Potts CFT with central charge $4/5$, and the theory hits the Lifshitz points at $\phi=\pi/6$ which is a scaling invariant theory without conformal symmetry. With certain parameters, the chiral clock model can also be mapped to $\IZ_3$-parafermion model \cite{paul2012chiralclock}. 

\begin{figure}
    \centering
    \includegraphics[width=0.6\linewidth]{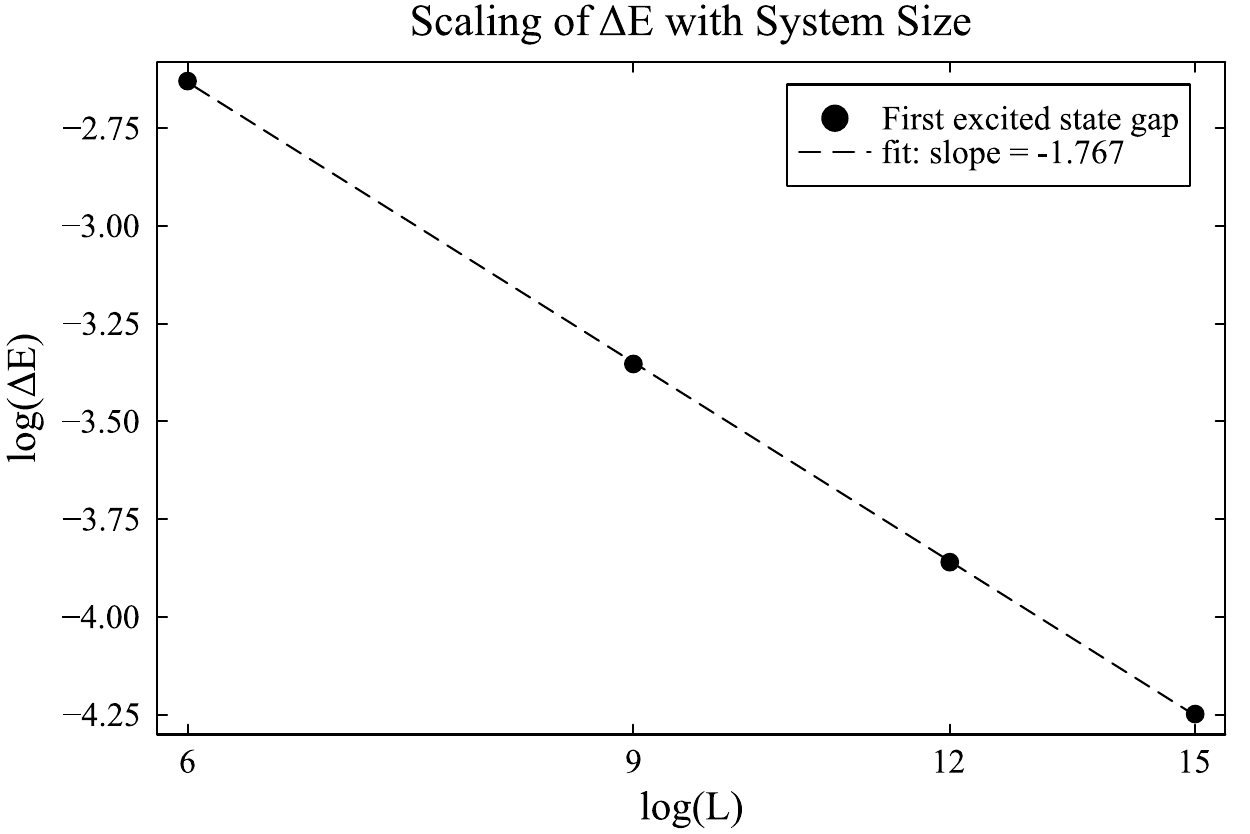}
    \caption{The first excited state gap scales with $L$ of \eqref{eq:igNISPT_IR}, which is presented using $\log(\Delta E)$ vs $\log(L)$ up to $L=15$. The fitting gives $\Delta E\sim1/L^{1.767}$, which indicates no emergent conformal symmetry but it is scaling invariant. The first excited state is identified with momentum $k=2\times 2\pi/3$.}
    \label{fig:scaling}
\end{figure}

The critical 3-state clock model is known to have a unique vacuum, whereas the vacuum structure of the critical chiral 3-state clock model remains to be determined. To this end, we performed exact diagonalization of the purely imaginary critical chiral clock model for finite system sizes,
\begin{equation}
    H_{\text{chiral clock}} = -\ii \sum_i Z_{i}^\dagger Z_{i+1}-Z_{i} Z_{i+1}^\dagger+X_i -X_i^\dagger~.
\end{equation}
 The low-lying spectrum shows that the first excited state appears at crystal momentum $k=2\times 2\pi/3$. The corresponding energy gap decreases with increasing system size as a power law of the form $\Delta E \sim 1/L^{1.767}$ as shown in \figref{fig:scaling}, rather than exponentially decays, indicating the absence of degenerate ground states in the thermodynamic limit. Also, it suggests a scaling invariant theory without the full conformal symmetry. In other words, \eqref{eq:igNISPT_IR} with certain $\phi$ realizes the scaling invariant theory with non-invertible symmetry $\TY(\doubleZ^r_3\times \doubleZ^s_3,\chi_d,+1)$, but no conformal symmetry. 

To conclude, we want to mention that while the above anomaly resolution works for any $\CN_d,\widetilde{C}$-symmetric seed Hamiltonians in a universal way, it is rather straightforward to generalize the procedure for $\CN_d$-symmetric seed Hamiltonians. After adding $ - \Delta \sum_{j\in \mathbb{Z}_2} \sigma_j^z \sigma^z_{j + \frac{1}{2}}$, to ensure the total Hamiltonian is invariant under the $U_t$ symmetry in \eqref{eq:igSPTUt}, one must couple the $\widetilde{X},\widetilde{Z}$ variables to $\sigma^z_j$'s while preserving $\CN_d$; therefore, the construction will then depend on details of the seed Hamiltonians.

\section{Constructing intrinsic NISPT in $(3+1)$-dim}\label{sec:4d}
In this section, we propose a construction of intrinsic NISPT in (3+1)-dim generalizing our previous approach. For simplicity, we will only consider adding duality and automorphism symmetry to $\mathbb{Z}_3^{[1]} \times \mathbb{Z}_3^{[1]}$ 1-form symmetry. We will also restrict ourselves to bosonic theories and neglect any potential contributions from the gravitational part. 

\subsection{A $\mathbb{Z}_4\times \mathbb{Z}_2$ extension of $\mathbb{Z}_3^{[1]}\times \mathbb{Z}_3^{[1]}$ with mixed anomaly}
Let us begin by specifying the anomalous fusion 3-category symmetry $\mathfrak{C}^{[3]}$ via the SymTFT. Similarly, we want to consider a $\mathbb{Z}_4 \times \mathbb{Z}_2$-graded extension of an anomaly-free $\mathbb{Z}_3^{[1]}\times \mathbb{Z}_3^{[1]}$ 1-form symmetry in 4-dim. The first factor is $\mathbb{Z}_4$ because the duality defect in 4-dim is actually an order-$4$ operation. The SymTFT of $\mathbb{Z}_3^{[1]}\times \mathbb{Z}_3^{[1]}$ is a 5-dim 2-form gauge theory with the action
\begin{equation}
    S = \frac{2\pi i}{3} \int a^{(2)}_1 \cup \delta \widehat{a}^{(2)}_1 + a^{(2)}_2 \cup \delta \widehat{a}^{(2)}_2 ~,
\end{equation}
where $a^{(2)}_1, \widehat{a}^{(2)}_1, a^{(2)}_2, \widehat{a}^{(2)}_2 \in C^2(\mathcal{M}_5,\mathbb{Z}_3)$.  The theory contains topological surface operators:
\begin{equation}
    S_{(q_1,q_2,\widehat{q}_1,\widehat{q}_2)}(\sigma) = e^{\frac{2\pi i}{3} q_1 \oint_\sigma a^{(2)}}_1 e^{\frac{2\pi i}{3} q_2 \oint_\sigma a^{(2)}_2} e^{\frac{2\pi i}{3} \widehat{q}_1 \oint_\sigma \widehat{a}^{(2)}_1} e^{\frac{2\pi i}{3} \widehat{q}_2 \oint_\sigma \widehat{a}^{(2)}_2} ~, \quad q_i,\widehat{q}_i \in \mathbb{Z}_3 ~,
\end{equation}
with the non-trivial commutation relations
\begin{equation}
\begin{aligned}
    e^{\frac{2\pi i}{3} \oint_\sigma a^{(2)}} e^{\frac{2\pi i}{3} \oint_{\sigma'} \widehat{a}^{(2)}} &= e^{-\frac{2\pi i}{3}\langle \sigma,\sigma'\rangle} e^{\frac{2\pi i}{3} \oint_{\sigma'} \widehat{a}^{(2)}} e^{\frac{2\pi i}{3} \oint_\sigma a^{(2)}} ~, \\
    e^{\frac{2\pi i}{3} \oint_\sigma b^{(2)}} e^{\frac{2\pi i}{3} \oint_{\sigma'} \widehat{b}^{(2)}} &= e^{-\frac{2\pi i}{3}\langle \sigma,\sigma'\rangle} e^{\frac{2\pi i}{3} \oint_{\sigma'} \widehat{b}^{(2)}} e^{\frac{2\pi i}{3} \oint_\sigma b^{(2)}} ~,
\end{aligned}
\end{equation}
where $\langle \sigma, \sigma'\rangle$ denotes the intersection form of 2-cycles on an equal time slice $\mathcal{M}_4$.

A bulk symmetry can be described as a matrix acting on the charge $(q_1,q_2,\widehat{q}_1,\widehat{q}_2)$ preserving the above commutation relations. In this case, we focus on the following two commuting symmetries given by
\begin{equation}
    U_{o} = \begin{pmatrix} 0 & 0 & 0 & 1 \\ 0 & 0 & 1 & 0 \\ 0 & -1 & 0 & 0 \\ -1 & 0 & 0 & 0\end{pmatrix} ~, \quad U_t = \begin{pmatrix} 0 & 1 & 0 & 0 \\ 1 & 0 & 0 & 0 \\ 0 & 0 & 0 & 1 \\ 0 & 0 & 1 & 0\end{pmatrix} ~.
\end{equation}
Here, $U_o$ generates $\mathbb{Z}_4$-symmetry and $U_t$ generates $\mathbb{Z}_2$-symmetry. In the 4-dim theory, $U_o$ becomes a non-invertible duality symmetry $\mathcal{N}_o$, while $U_t$ becomes the invertible 0-form symmetry $\mathcal{U}_t$ exchanging the two factors of $\mathbb{Z}_3^{[1]}$\footnote{Notice to fully specify the 3-category structure, after the bulk symmetries are chosen, one must also specify a higher symmetry fractionalization class in $H^3(\mathbb{Z}_4^o\times\mathbb{Z}_2^t,\mathbb{Z}_3^4)$ and a discrete torsion class in $H^5(\mathbb{Z}_4^o\times \mathbb{Z}_2^t)$. In our case, there is a unique choice of the fractionalization class because the corresponding cohomology group $H^3$ is trivial; and we choose the trivial discrete torsion in $H^5(\mathbb{Z}_4^o \times \mathbb{Z}_2^t,U(1))$.}. The global action of $\CN_o$ and $t$ on the 4-dim theory is given by\footnote{Our normalization for gauging 1-form symmetry $A^{[1]}$ in 4d is $\frac{1}{\sqrt{|H^2(\mathcal{M}_4,A)|}}$ following \cite{Apte:2022xtu}, and again we drop all the normalization factor for simplicity.}
\begin{equation}\label{eq:4d_ot_trans}
\begin{aligned}
    \mathcal{N}_o: & \quad Z_{\mathcal{X}}[A^{(2)},B^{(2)}] = \sum_{a^{(2)},b^{(2)}} Z_{\mathcal{X}}[a^{(2)},b^{(2)}] \exp\left(\frac{2\pi i}{3}\int_{\mathcal{M}_4} a^{(2)} \cup B^{(2)}+b^{(2)} \cup A^{(2)} \right) ~, \\
    \mathcal{U}_t: & \quad Z_{\mathcal{X}}[A^{(2)},B^{(2)}] = Z_{\mathcal{X}}[B^{(2)},A^{(2)}] ~.
\end{aligned}
\end{equation}
The fusion rules are given by
\begin{equation}
\begin{aligned}
    & \mathcal{N}_o \times \overline{\mathcal{N}}_o = \overline{\mathcal{N}}_o \times \mathcal{N}_o = \mathcal{C}_0 ~, \quad \mathcal{N}_o \times \mathcal{N}_o = \mathcal{C}_0 \times \mathcal{U}_c = \mathcal{C}_0 \times \mathcal{U}_c ~, \quad  \mathcal{N}_o \times \mathcal{U}_c = \mathcal{U}_c \times \mathcal{N}_o = \overline{\mathcal{N}}_o ~,  \\
    & \mathcal{N}_o \times \mathcal{U}_t = \mathcal{U}_t \times \mathcal{N}_o \equiv \mathcal{N}_d ~, \quad \mathcal{U}_t \times \mathcal{U}_c = \mathcal{U}_c \times \mathcal{U}_t ~,
\end{aligned}
\end{equation}
where $\mathcal{C}_0$ is the condensation defect of $\mathbb{Z}_3^{[1]}\times \mathbb{Z}_3^{[1]}$, $\mathcal{N}_d$ is the duality defect corresponding to the diagonal gauging, and $\mathcal{U}_c$ is the diagonal charge conjugation of $\mathbb{Z}_3^{[1]} \times \mathbb{Z}_3^{[1]}$:
\begin{equation}\label{eq:4d_dc_trans}
\begin{aligned}
    \mathcal{N}_d: & \quad Z_{\mathcal{X}}[A^{(2)},B^{(2)}] = \sum_{a^{(2)},b^{(2)}} Z_{\mathcal{X}}[a^{(2)},b^{(2)}] \exp\left(\frac{2\pi i}{3}\int_{\mathcal{M}_4} a^{(2)} \cup A^{(2)} + b^{(2)} \cup B^{(2)} \right) ~, \\
    \mathcal{U}_c: & \quad Z_{\mathcal{X}}[A^{(2)},B^{(2)}] = Z_{\mathcal{X}}[-A^{(2)},-B^{(2)}] ~.
\end{aligned}
\end{equation}
For later use, we will denote the fusion 3-category generated by $\mathbb{Z}_3^{[1]}\times \mathbb{Z}_3^{[1]}$ and $\CN_o$ (or $\CN_d$) as $3\TY_o$ (or $3\TY_d$). Clearly, $3\TY_o$ and $3\TY_d$ are fusion subcategories of $\mathfrak{C}^{[3]}$.

It is straightforward to show that there does not exist any Lagrangian algebra invariant under both $U_o$ and $U_t$ by explicitly enumerating all the Lagrangian algebras. It is then believed that the full symmetry category is not group-theoretical \cite{Sun:2023xxv}, and so is any symmetry category related to it via discrete gauging.

The fact that there is no Lagrangian algebra stable under $U_o$ and $U_t$ also implies that there will be no $\mathbb{Z}_3^{[1]}\times \mathbb{Z}_3^{[1]}$-SPT phase invariant under both transformations in \eqref{eq:4d_ot_trans}. This immediately implies that the full category is indeed anomalous \cite{Sun:2023xxv,Zhang:2023wlu,Cordova:2023bja,Antinucci:2023ezl}. However, the swap symmetry $\mathcal{U}_t$ is free of self-anomaly, and it is straightforward to check using \cite{Antinucci:2023ezl} that the duality defect $\mathcal{N}_o$ itself is also anomaly-free. Thus, we can interpret the anomaly of $\mathfrak{C}^{[3]}$ as a mixed anomaly, loosely speaking, between $\mathcal{N}_o$ and $\mathcal{U}_t$.

\subsection{Constructions of intrinsic NISPT phases}
To get the non-group-theoretical anomaly-free fusion $3$-category symmetry, we consider gauge the $\mathbb{Z}_2^{[0],t}$ symmetry in $\mathfrak{C}^{[3]}$ to get the dual symmetry $\widetilde{\mathfrak{C}}^{[3]}$. It is generated by the following
\begin{enumerate}
    \item 2-form symmetry $\mathbb{Z}_2^{[2],\widetilde{t}}$ generated by Wilson lines of $\mathbb{Z}_2^{[0],t}$ gauge field.
    \item Invertible 1-form symmetry $\mathbb{Z}_3^{[1]}$ coming from the diagonal subgroup of $\mathbb{Z}_3^{[1]}\times \mathbb{Z}_3^{[1]}$; and three non-invertible 1-form symmetries coming from the non-trivial orbit of $\mathbb{Z}_2^{[0],t}$.
    \item The 0-form duality defect $\widetilde{\CN}_o$\footnote{The duality defect $\mathcal{N}_o$ survives the gauging because it commutes with $\mathbb{Z}_2^{[0],t}$, and we add a tilde to denote that the surviving line lives in the dual category $\widetilde{\mathfrak{C}}^{[3]}$.} capturing the invariance under gauging the above 1-form symmetries. 
\end{enumerate}

Let us now demonstrate how to construct a NISPT phase protected by $\widetilde{\mathfrak{C}}^{[3]}$ via $\mathbb{Z}_2^{[0],t}$-gauging. Clearly, we need to find gapped phases of $\mathfrak{C}^{[3]}$ with two degenerate ground states (exchanged by $\mathcal{U}_t$) on any 3-manifold. This implies that it must be a direct sum of two $\mathbb{Z}_3^{[1]}\times \mathbb{Z}_3^{[1]}$-SPT. And a generic $\mathbb{Z}_3^{[1]}\times \mathbb{Z}_3^{[1]}$-SPT phase is parameterized by $(p,q,r) \in \mathbb{Z}_3\times \mathbb{Z}_3 \times \mathbb{Z}_3$ with the partition function
\begin{equation}\label{eq:4dSPT}
    Z_{\SPT_{(p,q,r)}}[A^{(2)},B^{(2)}] = \exp\left(\frac{2\pi i}{3} \int_{\mathcal{M}_4} p A^{(2)}\cup A^{(2)} + q B^{(2)} \cup B^{(2)} + r A^{(2)} \cup B^{(2)} \right) ~.
\end{equation}
And we find the following three pairs of $\mathbb{Z}_3^{[1]}\times \mathbb{Z}_3^{[1]}$-SPT phases possible\footnote{Notice that under the gauging transformation \eqref{eq:4d_ot_trans} and \eqref{eq:4d_dc_trans} in the $\SPT$ partition function \eqref{eq:4dSPT}, gravitational SPT terms will also be generated. However, we will neglect these as we restrict ourselves to only SPT from the group cohomology. For the computation and the interpretation of those gravitational SPT terms, see \cite{Apte:2022xtu,Gaiotto:2014kfa}.}
\begin{equation}\label{eq:4dSPTdiag}
    \begin{tikzpicture}[baseline=0, square/.style={regular polygon,regular polygon sides=4}]
	\filldraw[black] (-1,0) circle (2pt);
    \filldraw[black] (+1,0) circle (2pt);
	\node[above] at (-1.5,0) {\footnotesize $\SPT_{(1,-1,0)}$};
    \node[above] at (+1.5,0) {\footnotesize $\SPT_{(-1,1,0)}$};

    \draw[thick, ->-=1] (-0.8,0) -- (0.8,0);
    \draw[thick, -<-=0.1] (-0.8,0) -- (0.8,0);
    \node[above] at (0,0) {\scriptsize $\mathcal{N}_d,t$};

    \draw[thick, smooth, ->-=1] (-1.2,-0.2) arc (120:420:0.4);
    \node[below] at (-1,-1) {\scriptsize $\mathcal{N}_o$};

    \draw[thick, smooth, ->-=1] (0.8,-0.2) arc (120:420:0.4);
    \node[below] at (1,-1) {\scriptsize $\mathcal{N}_o$};

    \node[below] at (0,-1.5) {(I)};
	\end{tikzpicture} ~, \quad \begin{tikzpicture}[baseline=0, square/.style={regular polygon,regular polygon sides=4}]
	\filldraw[black] (-1,0) circle (2pt);
    \filldraw[black] (+1,0) circle (2pt);
	\node[above] at (-1.5,0) {\footnotesize $\SPT_{(1,-1,1)}$};
    \node[above] at (+1.5,0) {\footnotesize $\SPT_{(-1,1,1)}$};

    \draw[thick, ->-=1] (-0.8,0) -- (0.8,0);
    \draw[thick, -<-=0.1] (-0.8,0) -- (0.8,0);
    \node[above] at (0,0) {\scriptsize $\mathcal{N}_o,t$};

    \draw[thick, smooth, ->-=1] (-1.2,-0.2) arc (120:420:0.4);
    \node[below] at (-1,-1) {\scriptsize $\mathcal{N}_d$};

    \draw[thick, smooth, ->-=1] (0.8,-0.2) arc (120:420:0.4);
    \node[below] at (1,-1) {\scriptsize $\mathcal{N}_d$};
    \node[below] at (0,-1.5) {(II)};
	\end{tikzpicture} ~, \begin{tikzpicture}[baseline=0, square/.style={regular polygon,regular polygon sides=4}]
	\filldraw[black] (-1,0) circle (2pt);
    \filldraw[black] (+1,0) circle (2pt);
	\node[above] at (-1.5,0) {\footnotesize $\SPT_{(1,-1,-1)}$};
    \node[above] at (+1.5,0) {\footnotesize $\SPT_{(-1,1,-1)}$};

    \draw[thick, ->-=1] (-0.8,0) -- (0.8,0);
    \draw[thick, -<-=0.1] (-0.8,0) -- (0.8,0);
    \node[above] at (0,0) {\scriptsize $\mathcal{N}_o,t$};

    \draw[thick, smooth, ->-=1] (-1.2,-0.2) arc (120:420:0.4);
    \node[below] at (-1,-1) {\scriptsize $\mathcal{N}_d$};

    \draw[thick, smooth, ->-=1] (0.8,-0.2) arc (120:420:0.4);
    \node[below] at (1,-1) {\scriptsize $\mathcal{N}_d$};
    \node[below] at (0,-1.5) {(III)};
	\end{tikzpicture} ~.
\end{equation}

Let us first consider case (I). In this case, on each ground state we must enrich the $\mathbb{Z}_3^{[1]} \times \mathbb{Z}_3^{[1]}$-SPT to a $\mathcal{N}_o$-SPT. To determine the possible higher data, one can consider the gauging:
\begin{equation}\label{eq:4dgaugingmap}
\begin{aligned}
    Z_{\widetilde{\mathcal{X}}}[A^{(2)},B^{(2)}] := & \sum_{b^{(2)}\in H^2(\mathcal{M}_4,\mathbb{Z}_3)} Z_{\mathcal{X}}[A^{(2)},b^{(2)}] \exp\left(+\frac{2\pi i}{3}\int_{\mathcal{M}_4} b^{(2)} \cup B^{(2)} \right) ~, \\
    Z_{\mathcal{X}}[A^{(2)},B^{(2)}] =& \sum_{b^{(2)} \in H^2(\mathcal{M}_4,\mathbb{Z}_3)} Z_{\widetilde{\mathcal{X}}}[A^{(2)},b^{(2)}]  \exp\left(-\frac{2\pi i}{3}\int_{\mathcal{M}_4} b^{(2)} \cup B^{(2)} \right) ~,
\end{aligned}
\end{equation}
under which the $\mathcal{N}_o$ non-invertible symmetry in a theory $\mathcal{X}$ is mapped to an invertible automorphism in theory $\widetilde{\mathcal{X}}$:
\begin{equation}\label{eq:2groupaut}
    Z_{\widetilde{\mathcal{X}}}[A^{(2)},B^{(2)}] = Z_{\widetilde{\mathcal{X}}}[-B^{(2)},A^{(2)}] ~.
\end{equation}
Therefore, $\mathbb{Z}_3^{[1]}\times \mathbb{Z}_3^{[1]}$ with $\mathcal{N}_o$ is mapped to the invertible symmetry $(\mathbb{Z}_3^{[1]}\times \mathbb{Z}_3^{[1]})\rtimes \mathbb{Z}_4^{[0]}$ under the gauging \eqref{eq:4dgaugingmap}. Furthermore, the $\SPT_{1,-1,0}$ ($\SPT_{-1,1,0}$) is mapped to $\widetilde{\SPT}_{(1,1,0)}$ ($\widetilde{\SPT}_{(-1,-1,0)}$) of the dual 1-form symmetry. Thus, the enrichment question for $\mathcal{N}_o$ becomes the question of enriching $\mathbb{Z}_3^{[1]}\times \mathbb{Z}_3^{[1]}$-SPT phase to $(\mathbb{Z}_3^{[1]}\times \mathbb{Z}_3^{[1]})\rtimes \mathbb{Z}_4^{[0]}$-SPT phase, which is classified by the cohomology of the 2-group. Indeed, using the spectral sequence in \cite{Kapustin:2013uxa}, one can show that the enrichment exists and is unique for both\footnote{To see this, let 2-group $\mathbb{G} = (\mathbb{Z}_3^{[1]} \times \mathbb{Z}_3^{[1]})\rtimes \mathbb{Z}_4^{[0]}$ and denote the automorphism action \eqref{eq:2groupaut} of $\mathbb{Z}_4^{[0]}$ on $\mathbb{Z}_3^{[1]} \times \mathbb{Z}_3^{[1]}$ to be $\rho$. The 4d SPT phase of $\mathbb{G}$ is classified by $H^5(B\mathbb{G},\mathbb{Z})$ which can be computed via Serre spectral sequence \cite{Kapustin:2013uxa}. In the spectral sequence, the $\mathrm{E}^2$ page is given by $\mathrm{E}^2_{p,q} \simeq H^p_\rho(B\mathbb{Z}_4,H^q(B^2(\mathbb{Z}_3 \times\mathbb{Z}_3),\mathbb{Z}))$ and the relevant data for 4d SPT corresponds to $p+q = 5$. It is straightforward to check that the only non-vanishing one is $\mathrm{E}^2_{0,5} = H_\rho^0(B\mathbb{Z}_4,H^5(B^2(\mathbb{Z}_3 \times\mathbb{Z}_3),\mathbb{Z}))$ which is nothing but the group of the $\mathbb{Z}_4^{[0]}$-invariant $\mathbb{Z}_3^{[1]} \times \mathbb{Z}_3^{[1]}$ SPT phases. Hence, we conclude there is a unique way of extending  each $\mathbb{Z}_4^{[0]}$-invariant $\mathbb{Z}_3^{[1]} \times \mathbb{Z}_3^{[1]}$ SPT phase to a $\mathbb{G}$-SPT.}. To summarize, we conclude that the two ground states in (I) of \eqref{eq:4dSPTdiag} have two distinct $3\TY_o$-SPT phases exchanged by $\mathcal{U}_t$. This is a partial SSB phase of $\mathfrak{C}^{[3]}$ with two ground states swapped by $\mathcal{U}_t$\footnote{Here, we expect there is no higher data and obstruction in gluing two phases exchanged by $\mathcal{U}_t$ to get a phase with $\mathcal{U}_t$-symmetry as in 2-dim.}.

Next, let's consider the case (II) in \eqref{eq:4dSPTdiag}, where each vacuum must now realize a $3\TY_d$-SPT phase. In this case, we need to enrich $\SPT_{(1,-1,1)}$ and $\SPT_{(-1,1,1)}$ to $3\TY_d$-SPT phases. The enrichment can be analyzed by the discrete gauging
\begin{equation}\label{eq:4dIIgauge}
    Z_{\widetilde{\mathcal{X}}'}[A^{(2)},B^{(2)}] = \sum_{a^{(2)},b^{(2)}} Z_{\mathcal{X}}[a^{(2)},b^{(2)}] e^{\frac{2\pi i}{3}\int_{\mathcal{M}_4} a^{(2)}\cup a^{(2)} - b^{(2)} \cup b^{(2)} + a^{(2)} \cup b^{(2)} + a^{(2)} \cup A^{(2)} + b^{(2)} \cup B^{(2)} - A^{(2)}\cup B^{(2)}} ~,
\end{equation}
under which the diagonal duality defect $\mathcal{N}_d$ in $\mathcal{X}$ is mapped to an invertible automorphism symmetry in the dual theory $\widetilde{\mathcal{X}}'$ which acts as
\begin{equation}
    Z_{\widetilde{\mathcal{X}}}[A^{(2)},B^{(2)}] = Z_{\widetilde{\mathcal{X}}}[-A^{(2)} + B^{(2)}, A^{(2)} + B^{(2)}] ~.
\end{equation}
Under the gauging \eqref{eq:4dIIgauge}, the $\SPT_{(1,-1,1)}$ and $\SPT_{(-1,1,1)}$ are mapped to $\widetilde{\SPT}'_{(-1,1,1)}$ and $\widetilde{\SPT}'_{(0,0,0)}$ of the dual $\mathbb{Z}_3^{[1]} \times \mathbb{Z}_3^{[1]}$-symmetry respectively; and the same analysis shows the enrichment is unique for both using the same group cohomology argument. Hence, case (II) leads to a single desired partial SSB phase.

Finally, let's consider the case (III) in \eqref{eq:4dSPTdiag}.  The analysis is completely the same as the case (II), except it is convenient to consider an alternative discrete gauging 
\begin{equation}\label{eq:4dIIIgauge}
    Z_{\widetilde{\mathcal{X}}''}[A^{(2)},B^{(2)}] = \sum_{a^{(2)},b^{(2)}} Z_{\mathcal{X}}[a^{(2)},b^{(2)}] e^{\frac{2\pi i}{3}\int_{\mathcal{M}_4} a^{(2)}\cup a^{(2)} - b^{(2)} \cup b^{(2)} - a^{(2)} \cup b^{(2)} + a^{(2)} \cup A^{(2)} + b^{(2)} \cup B^{(2)} + A^{(2)} \cup B^{(2)}} ~,
\end{equation}
which turns the non-invertible symmetry $\CN_d$ in $\mathcal{X}$ to be an invertible automorphism symmetry of the dual $\mathbb{Z}_3^{[1]}\times \mathbb{Z}_3^{[1]}$-symmetry acting as
\begin{equation}
    Z_{\widetilde{\mathcal{X}}''}[A^{(2)}, B^{(2)}] = Z_{\widetilde{\mathcal{X}}''}[-A^{(2)}-B^{(2)},-A^{(2)}+B^{(2)}] ~.
\end{equation}
Under the gauging \eqref{eq:4dIIIgauge}, $\SPT_{(1,-1,-1)}$ and $\SPT_{(-1,1,-1)}$ are mapped to SPT phases $\widetilde{\SPT}''_{(-1,1,-1)}$ and $\widetilde{\SPT}''_{(0,0,0)}$ of the dual $\mathbb{Z}_3^{[1]} \times \mathbb{Z}_3^{[1]}$-symmetry respectively; and the same analysis shows the enrichment is unique for both using the same group cohomology argument. Hence, case (III) leads to a single desired partial SSB phase.

To summarize, via discrete gauging $\mathbb{Z}_2^{[0],t}$, we find three $\widetilde{\mathfrak{C}}^{[3]}$-SPT phases. It would be interesting to study their interfaces, which we leave for future study.

\section*{Acknowledgments}
Z.S. thanks Theo Jacobson for discussions. D.C.L. thanks Arkya Chatterjee, Saranesh Prembabu, Ashvin Vishwanath, and Xiao-Gang Wen for discussions. This research was supported in part by grant NSF PHY-2309135 to the Kavli Institute for Theoretical Physics (KITP). Research of D.C.L. is supported by the Simons Collaboration on Ultra-Quantum Matter, which is a grant from the Simons Foundation 651440. Z.S. is supported by the Simons Collaboration on Global Categorical Symmetries.

\appendix

\section{SPT phases of certain $\TY(\doubleZ_p\times \doubleZ_p)$-fusion categories}\label{app:ff}
In this appendix, we list the details on acquiring fiber functors of the two $\TY$-fusion categories used in this paper. We will first review the classification of SPT phases of a group-theoretical fusion category in Appendix \ref{app:gptSPT} and the result acquired by direct computation in $\TY$-fusion categories \cite{tambara2000representations,meir2012module,Thorngren:2019iar} in Appendix \ref{app:TYff}. Then we will apply the above two alternative methods to $\TYpo$ in Appendix \ref{app:TYo} and $\TYpd$ in Appendix \ref{app:TYd}.

\subsection{Classification of SPT phases of a group-theoretical fusion category}\label{app:gptSPT}
For a categorical symmetry $\mathcal{C}$, if it is mapped to an invertible symmetry $\VEC_{G}^\omega$ under some discrete gauging, then the classification of $\mathcal{C}$-SPT phases can be mapped to the classification of certain gapped phases of $\VEC_G^\omega$, and the latter is well established. The classification is determined in \cite{ostrik2002module,natale2017grptheo} via this, and to state the result, let us notice that one can always construct such symmetry $\mathcal{C}$ as the dual symmetry of gauging an anomaly-free $H$ subgroup with discrete torsion $[\psi] \in H^2(H,U(1))$ in $\VEC_G^{\omega}$. Such $\mathcal{C}$ is known as a group-theoretical fusion category, and we denote it as $\mathcal{C}(G,\omega;H,\psi)$. 

The SPT phases of the group-theoretical fusion category $\mathcal{C}(G,\omega;H,\psi)$ are labeled by $(K,\psi_K)$ such that
\begin{enumerate}
    \item $K$ is an anomaly-free subgroup of $G$; 
    \item $\psi_K \in C^2(K,U(1))$ such that $d\psi_K = \omega|_K$;
    \item The 2-cocycle $\psi_{\dsi} = (\psi_{H\cap K})(\psi_K |_{H\cap K})^{-1}$ is non-degenerate, which implies there’s only a single irreducible projective representation for the little group $H^\dsi \cong H\cap K$.
\end{enumerate}
Furthermore, $(K,\psi_K)$ and $(\widetilde{K},\psi_{\widetilde{K}})$ parameterize the same $\mathcal{C}(G,\omega;H,\psi)$ SPT phase if there exists $g \in G$ such that $K = {}^g K$ (where we use ${}^g x$ to denote $g x g^{-1}$) and the following 2-cocycle in $H^2(\widetilde{K},U(1))$
\begin{equation}
    \frac{\psi_K({}^g \widetilde{k}_1, {}^g \widetilde{k}_2)}{\psi_{\widetilde{K}}(\widetilde{k}_1,\widetilde{k}_2)} \frac{\omega({}^g \widetilde{k}_1, {}^g \widetilde{k}_2, g)\omega(g,\widetilde{k}_1,\widetilde{k}_2)}{\omega({}^g \widetilde{k}_1,g,\widetilde{k}_2)}
\end{equation}
is cohomologically trivial.

The above classification is acquired by the technique of discrete gauging mentioned above. First, any $(K,\psi_K)$ satisfying the first two conditions labels a $\VEC_G^\omega$-symmetric gapped phase where an anomaly-free subgroup $K$ is unbroken and the $K$-SPT phase $\psi_K$ is realized on some vacuum. Furthermore, $(K,\psi_K)$ and $(\widetilde{K},\psi_{\widetilde{K}})$ label equivalent $\VEC_G^\omega$-symmetric gapped phases if they satisfy the above conditions\cite{natale2017grptheo}. Finally, the third condition on $(K,\psi_K)$ ensures that gauging $H$ with discrete torsion $\psi$ completely removes the ground state degeneracy, thus leading to an SPT phase.

\subsection{Classification of SPT phases in the $\TY$ frame}\label{app:TYff}
The SPT phases of a generic Tambara-Yamagami fusion category $\TY(A,\chi,\epsilon)$ can be  classified directly in the $\TY$ frame. First, $A$-SPT phases are labeled by $\alpha \in H^2(A,U(1))$, and $\SPT_\alpha$ is invariant under gauging specified by $\chi$ if 
\begin{equation}\label{eq:TYFF1}
    \frac{\alpha(g,\sigma(h))}{\alpha(\sigma(h),g)} = \chi(g,h) ~, \quad \forall g,h \in A ~,
\end{equation}
for some order-$2$ automorphism $\sigma$ of $A$, which is unique when it exists. Physically, this means in the putative SPT phase, bringing a $g$-defect across $\mathcal{N}$ will turn it into the $\sigma(g)$-defect. To enrich the $\SPT_\alpha$ into a $\TY$-SPT phase, one must specify the phase factor $\alpha(g)$:
\begin{equation}\label{eq:TYnu}
    \begin{tikzpicture}[baseline={([yshift=+.5ex]current bounding box.center)},vertex/.style={anchor=base,
    circle,fill=black!25,minimum size=18pt,inner sep=2pt},scale=0.7]
    \filldraw[grey] (-2,-2) rectangle ++(4,4);
    \draw[black] (-2,-2) rectangle ++(4,4);
    \draw[blue, line width = 0.4mm] (0,-2) -- (0,2);
    \node[below, blue] at (0,-2) {$\mathcal{N}$};
    \draw[black, line width = 0.4mm, ->-=0.5] (-1,-2) arc(180:90:1 and 2);
    \node[below, black] at (-1,-2) {$g$};
\end{tikzpicture} \quad = \nu(g) \quad \begin{tikzpicture}[baseline={([yshift=+.5ex]current bounding box.center)},vertex/.style={anchor=base,
    circle,fill=black!25,minimum size=18pt,inner sep=2pt},scale=0.7]
    \filldraw[grey] (-2,-2) rectangle ++(4,4);
    \draw[black] (-2,-2) rectangle ++(4,4);
    \draw[blue, line width = 0.4mm] (0,-2) -- (0,2);
    \node[below, blue] at (0,-2) {$\mathcal{N}$};
    \draw[black, line width = 0.4mm, ->-=0.5] (1,-2) arc(0:90:1 and 2);
    \node[below, black] at (1,-2) {$\sigma(g)$};
\end{tikzpicture} ~,
\end{equation}
which subject to the following constraints
\begin{equation}\label{eq:TYFF2}
    \frac{\nu(g)\nu(h)}{\nu(gh)} = \frac{\alpha(g,h)}{\alpha(\sigma(h),\sigma(g))} ~, \quad \nu(g) \nu(\sigma(g)) = 1 ~, \quad sgn\left(\sum_{g = \sigma(g)} \nu(g) \right) = \epsilon ~.
\end{equation}

Furthermore, given two pairs of $(\alpha,\sigma)$ and $(\alpha',\sigma')$, they give rise to equivalent fiber functors if there exists a gauge transformation $\phi \in C^1(A,U(1))$ relating the two
\begin{equation}
    \alpha'(g,h) = \alpha(g,h) \frac{\phi(g) \phi(h)}{\phi(gh)} ~, \quad \nu'(g) = \nu(g) \frac{\phi(g)}{\phi(\sigma(g))} ~.
\end{equation}
The transformation on $\alpha$ implies that only the cohomology class of $\alpha$ matters, as it should since only the $A$-SPT phases are labeled by the cohomology class. Fixing a representative $\alpha$ in the class, then $\nu$ and $\nu'$ parameterize the same $\TY$-SPT phase if there exists $\phi \in Z^1(A,U(1))$ such that $\nu'(g) = \nu(g) \frac{\phi(g)}{\phi(\sigma(g))}$.

\subsection{$\TY(\mathbb{Z}_p\times \mathbb{Z}_p,\chi_{o},+1)$ when $p$ is an odd prime}\label{app:TYo}
\subsubsection*{Group-theoretical method}

The $\TY(\IZ_p\times \IZ_p,\chi_{o},+1)$ can be acquired by starting with the finite group symmetry $G = (\IZ_p^r\times \IZ_p^s)\rtimes\IZ_2^t$ (where $t$ swaps the generators $r$ and $s$) with the trivial anomaly $\omega = 1$ and gauge the subgroup $H = \mathbb{Z}_p^r$. This can be seen as follows. The duality defect $\mathcal{N}_{o}$ in $\TY(\IZ_p\times \IZ_p,\chi_{o},+1)$ implies that any 2d QFT $\mathcal{X}$ admitting $\TY(\IZ_p\times \IZ_p,\chi_{o},+1)$ is invariant under the following discrete gauging
\begin{equation}\label{eq:o_gg}
    Z_{\mathcal{X}}[A^{(1)},B^{(1)}] = \sum_{a^{(1)},b^{(1)}} Z_{\mathcal{X}}[a^{(1)},b^{(1)}] \exp\left(\frac{2\pi i}{p}\int_{\mathcal{M}_2} a^{(1)}\cup B^{(1)} + b^{(1)} \cup A^{(1)} \right) ~.
\end{equation}
Consider the new theory $\widetilde{\mathcal{X}}$ acquired by discrete gauging $\mathbb{Z}_p$:
\begin{equation}
\begin{aligned}
    Z_{\widetilde{\mathcal{X}}}[A^{(1)},B^{(1)}] &= \sum_{a^{(1)}} Z_{\mathcal{X}}[a^{(1)},B^{(1)}] \exp\left(\frac{2\pi i}{p}\int_{\mathcal{M}_2} a^{(1)}\cup A^{(1)} \right) ~, \\
    Z_{\mathcal{X}}[A^{(1)},B^{(1)}] &= \sum_{a^{(1)}} Z_{\widetilde{\mathcal{X}}}[a^{(1)},B^{(1)}] \exp\left(\frac{2\pi i}{p}\int_{\mathcal{M}_2} a^{(1)}\cup A^{(1)} \right) ~, \\
\end{aligned}
\end{equation}
one can show that it is invariant under the automorphism swapping the two factors of $\mathbb{Z}_p$'s, as
\begin{equation}
\begin{aligned}
    Z_{\widetilde{\mathcal{X}}}[A^{(1)},B^{(1)}] 
    = & \sum_{a^{(1)}} Z_{\mathcal{X}}[a^{(1)},B^{(1)}] \exp\left(\frac{2\pi i}{p}\int_{\mathcal{M}_2} a^{(1)}\cup A^{(1)} \right) \\
    = & \sum_{a^{(1)}} Z_{\mathcal{X}}[\widetilde{a}^{(1)},b^{(1)}] \exp\left(\frac{2\pi i}{p}\int_{\mathcal{M}_2} \widetilde{a}^{(1)} \cup B^{(1)} + b^{(1)} \cup a^{(1)} + a^{(1)}\cup A^{(1)} \right) \\
    = & \sum_{a^{(1)}} Z_{\mathcal{X}}[\widetilde{a}^{(1)},A^{(1)}] \exp\left(\frac{2\pi i}{p}\int_{\mathcal{M}_2} \widetilde{a}^{(1)} \cup B^{(1)} \right) \\
    = & Z_{\widetilde{\mathcal{X}}}[B^{(1)},A^{(1)}] ~,
\end{aligned}
\end{equation}
where we have suppressed the normalization factor in the intermediate step. That the FS indicator of $\mathcal{N}_{od}$ is $+1$ implies that there is no self-anomaly on $\mathbb{Z}_2^t$. On the other hand, $\gcd(p,2) = 1$ implies that there cannot be a mixed anomaly between $\mathbb{Z}_2^t$ and $\mathbb{Z}_p^r \times \mathbb{Z}_p^{s}$ in the first place. Then we conclude that the anomaly of $(\mathbb{Z}_p^r \times \mathbb{Z}_p^{s})\rtimes \mathbb{Z}_2^t$ is trivial.

Having determined $\TY(\mathbb{Z}_p\times \mathbb{Z}_p,\chi_o,+1) = \mathcal{C}\big((\mathbb{Z}_p^r \times \mathbb{Z}_p^{s})\rtimes \mathbb{Z}_2^t,1;\mathbb{Z}_p^r,1\big)$ in the group-theoretical fusion category parameterization, it is straightforward to apply the classification of SPT phases in Section \ref{sec:SPTgrpt}. It is straightforward to check that there are $p+1$ subgroups $K$'s satisfying the three conditions:
\begin{equation}
    \langle rs^{-1},t \rangle ~, \quad \langle rst \rangle ~, \quad \langle s^k t\rangle ~, \quad k = 1,\cdots,p-1 ~,
\end{equation}
where the first subgroup is isomorphic to $D_{2p}$ while the last $p$ subgroups are isomorphic to $\mathbb{Z}_{2p}$. On the other hand, it is straightforward to check that all the $\mathbb{Z}_{2p}$ subgroups are related to each other by conjugation. Hence, we conclude that in total there are only two $\TY(\mathbb{Z}_p\times \mathbb{Z}_p,\chi_{o},+1)$-SPT phases.

\subsubsection*{Direct computation}
Apply Appendix \ref{app:TYff} to $A = \mathbb{Z}_p\times \mathbb{Z}_p$, $\chi_o(g,h) = \omega^{\mathrm{g}_1 \mathrm{h}_2 + \mathrm{g}_2 \mathrm{h}_1}$ and $\epsilon = 1$, we first solve for allowed $\alpha$ (up to cohomology class) and all possible $\nu$'s. We find
\begin{equation}
\begin{aligned}
    & \alpha_+(g,h) = \omega^{+\mathrm{g}_1 \mathrm{h}_2} ~, \quad \sigma_+: (\mathrm{g}_1,\mathrm{g}_2) \mapsto (-\mathrm{g}_1,\mathrm{g}_2) ~, \quad  \nu_{+,\beta}(g) = \omega^{-\mathrm{g}_1 \mathrm{g}_2 + \beta \mathrm{g}_1} ~, \\
    & \alpha_-(g,h) = \omega^{-\mathrm{g}_1 \mathrm{h}_2} ~, \quad \sigma_-: (\mathrm{g}_1,\mathrm{g}_2)\mapsto (\mathrm{g}_1, -\mathrm{g}_2) ~, \quad \nu_{-,\beta}(g) = \omega^{+\mathrm{g}_1 \mathrm{g}_2 + \beta \mathrm{g}_2} ~,
\end{aligned}
\end{equation} 
where $\beta \in \mathbb{Z}_p$. Next, we consider identifications of $\nu$'s. The allowed gauge transformation parameter $\phi \in Z^1(A,U(1))$ is parameterized by $\phi(g) = \omega^{a_1 g_1 + a_2 g_2}$. For $\alpha_+$, $\nu_{+,\beta}$ and $\nu_{+,\beta'}$ is identified if there exists $\alpha_i$ such that 
\begin{equation}
    \omega^{(\beta'-\beta)g_1} = \omega^{2a_1 g_1} ~,
\end{equation}
which always admits a solution $a_1 = \frac{p+1}{2} (\beta'-\beta) \mod p$ as $p$ is an odd prime. Thus, all the $\nu_{+,\beta}$'s are equivalent to each other. And similarly, all the $\nu_{-,\beta}$'s are equivalent to each other. And we rediscover the two inequivalent fiber functors and we use $\mathcal{F}_{o,\pm}$ to denote them. 

\

In Section \ref{sec:intrinNISPT}, we need the action of the swap symmetry $t$ on the $\TY$-SPT phases. This can be easily checked in the $\TY$-symmetry frame instead of in the invertible symmetry frame, because the discrete gauging making $\TY$ invertible will make the swap symmetry $t$ non-invertible. Because $t$ swaps the two $\mathbb{Z}_p\times \mathbb{Z}_p$-SPT $\SPT_{\pm}$, it must also swap the two $\TY$-SPT $\mathcal{F}_{o,\pm}$.

\subsection{$\TY(\mathbb{Z}_p\times \mathbb{Z}_p,\chi_d,+1)$ when $p$ is an odd prime}\label{app:TYd}
\subsubsection*{Group-theoretical method}
For the duality defect associated with the diagonal bicharacter $\chi_d$, whether it is group-theoretical depends on the odd prime number $p$. When $p = 3 \mod 4$, $\TY(\mathbb{Z}_p\times \mathbb{Z}_p,\chi_d,+1)$ is not group-theoretical and is automatically anomalous; but when $p = 1 \mod 4$, the category is group-theoretical and can be acquired by gauging the $H = \mathbb{Z}_p^r$ subgroup in $G = (\mathbb{Z}_p^r\times\mathbb{Z}_p^s)\rtimes \mathbb{Z}_2^{\widetilde{t}}$ where the $\rtimes$ is given by $\widetilde{t}r\widetilde{t} = s^x, \widetilde{t}s\widetilde{t} = r^{-x}$ and $x$ satisfies $x^2 = -1 \mod p$ and exists only when $p = 1 \mod 4$ (assuming already $p$ is an odd prime).

Again, to see this, we consider the following gauging in a theory $\mathcal{X}$ with $\TY(\mathbb{Z}_p\times \mathbb{Z}_p,\chi_d,+1)$ to get a new theory $\widetilde{\mathcal{X}}$:
\begin{equation}\label{eq:TYd_dgg}
\begin{aligned}
    Z_{\widetilde{\mathcal{X}}}[A^{(1)},B^{(1)}] &= \sum_{a^{(1)}}Z_{\mathcal{X}}[a^{(1)},x a^{(1)} + B^{(1)}]\exp\left(\frac{2\pi i}{p} \int_{\mathcal{M}_2} a^{(1)} \cup A^{(1)} + \frac{p+1}{2}x A^{(1)}\cup B^{(1)}\right) ~, \\
    Z_{\mathcal{X}}[A^{(1)},B^{(1)}] &= \sum_{a^{(1)}} Z_{\widetilde{\mathcal{X}}}[a^{(1)},-xA^{(1)}+B^{(1)}] \exp\left(\frac{2\pi i}{p}\int_{\mathcal{M}_2} a^{(1)} \cup \left(\frac{p+1}{2} A^{(1)} - \frac{p+1}{2}xB^{(1)}\right)\right) ~,
\end{aligned}
\end{equation}
It is straightforward to check the invariance under the diagonal gauging of $\mathcal{X}$:
\begin{equation}\label{eq:TYd_inv_gg}
    Z_{\mathcal{X}}[A^{(1)},B^{(1)}] = \sum_{a^{(1)},b^{(1)}} Z_{\mathcal{X}}[a^{(1)},b^{(1)}] \exp\left(\frac{2\pi i}{p}\int_{\mathcal{M}_2} a^{(1)}\cup A^{(1)} + b^{(1)} \cup B^{(1)} \right)~,
\end{equation}
is mapped to the invariance under the following automorphism of $\mathbb{Z}_p\times \mathbb{Z}_p$ symmetry in $\widetilde{\mathcal{X}}$: 
\begin{equation}
    Z_{\widetilde{\mathcal{X}}}[A^{(1)},B^{(1)}] = Z_{\widetilde{\mathcal{X}}}[x B^{(1)},-xA^{(1)}] ~.
\end{equation}
This, together with the gauging relation given by the second line in \eqref{eq:TYd_dgg} determines the group-theoretical construction of $\TY(\mathbb{Z}_p \times \mathbb{Z}_p, \chi_d,+1)$ is parameterized by
\begin{equation}
    G = (\mathbb{Z}_p^r \times \mathbb{Z}_p^s) \rtimes' \mathbb{Z}_2^{\widetilde{t}} = \langle r,s,t| r^p = s^p = \widetilde{t}^2 = 1, rs = sr, \widetilde{t}s\widetilde{t} = r^{-x}\rangle ~, \quad \omega = 1 ~, \quad H = \mathbb{Z}_p^r ~.
\end{equation}
Applying the group-theoretical method, we find the subgroups $K$ satisfying the three conditions are given by
\begin{equation}
    \langle r^x s, \widetilde{t}\rangle ~, \quad \langle r^{-x}s\widetilde{t}\rangle ~, \quad \langle s^k \widetilde{t} \rangle ~, \quad k = 1,\cdots,p-1 ~,
\end{equation}
where the first subgroup is isomorphic to $D_{2p}$ while the last $p$ subgroups are all isomorphic $\mathbb{Z}_{2p}$. However, all the $\mathbb{Z}_{2p}$ subgroups are related to each other by conjugation. Therefore, we conclude there are only two inequivalent partial SSB phases, with $\langle r^x s, \widetilde{t}\rangle$ and $\langle r^{-x}s \widetilde{t} \rangle $ unbroken respectively, that lead to $\TY$-SPT phases after gauging.

\subsubsection*{Direct computation}
Apply Appendix \ref{app:TYff} to $A = \mathbb{Z}_p\times \mathbb{Z}_p$, $\chi_d(\mathrm{g},\mathrm{h}) = \omega^{\mathrm{g}_1 \mathrm{h}_1 + \mathrm{g}_2 \mathrm{h}_2}$ and $\epsilon = 1$, we first solve for allowed $\alpha$ (up to cohomology class) and all possible $\nu$'s. The solutions exist only when $p = 1 \mod 4$, and a way to understand this is that only when $p = 1 \mod 4$, there are $\mathbb{Z}_p\times \mathbb{Z}_p$-SPT phases invariant under gauging \eqref{eq:TYd_inv_gg}: the $\SPT_k$ will be mapped to $\SPT_{-k^{-1}}$ (where $k^{-1}$ is the mod $p$ inverse of $k$), and \eqref{eq:TYd_inv_gg} implies that
\begin{equation}
    k^2 = -1 \mod p ~,
\end{equation}
which only admit solutions when $p = 1 \mod 4$ and they are given by
\begin{equation}
    \pm x \equiv \pm \left(\frac{p-1}{2}\right)! ~.
\end{equation}

The allowed $\alpha$'s and $\nu$'s when $p = 1 \mod 4$ are given by
\begin{equation}
\begin{aligned}
    & \alpha_{+x}(g,h) = \omega^{+x\mathrm{g}_1 \mathrm{h}_2} ~, \quad \sigma_{+x}(g) = (x\mathrm{g}_2, -x\mathrm{g}_1) ~, \quad \nu_{+x}(g) = 1 ~, \\
    & \alpha_{-x}(g,h) = \omega^{-x\mathrm{g}_1 \mathrm{h}_2} ~, \quad \sigma_{-x}(g) = (-x\mathrm{g}_2,x \mathrm{g}_1) ~, \quad \nu_{-x}(g) = 1 ~.
\end{aligned}
\end{equation}
Here, for each given $\alpha$, there is a unique $\nu(g)$ satisfying \eqref{eq:TYFF2} which directly leads to a single SPT phase. In total, there are two inequivalent SPT phases from direct computation, matching the group-theoretical method. And we denote the two SPT phases as $\mathcal{F}_{d,\pm}$ corresponding to $\alpha_{\pm x}$ respectively. And it is straightforward to check that the swap symmetry $t$ does swap the two $\TY$-SPT $\mathcal{F}_{d,\pm}$ here.

\section{Details on gauging $\IZ_2^t$ on the lattice}\label{app:gauging}
Here, we briefly explain the procedure of gauging the $\mathbb{Z}_2^t$ symmetry generated by $U_t$ used in Section \ref{sec:iNISPT_lattice} and Section \ref{sec:igNISPTlatt}, following \cite{xgw2024reps3}. The $S_3$ symmetry in \eqref{eq:TYxz2ham2} and \eqref{eq:TYdZ2Ham1} is generated by
\begin{align}
    U_r = \prod \tX_i, \quad U_t=\prod_j \tC_j \sigma^x_j ~.
\end{align}
Both Hamiltonians are generated by the following $S_3$-invariant terms (plus other variables uncharged under the $S_3$ symmetry) which form the so-called bond algebra for $S_3$:
\begin{equation}\label{eq:S3ba}
\begin{aligned}
    \langle \sigma^x_i,\quad \sigma_i^z\sigma_{i+1}^z, &\quad(\tX_i+\tX_i^\dagger),\quad (\tZ_i \tZ_{i+1}^\dagger + \tZ_i^\dagger \tZ_{i+1}), \\
    & \quad \quad \quad \quad \quad \quad \quad \quad \quad \quad \quad \quad \sigma_i^z(\tX_i-\tX_i^\dagger),\quad \sigma^z_{i} (\tZ_i \tZ_{i+1}^\dagger - \tZ_i^\dagger \tZ_{i+1})\rangle ~.
\end{aligned}
\end{equation}

To gauge $\mathbb{Z}_2^t$ symmetry, we extend the Hilbert space by introducing a spin variable on each link and the $\mathbb{Z}_2$ gauge field is given by the Pauli matrix $\mu_{i+\frac{1}{2}}^x$. The Gauss law operators which generate local gauge transformations for the $\IZ_2^t$ symmetry are
\begin{equation}
    G^t_i=\mu_{i-1/2}^z  \tC_i\sigma^x_i \mu_{i+1/2}^z ~.
\end{equation}
The Hamiltonians on the extended Hilbert space must be gauge invariant and thus commute with the Gauss law operators $G^t_i$. This can be achieved by coupling each term in the bond algebra \eqref{eq:S3ba} minimally to the $\mathbb{Z}_2^t$ gauge fields $\mu_{i + \frac{1}{2}}^x$, and the resulting gauge-invariant terms generate the $S^3$ bond algebra on the extended Hilbert space:
\begin{equation}\label{eq:eba}
\begin{aligned}
    & \langle \sigma^x_i,\quad \sigma_i^z\mu_{i+1/2}^x\sigma_{i+1}^z, \quad(\tX_i+\tX_i^\dagger),\quad (\tZ_i^{\mu_{i+1/2}^x} \tZ_{i+1}^\dagger + \tZ_i^{-\mu_{i+1/2}^x} \tZ_{i+1}), \\
    & \quad \quad \quad \quad \quad \quad  \quad \quad \quad \quad \quad \quad  \quad \quad \sigma_i^z(\tX_i-\tX_i^\dagger),\quad \sigma^z_{i} \mu_{i+1/2}^x(\tZ_i^{\mu_{i+1/2}^x} \tZ_{i+1}^\dagger - \tZ_i^{-\mu_{i+1/2}^x} \tZ_{i+1})\rangle ~.
\end{aligned}
\end{equation}

The physical Hilbert space is the $G^t_i$-invariant subspace of the extended Hilbert space, and oftentimes one can solve the Gauss law constraints and restrict the gauged Hamiltonian on the physical Hilbert space to simplify its form. The procedure involves a change of variables simplifying the form of $G_i^t$, and can be captured by conjugating every operator with some unitary operator $U$ such that $U G^t_i U^\dagger =\sigma^x_i$. 

To be specific on the unitary transformation, we first apply the onsite unitary transformation $U_\text{cond}=\prod_i C^\sigma\tC_{i}$, where
\begin{equation}
    C^\sigma\tC_i = \frac{1+\sigma^z_i}{2} I_i +\frac{1-\sigma^z_i}{2} \tC_i ~,
\end{equation}
such that
\begin{equation}
G_i^t=\mu_{i-1/2}^z  \tC_i\sigma^x_i \mu_{i+1/2}^z\xrightarrow{U_\text{cond}} \mu_{i-1/2}^z  \sigma^x_i \mu_{i+1/2}^z ~,
\end{equation}
also it transforms,
\begin{equation}
\begin{aligned}
    \sigma^x\tC \leftrightarrow \sigma^x I,\quad  I \tX \leftrightarrow \tX^{\sigma^z},\quad I \tZ \leftrightarrow \tZ^{\sigma^z}, \\
    I(\tX-\tX^\dagger) \leftrightarrow\sigma^z (\tX-\tX^\dagger) ,\quad I(\tZ-\tZ^\dagger) \leftrightarrow\sigma^z (\tZ-\tZ^\dagger) ~.
\end{aligned}
\end{equation}
We then apply another unitary transformation $\prod_i\mathrm{CZ}_{i,i-1/2}\prod_i\mathrm{CZ}_{i,i+1/2}$ to get $\mu_{i-1/2}^z  \sigma^x_i \mu_{i+1/2}^z \rightarrow \sigma_i^x$. The combined transformation then maps \eqref{eq:eba} to 
\begin{equation}
\begin{aligned}
    & \langle \mu^z_i \sigma^x_i \tC_i\mu_{i+1}^z,\quad \mu_{i+1/2}^x, \quad(\tX_i+\tX_i^\dagger),\quad (\tZ_i^{\mu_{i+1/2}^x} \tZ_{i+1}^\dagger + \tZ_i^{-\mu_{i+1/2}^x} \tZ_{i+1}), \\
    & \quad \quad \quad \quad \quad \quad \quad \quad \quad \quad  \quad \quad \quad \quad \quad  \quad \quad \quad \quad \quad  (\tX_i-\tX_i^\dagger),\quad (\tZ_i\tZ_{i+1}^{-\mu^x_{i+1/2}}-\tZ_i^\dagger\tZ_{i+1}^{\mu^x_{i+1/2}})\rangle ~.
\end{aligned}
\end{equation}
Finally, we shift the link $i + \frac{1}{2} \mapsto i + 1$ and project to the $\sigma_i^x=1$ physical subspace, and obtain the gauged bond algebra:
\begin{equation}\label{eq:gba}
\begin{aligned}
    & \langle \mu^z_i \tC_i\mu_{i+1}^z,\quad \mu_{i+1}^x, \quad(\tX_i+\tX_i^\dagger),\quad (\tZ_i^{\mu_{i+1}^x} \tZ_{i+1}^\dagger + \tZ_i^{-\mu_{i+1}^x} \tZ_{i+1}), \\
    & \quad \quad \quad \quad \quad \quad \quad \quad \quad \quad \quad \quad  \quad \quad \quad \quad \quad \quad \quad \quad (\tX_i-\tX_i^\dagger),\quad  (\tZ_i\tZ_{i+1}^{-\mu^x_{i+1}}-\tZ_i^\dagger\tZ_{i+1}^{\mu^x_{i+1}})\rangle ~.
\end{aligned}
\end{equation}

To summarize, to implement the $\mathbb{Z}_2^t$ gauging, one only needs to replace every $U_t$-invariant term in \eqref{eq:S3ba} with the corresponding term in \eqref{eq:gba}, and we will refer to this as the $\mathbb{Z}_2^t$ gauging map. It is straightforward to check that \eqref{eq:TYxz2ham} is mapped to \eqref{eq:nongrpham} under the $\IZ_2^t$ gauging map.

\bibliographystyle{utphys}
\bibliography{main}

\end{document}